\documentclass[preprint,showpacs,preprintnumbers, amsmath,amssymb,superscriptaddress,nofootinbib]{revtex4-1}
\usepackage{graphicx,color}
\usepackage{amsmath,amssymb}
\usepackage{url}
\usepackage{epstopdf}

\newcommand{\hs}{\hspace*{0.5cm}}

\newcommand{\eq}[1]{Eq.~(\ref{#1})}

\newcommand{\be}{\begin{equation}}
\newcommand{\ee}{\end{equation}}
\newcommand{\bea}{\begin{eqnarray}}
\newcommand{\eea}{\end{eqnarray}}
\newcommand{\nn}{\nonumber}
\newcommand{\crn}{\nonumber \\}

\newcommand{\fr}{\frac}
\newcommand{\bc}{\begin{center}}
\newcommand{\ec}{\end{center}}

\newcommand {\ba}{\begin{array}}
\newcommand {\ea}{\end{array}}
\newcommand{\ben}{\begin{enumerate}}
\newcommand{\een}{\end{enumerate}}

\usepackage{epsfig,graphicx}
\usepackage{bm}
\usepackage{dcolumn}

\begin{document}

\title{An explanation of  experimental data of  $(g-2)_{e,\mu}$  in 3-3-1 models with inverse seesaw neutrinos}
\author{L. T. Hue}
\email{lethohue@vlu.edu.vn}
%-
\affiliation{Subatomic Physics Research Group, Science and Technology Advanced Institute, Van Lang University, Ho Chi Minh City 70000, Vietnam}
\affiliation{Faculty of Applied Technology, School of Engineering and Technology, Van Lang University, Ho Chi Minh City 70000, Vietnam}
\author{Khiem Hong Phan} \email{phanhongkhiem@duytan.edu.vn}
\affiliation{Institute of Fundamental and Applied Sciences, 
	Duy Tan University, Ho Chi Minh City 700000, Vietnam}
\affiliation{Faculty of Natural Sciences, Duy Tan University, 
	Da Nang City 550000, Vietnam}
\author{T. Phong Nguyen} \email{thanhphong@ctu.edu.vn }
\affiliation{Department of Physics, Can Tho University, 3/2 Street, Ninh Kieu, Can Tho City 94000, Vietnam}
\author{H. N. Long }\email{hnlong@iop.vast.vn}

\affiliation{Institute of Physics,   Vietnam Academy of Science and Technology, 10 Dao Tan, Ba Dinh, 10000 Hanoi, Vietnam}
%-
\author{H. T. Hung } \email{hathanhhung@hpu2.edu.vn (corresponding author)}
\affiliation{Department of Physics, Hanoi Pedagogical University 2, Phuc Yen, Vinh Phuc 15000, Vietnam}
\begin{abstract}
We show that the  anomalous magnetic moment experimental data of 
muon   and electron $(g-2)_{\mu,e}$ can be explained simultaneously in simple extensions of the 3-3-1  models consisting  of new heavy neutrinos  and a singly charged Higgs boson. The  heavy neutrinos generate  active neutrino masses and mixing through the general seesaw  mechanism.  They also  have non-zero Yukawa couplings with singly charged Higgs bosons and  right-handed charged leptons, which result in  large one-loop contributions known as \emph{chirally-enhanced} ones.  Numerical investigation  confirms a conclusion indicated previously that these contributions are the key point to explain the  large $(g-2)_{\mu,e}$ data, provided that  the inverse seesaw mechanism is necessary to allow both conditions that heavy neutrino masses are above few hundred GeV and  non-unitary part of the active neutrino mixing matrix must be large enough. 
\end{abstract}
%\pacs{{\bf Last updated: \today}
%11.15.Ex  Supersymmetric models,
%12.60.Fr  Extensions of electroweak Higgs sector,
 %13.66.Fg Gauge and Higgs boson production in $e^-\,  e^+$ interactions
 %}
\maketitle
%%%%%%%%%%%%%%%%%%%
\section{Introduction}
\label{sec:intro}
\allowdisplaybreaks
Recently, anomalous magnetic moments (AMM) of charged leptons $a_{e_a} \equiv (g-2)_{e_a}/2$ have been studied widely because the recent experimental data shows  large deviations from the 
Standard Model (SM)  predictions. The recent improved AMM value  of muon $a_{\mu}$ predicted by the SM   is  accepted widely as \cite{Aoyama:2020ynm} $a^{\mathrm{SM}}_{\mu}= 116591810(43)\times 10^{-11}$,  which is derived from the combination of various contributions using   the dispersion approach~\cite{Davier:2010nc, Davier:2017zfy, Keshavarzi:2018mgv, Colangelo:2018mtw, Hoferichter:2019mqg, Davier:2019can, Keshavarzi:2019abf, Kurz:2014wya, Melnikov:2003xd, Masjuan:2017tvw, Colangelo:2017fiz, Hoferichter:2018kwz, Gerardin:2019vio, Bijnens:2019ghy, Colangelo:2019uex, Colangelo:2014qya, Blum:2019ugy, Aoyama:2012wk, Aoyama:2019ryr, Czarnecki:2002nt, Gnendiger:2013pva}.  However, this results is inconsistent with the lattice-QCD calculation \cite{Borsanyi:2020mff}, which  is closer to the  recent experimental data. In our work, we will use the  larger discrepancy between theoretical and experimental results, because it is more interesting for theoretical discussions and the allowed regions of parameter space are still applicable for the smaller deviation reported in Ref. \cite{Borsanyi:2020mff}.  The latest experimental measurement  has been reported from Fermilab~\cite{Abi:2021gix} and  is also in agreement with previous experimental result measured by Brookhaven National Laboratory (BNL) E82~\cite{Muong-2:2006rrc}. A combination of these  results  in  the  new average value of $a^{\mathrm{exp}}_{\mu}=116592061(41)\times 10^{-11}$, which leads to the  improved standard deviation of 4.2 $\sigma$ from the SM prediction, namely 
\be\label{eq_damu}
	\Delta a^{\mathrm{NP}}_{\mu}\equiv  a^{\mathrm{exp}}_{\mu} -a^{\mathrm{SM}}_{\mu} =\left(2.51\pm 0.59 \right) \times 10^{-9}\,.
\ee 

The recent experimental AMM values of  electron $a_e$  were reported from different groups~\cite{Hanneke:2008tm, Parker:2018vye, Morel:2020dww} (for calculation of  $a_e$  in the SM,  see 
   Refs. \cite{Aoyama:2012wj, Aoyama:2012wk, Laporta:2017okg, Aoyama:2017uqe,  Terazawa:2018pdc, Volkov:2019phy,Gerardin:2020gpp}).  In our numerical  discussion,   we adopt the experimental values of $a_e$  corresponding to the  following  standard deviation of $2.5\sigma$  from the SM one: 
\begin{equation}\label{eq_dae}
	\Delta a^{\mathrm{NP}}_{e}\equiv  a^{\mathrm{exp}}_{e} -a^{\mathrm{SM}}_{e} = \left( -8.7\pm 3.6\right) \times 10^{-13}.
\end{equation} 
Many  models beyond the SM (BSM) have been constructed to explain the  experimental data of $(g-2)_{\mu,e}$, such as models adding vector-like lepton multiplets~\cite{Dermisek:2013gta, Crivellin:2018qmi, Escribano:2021css, Hernandez:2021tii, Crivellin:2021rbq, Dermisek:2021ajd, Chun:2020uzw, Frank:2020smf, Endo:2020tkb, Cogollo:2020nrc, Chen:2020tfr, Bharadwaj:2021tgp}, leptoquarks \cite{Crivellin:2020tsz}, both neutral and charged  Higss bosons as $SU(2)_L$ singlets \cite{Mondal:2021vou},  $SU(2)_L$ triplets of leptons and scalars \cite{Arbelaez:2020rbq}.  The minimal supersymmetric standard  model can explain both $(g-2)_{e,\mu}$ data in the regions of the parameter space with light slepton masses below a few hundred GeVs~\cite{Badziak:2019gaf, Li:2021koa}.  Some two Higgs doublet  models (THDM) adding  new $SU(2)_L$ Higgs doublets can give  large  two-loop contributions to $\Delta a_{\mu}$ \cite{Li:2020dbg, DelleRose:2020oaa, Botella:2020xzf, Han:2018znu, Han:2021gfu}, provided  that the masses of  the  new neutral and/or charged Higgs bosons  must be light  at a few hundred GeVs.  

In this work, we will focus on the AMM problems predicted by a BSM  class called as 3-3-1 models, constructed based on the $SU(3)_C\times SU(3)_L\times U(1)_X$ group~\cite{Singer:1980sw, Pisano:1992bxx, Frampton:1992wt, Foot:1992rh, Pleitez:1992xh, Foot:1994ym, Ozer:1995xi, Diaz:2004fs, Buras:2012dp, Hue:2015mna, Fonseca:2016tbn}. It was shown that the  early 3-3-1 versions cannot predict large  $\Delta a_{\mu}$  given by the experimental data \cite{Ky:2000ku, Kelso:2014qka, Binh:2015jfz, DeConto:2016ith, deJesus:2020upp, deJesus:2020ngn, Lindner:2016bgg}. Extended versions were introduced to solve this problem,  such as 3-3-1 models with new vector-like leptons or inert $SU(3)_L$ Higgs triplets ~\cite{deJesus:2020ngn, deJesus:2020upp}, the models with new singly charged Higgs couplings with heavy  neutrinos generating neutrino masses through the inverse seesaw (ISS) mechanism~\cite{Hue:2020wnn, Hue:2021xap}, and 3-3-1 models with discrete symmetries  containing a  rather large number of new particles needed to explain the hierarchy problems of  fermion masses \cite{CarcamoHernandez:2019lhv, CarcamoHernandez:2020pxw}. In  Ref.~\cite{Crivellin:2018qmi},   the very precise analytic formulas applicable to calculate general one-loop contributions to AMM in a wide class of  BSM were presented.  Using these formulas,  we can estimate  again the previous results available in all current 3-3-1 models. These analytic formulas are consistent with those calculated previously for 3-3-1 models \cite{Lavoura:2003xp, Hue:2017lak}. More importantly, we  will show that  the 3-3-1 models can give large one-loop contributions to AMM by adding new $SU(3)_L$ singlets such as   singly charged Higgs bosons $h^\pm$ and heavy neutrinos, a similar way that was applied to  the 3-3-1 model with right-handed (RH) neutrinos  (331RHN). On the other hand,  Higgs triplets  and their Yukawa couplings needed  to generate masses for  charged leptons, quarks, and neutrinos  in many 3-3-1 models may have different features from those in the 331RHN, leading to  new predictions of the allowed regions of parameter space satisfying the AMM experimental data predicted by different 3-3-1 models. Heavy neutrinos are needed to generate active neutrino masses and mixing through the general seesaw (GSS) mechanism, and  Yukawa couplings of singly charged Higgs boson and RH charged leptons.   Hence the new particles result in Yukawa terms like  $N\left( \lambda ^L P_L +\lambda^R P_L\right) e_a H^+$ corresponding to the presence of  the so called \emph{chirally-enhanced} one-loop contributions proportional to $ \lambda ^{L*}\lambda ^{R}$, where $N$ and $H^+$ denote two physical states of  a neutrino and a singly charged Higgs boson.  They are the most important terms that can be large enough to explain the recent AMM data \cite{Crivellin:2018qmi}. Other chirally-enhanced one-loop contributions originating from the 3-3-1 models will be also mentioned.  For convenience, the 3-3-1 models discussed in our work will be generalized in the  form of the 3-3-1 model  with an arbitrary parameter $\beta$ (331$\beta$)  defining the electric charge operator as follows ~\cite{Diaz:2004fs, Buras:2012dp}:
\begin{equation}
Q =  T_3 + \beta T_8 + X.
\label{eq:charge_Q}
\end{equation}
We have introduced the $SU(3)$ generators $T_a$, $a=1,\ldots 8$ and $X$ is  the new quantum charge corresponding to  the group $U(1)_X$. Thus, the charge operator $Q$ depends on two parameters $\beta$ and $X$. The 3-3-1 models  corresponding to different  $\beta$  distinguish  each other by new heavy leptons and quarks  arranged in the third components of  fermion (anti-)triplets,  for example, $\beta=-\sqrt{3}, \; \frac{1}{\sqrt{3}},\;-\frac{1}{\sqrt{3}},$ and $0$   for the minimal 3-3-1 model \cite{Pisano:1992bxx},  the 331RN \cite{Foot:1994ym},  with  heavy singly charged leptons \cite{Ozer:1995xi},  and the simplest 3-3-1 model \cite{Hue:2015mna}, respectively. These models result in  distinguishable consequences for  many  interesting processes \cite{CarcamoHernandez:2005ka, Buras:2014yna, Buras:2016dxz, Buras:2015kwd, Long:2018fud}. Hence,  successful solutions for AMM problems  in 3-3-1 models will guarantee their realities.

 The explanation of AMM data in Refs. \cite{Hue:2020wnn, Hue:2021xap} was just valid for the specific 331RHN model corresponding to $\beta=\pm \frac{1}{\sqrt{3}}$, which  results in a special case that heavy $SU(3)_L$ leptons in the third components of the lepton (anti) triplets are  exotic  heavy neutrinos   $\psi_{aL}\sim (e_{a}, \nu_{a}, N_{a} )^T_L$. They play roles of right-handed neutrinos $N_{aR}\equiv(N_{aL})^c$,  and mix with SM neutrinos through  a very special form of the total antisymmetric $3\times3$ neutrino Dirac mass matrix $m_D$.  As a result,  strict relations between parameters  are necessary  to explain simultaneously all  neutrino oscillation data, AMM $(g-2)_{\mu}$, and constraints of lepton flavor violating  (cLFV)  decays $e_b\to e_a \gamma$   that must be consistent with experiments. In addition, the 331RHN model  needs   three more neutrino singlets for generating the ISS neutrino mass matrix (331ISS), and a singly charged scalar singlet to give large one-loop contributions to AMM of muon, while  the destructive interference among different one-loop contributions gives a small total one loop contribution to every decay amplitude  $e_b\to e_a \gamma$.

The models 331$\beta$ and 331ISS need three $SU(3)_L$ Higgs triplets for generating non-zero  masses of  all quarks and leptons at the tree level, including  active Dirac and Majorana neutrino masses. Two of these Higgs triplets give masses for SM fermions and gauge bosons, therefore they play the same role as the ones   well-known in the two Higgs doublet models. They contain two neutral Higgs components with two vacuum expectation values (VEVs) denoted as $v_{1,2}$  satisfying  $v_1^2+v_2^2\simeq (246\; \mathrm{GeV})^2$. The important parameter  $t_{\beta}\equiv v_2/v_1$, where $v_2$ is always chosen to generate the top quark mass, must have a lower bound $t_{\beta}\ge 0.33$ from the perturbative limit of the top quark Yukawa coupling. In the  331ISS model, charged lepton masses  and the neutrino Dirac mass term are originated from $v_1$, therefore  large $t_{\beta}>30$ is the necessary condition to give large one-loop contributions to AMM of muon consistent with experimental data \cite{Hue:2021xap}. In contrast,  the neutrino Dirac mass term comes from the Yukawa couplings of the neutral Higgs component with VEV $v_2$ in the  331$\beta$ model. The same property also happens for the Yukawa couplings between leptons and the  singly charged Higgs bosons.  In addition, no $SU(3)_L$ neutral leptons are available, therefore the  331$\beta$ model needs six exotic neutrino singlets for the ISS mechanism.   Therefore, the allowed regions of the parameter space give large one-loop contributions to $(g-2)_{e,\mu}$ cannot  be generalized qualitatively from previous results given in Ref. \cite{Hue:2021xap}, which  requires  large $t_{\beta}$.   A first derivation may start from the most important property that  the Yukawa couplings of singly charged Higgs singlet  and $m_D$ relate to the Higgs triplet containing VEV $v_2$ instead of $v_1$. Therefore, the proper values of $t_{\beta}$  may be small enough to explain the experimental AMM data, leading to the existence of an upper bound for  $t_{\beta}$.  This may be conflict with the perturbative limit $t_{\beta}\ge 0.33$. This problem will be addressed in this work.

Before coming to detailed analysis, we emphasize our works is helpful because of the following reasons.  First, we will see that  the $331\beta$ model considered in this work  explain simultaneously both experimental data of $(g-2)_{e,\mu}$  only when the mixing  between $h^\pm$  and    singly charged components of the $SU(3)_L$ Higgs triplets is non-zero. This non-zero mixing  implies the existence of  a non-trivial coupling of  $h^\pm$ with two $SU(3)_L$ Higgs  triplets, $f_h \left( \rho^\dagger \eta h^+ +\mathrm{h.c.}\right)$, which was not introduced previously. This may be an indirect link between  $SU(3)_L$ Higgs triplets and the ISS  neutral lepton singlets, apart from the  small ISS mixing among $X_{aR}$ and $\nu_{bL}$. In other words,  the existence of the $SU(3)_L$ Higgs triplet components can be detected through their decays to leptons.    The second reason, many previous discussions on 3-3-1 models showed clearly that they did not accommodate the $(g-2)_{\mu}$ data unless adding some other new particles, such as vector-like fermions,\dots.  Our qualitative estimation in this work provides another interesting approach to confirm this conclusion. Finally, the original appearance of the 3-3-1 models solved some interesting questions, such as the answer to the question of three fermion flavors confirmed by experiments, .... New improved versions of 3-3-1 models have been introduced to explain successfully the latest experimental results.  Our model is one of them  constructed to explain dark matter data, the hierarchies problems of fermion masses,...Many of them contain complicated particle contents including new singly charged Higgs scalars and neutral leptons. Our discussion on AMM data with a very simple Higgs sector will be helpful for further realization solutions for AMM data in these models.
 
 Our work is arranged as follows.  In Sec.~\ref{sec_331beta}, the  331$\beta$ model will be reviewed, where we pay attention to the leptons, gauge bosons, and Higgs sectors, giving all physical states as well as the couplings that may give large one-loop contributions to AMM. In Sec. \ref{sec_SS}, the $331\beta$ model with the GSS will be presented along with the two particular frameworks of the minimal seesaw (MSS) and  simple ISS. In Sec.~\ref{eq_discussion},  analytic formulas for one-loop contributions to AMM are constructed. Numerical discussions for both MSS and ISS will be shown in detail.  Our main results are collected in Sec.~\ref{eq_conclusion}. There are three appendices listing master functions for  one-loop contributions to AMM given in Ref.~\cite{Crivellin:2018qmi},  analytic formulas for one-loop contributions from the singly charged Higgs bosons to AMM, and a detailed  discussion on the masses and mixing of the singly charged bosons.

%%%
\section{ \label{sec_331beta} The 3-3-1 model with arbitrary $\beta$}
Let us review the $331\beta$ model. Left-handed leptons are assigned to anti-triplets and RH leptons are  singlets: 
\bea && L'_{aL}=\left(
       \begin{array}{c}
         e'_a \\
         -\nu'_{a} \\
         E'_a \\
       \end{array}
     \right)_L \sim \left(3^*~, -\frac{1}{2}+\frac{\beta}{2\sqrt{3}}\right), \hs a=1,2,3,\crn
     && e'_{aR}\sim   \left(1~, -1\right)  , \hs X_{IR}\sim  \left(~1~, 0\right) ,\hs E'_{aR} \sim   \left(~1~, -\frac{1}{2}+\frac{\sqrt{3}\beta}{2}\right).  \label{lep}\eea
  The model includes  $K$ RH neutrinos $X_{IR}$, $I=1,2,...,K$,  and three exotic leptons $E'^a_{L,R}$ which are much heavier than the normal leptons. 
  The prime denotes flavor states to be distinguished with mass eigenstates introduced later. 
  The numbers in the parentheses are to label the representation of $SU(3)_L\otimes U(1)_X$ group. 
  The quark sector is ignored here because it is irrelevant to our present work.  We note that our result will be true for 3-3-1 models consisting left-handed lepton triplets because they are equivalent with the models with lepton sector defined in Eq.~\eqref{lep} through a transformation keeping physical results unchanged \cite{Descotes-Genon:2017ptp, Hue:2018dqf}.

The model has  nine  electroweak gauge bosons, included in the following covariant derivative
 \be  D_{\mu}\equiv \partial_{\mu}-i g T^a W^a_{\mu}-i g_X X T^9X_{\mu},  \label{coderivative1}\ee
where $T^9=1/\sqrt{6}$, $g$ and $g_X$ are gauge couplings of  the two groups $SU(3)_L$ and $U(1)_X$, respectively. 
The matrix $W^aT^a$, where $T^a =\lambda_a/2$ corresponding to a triplet representation, is  written as
 \bea W^a_{\mu}T^a=\frac{1}{2}\left(
                     \begin{array}{ccc}
                       W^3_{\mu}+\frac{1}{\sqrt{3}} W^8_{\mu}& \sqrt{2}W^+_{\mu} &  \sqrt{2}Y^{A}_{\mu} \\
                        \sqrt{2}W^-_{\mu} &  -W^3_{\mu}+\frac{1}{\sqrt{3}} W^8_{\mu} & \sqrt{2}V^{B}_{\mu} \\
                       \sqrt{2}Y^{-A}_{\mu}& \sqrt{2}V^{-B}_{\mu} &-\frac{2}{\sqrt{3}} W^8_{\mu}\\
                     \end{array}
                   \right),
  \label{wata}\eea
where we have defined the mass eigenstates of the charged gauge bosons as
\bea W^{\pm}_{\mu}=\frac{1}{\sqrt{2}}\left( W^1_{\mu}\mp i W^2_{\mu}\right),\crn
Y^{\pm A}_{\mu}=\frac{1}{\sqrt{2}}\left( W^4_{\mu}\mp i W^5_{\mu}\right),\crn
V^{\pm B}_{\mu}=\frac{1}{\sqrt{2}}\left( W^6_{\mu}\mp i W^7_{\mu}\right).
   \label{gbos}\eea 
From \eq{eq:charge_Q}, the electric charges of the gauge bosons are calculated as 
\begin{equation}
	A=\fr{1}{2}+\fr{\sqrt{3}\beta}{2}, \quad 
	B=-\fr{1}{2}+\fr{\sqrt{3}\beta}{2}.\label{eq_chargeAB}
\end{equation}
To generate masses for gauge bosons and fermions, the model has three scalar triplets 
defined as
  \bea && \chi=\left(
              \begin{array}{c}
                \chi_{A} \\
                \chi_{B} \\
                \chi^0 \\
              \end{array}
            \right)\sim \left(3~, \frac{\beta}{\sqrt{3}}\right), \hs  \rho=\left(
              \begin{array}{c}
                \rho^+ \\
                \rho^0 \\
                \rho^{-B} \\
              \end{array}
            \right)\sim \left(3~, \frac{1}{2}-\frac{\beta}{2\sqrt{3}}\right)\crn
  && \eta=\left(
              \begin{array}{c}
                \eta^0 \\
                \eta^- \\
                \eta^{-A} \\
              \end{array}
            \right)\sim \left(3~, -\frac{1}{2}-\frac{\beta}{2\sqrt{3}}\right), \; h^+\sim (1,1,1).
    \label{higgsc}
  \eea
where $A,B$ denote electric charges as defined in \eq{eq_chargeAB}; and $h^+$ is a new singly charged Higgs boson needed for giving large one-loop contributions  to AMM.  
These  Higgs bosons   develop the following  non-zero VEVs  
\begin{align}
	\langle  \chi^0\rangle = \frac{u}{\sqrt{2}},\; \langle  \rho^0 \rangle= \frac{v_2}{\sqrt{2}}, \;  \langle   \eta^0 \rangle= \frac{v_1}{\sqrt{2}}.  \label{vevhigg1}
\end{align}
 This VEV configuration of the $331\beta$ model without $h^\pm$ was  shown to be valid  in Ref. \cite{Costantini:2020xrn}. This is also true for the $331\beta$ adding new singly charged Higgs boson $h^\pm$ we consider here. For convenience, we will use the following notations:
\begin{align}
	\label{eq_tb}
	t_\beta&\equiv \frac{v_2}{v_1},\;  \rightarrow s_{\beta}= \frac{v_2}{v},\; c_{\beta}=\frac{v_1}{v}, 
\end{align}
where $v^2\equiv v_1^2+v_2^2$, and $s^2_{\beta}+c^2_{\beta}=1$. 

 The symmetry breaking happens in two steps:  
$SU(3)_L\otimes U(1)_X\xrightarrow{u} SU(2)_L\otimes U(1)_Y\xrightarrow{v_1,v_2} U(1)_Q$, leading to  
 the condition that  $u\gg v_1,v_2$. After the first step, five gauge bosons will be massive and the remaining four massless ones  
can be identified with the before-symmetry-breaking SM gauge bosons, resulting in  the following important equation: 
\bea
\fr{g_X^2}{g^2} = \fr{6s_W^2}{1-(1+\beta^2)s_W^2}, \; g=g_2,
\label{eq_beta_coupl}
\eea
where the weak mixing angle is defined as $t_W = \tan\theta_W = g_1/g_2$, $g_{1,2}$ are the gauge couplings of  the SM gauge groups $U(1)_Y$ and $SU(2)_L$, respectively.  We  denote $s_W = \sin\theta_W$ and  $c_W = \cos \theta_W$. 
Putting in the value of $s_W$, we get approximately
\bea
|\beta| \le \sqrt{3}.
\label{eq_beta_constraint}
\eea
The masses of the charged gauge bosons are
\bea 
m^2_{Y^{\pm A}} = \frac{g^2}{4}(u^2+v^{2}_1),\quad 
m^2_{V^{\pm B}}=\frac{g^2}{4}(u^2+v^2_2),\quad
m^2_{W^\pm} = \fr{g^2}{4}(v^2_1 + v^{2}_2),
\label{masga}\eea 
where the gauge boson $W^\pm$ is identified with the SM one, implying that
\begin{equation}\label{eq_SMWmass}
	v^2\equiv v_1^2+v_2^2=\frac{4 m_W^2}{g^2} \simeq (246\;\mathrm{ GeV})^2.
\end{equation}

The above Higgs bosons are enough to generate all SM quark masses and heavy new quark masses \cite{Buras:2012dp, Hung:2019jue}. In addition, the Yukawa term $ Y^u_{3a}\overline{Q_{3L}}\rho^*u_{aR}\to  \frac{Y^u_{3a}v_2}{\sqrt{2}} \overline{u_{3L}}u_{aR}$ mainly contributes to the top quark mass, $m_t\simeq \frac{Y^u_{33}v_2}{\sqrt{2}} \leq \sqrt{4\pi} v_2/\sqrt{2}$, equivalently $s_{\beta}\ge \sqrt{2}m_t/(\sqrt{4\pi}v) \to t_{\beta}\ge0.3$.

 In general, the mixing between a SM lepton and a new lepton is allowed 
if they have the same electric charge in some particular values of $\beta$.  This 
mixing effect will be neglected in the $331\beta$ model under consideration.  The Yukawa Lagrangian now is 
\begin{align}
\label{eq_ylepton1}
-\mathcal{L}^\text{yuk}_{\mathrm{lepton}} &=  Y^e_{ab} \overline{e'_{aR}} \eta^TL'_{bL}  + Y^E_{ab} \overline{E'_{aR}} \chi^T L'_{bL} + Y^X_{Ib} \overline{X_{IR}} \rho^TL'_{bL} + \frac{1}{2}M_{N,IJ} \overline{X_{IR}}(X_{JR})^c \crn 
& + Y^{h}_{Ib}\overline{(X_{IR})^c} e'_{bR} h^+ +\mathrm{h.c.}, 	
\end{align}
where $a,b=1,2,3$ are family indices, and $I=1,2,3,...K$ are the number of new neutral lepton singlets.  The perturbative limit of the $Y^{h}$ is important in this work, which should satisfy  $|Y^{h}_{Ia}|<\sqrt{4\pi}$. In fact,   the trust values of $|Y^{h}_{Ia}|$ may be smaller~\cite{Allwicher:2021rtd}.  In  Lagrangian \eqref{eq_ylepton1},  the neutrino Dirac mass matrix comes from the third term including the Higgs triplet $\rho$ which also generates the top quark mass. This important property is different from that given in Ref. \cite{Hue:2021xap}, hence the dependence of  the Dirac mass term, the Yukawa couplings of singly charged Higgs bosons, and the perturbative condition of  the Yukawa couplings with heavy neutrinos $Y^X_{Ib}$  on $t_{\beta}$ will be different between two models 331$\beta$ and 331ISS discussed in Ref. \cite{Hue:2021xap}. In later discussions, we will set $K=3$ and $K=6$ for the respective MSS and ISS mechanisms considered in this work. The corresponding mass terms are:
\begin{align}
	-\mathcal{L}^\text{mass}_{\mathrm{lepton}}&=  \frac{Y^e_{ab}v_1}{\sqrt{2}} \overline{e'_{aR}}e'_{bL} +  \frac{Y^E_{ab}u}{\sqrt{2}} \overline{E'_{aR}} E'_{bL}
	\crn & + \frac{1}{2} \begin{pmatrix}
		\overline{(\nu'_L)^c}& \overline{X_R}  
	\end{pmatrix}
	\mathcal{M}^{\nu} 
	\begin{pmatrix}
\nu'_L		\\
(X_R)^c		
	\end{pmatrix}
	   +\mathrm{h.c.},  	\mathcal{M}^{\nu}  = \begin{pmatrix}
	 0_{3\times 3}& M^T_D \\
	 M_D&  M_N
	   \end{pmatrix}
	 	   \; ,\label{eq_mlepton1}
\end{align}
where $ (M_D)_{Ib}\equiv  M_{D,Ib}= \frac{-Y^X_{Ib} v_2}{\sqrt{2}}$, $\nu'_L=(\nu'_1, \nu'_2, \nu'_3)_L^T$ and $X_R=(X_1,X_2,...,X_K)_R^T$. Note that, here, charged lepton masses and $M_D$ come from different Higgs triplets, while these mass terms discussed in Ref.~\cite{Hue:2021xap} come from the same  Higgs triplet. Therefore, the effects on $\Delta a_{\mu,e}$ relating to the relevant Yukawa couplings in the model under consideration will be different from those discussed in Ref.~\cite{Hue:2021xap}. At present, the active neutrino masses and mixing are still generated from the GSS mechanism.  The total mixing matrix is defined as 
\begin{align}
	\label{eq_Unu}
	U^{\nu T} 	\mathcal{M}^{\nu} U^{\nu } &= \hat{\mathcal{M}}^{\nu}=\mathrm{diag}(m_{n_1},\;m_{n_1},\;...,m_{n_{K+3}}),
	\crn 	\begin{pmatrix}
		\nu'_L		\\
		(X_R)^c		
	\end{pmatrix} &= U^{\nu} n_{L},\; 
\begin{pmatrix}
(\nu'_L)^c		\\
X_R		
\end{pmatrix} = U^{\nu*} n_{R}= U^{\nu*} (n_{L})^c,\; 
\end{align} 
where $ n_{L,R}=(n_{1},n_{2},..., n_{(K+3)})_{L,R}$ are Majorana neutrino mass eigenstates satisfying $n_{iL,R}=(n_{iR,L})^c$, and the four-component forms are  $n_i=( n_{iL}, n_{iR})^T$.

From now on we will work in the basis where the SM charged leptons are in their mass eigenstates, namely $Y^e_{ab} = Y^e_{ab} \delta_{ab}$ and $e'_a= e_a$ in Eqs.~\eqref{eq_ylepton1} and \eqref{eq_mlepton1}. This can always be done without loss of generality. The transformations from the flavor states to mass eigenstates of the heavy lepton $E_a$ are defined as
\be \label{eq_Emass}
	E'_{aL} = V^L_{ab} E_{bL}, \quad E'_{aR} = V^R_{ab} E_{bR}, 
\ee 
where $V^{L,R}$ is a  $3\times 3$ unitary mixing matrix for new charged  leptons.

For the Higgs sector, the ratios between  VEVs are used to define two mixing angles:
\begin{equation} \label{eq_vevang}	
  s^2_{iu}=\sin^2\beta_{iu}= \frac{v_i^2}{u^2+v_i^2}, \; i=1,2.  
\end{equation}
We will also use the following notations  $t_{iu}=s_{iu}/c_{iu}$.
The scalar potential is
\begin{align}
V_{\mathrm{h}}&=\mu_1^2 \eta^{\dagger}\eta+\mu_2^2\rho^{\dagger}\rho+\mu_3^2\chi^{\dagger}\chi
+\lambda_1 \left(\eta^{\dagger}\eta\right)^2
+\lambda_2\left(\rho^{\dagger}\rho\right)^2
+\lambda_3\left(\chi^{\dagger}\chi\right)^2\crn
&+ \lambda_{12}(\eta^{\dagger}\eta)(\rho^{\dagger}\rho)
+\lambda_{13}(\eta^{\dagger}\eta)(\chi^{\dagger}\chi)
+\lambda_{23}(\rho^{\dagger}\rho)(\chi^{\dagger}\chi)\crn
&+\tilde{\lambda}_{12} (\eta^{\dagger}\rho)(\rho^{\dagger}\eta) 
+\tilde{\lambda}_{13} (\eta^{\dagger}\chi)(\chi^{\dagger}\eta)
+\tilde{\lambda}_{23} (\rho^{\dagger}\chi)(\chi^{\dagger}\rho)\crn
&+ \sqrt{2} f\left(\epsilon_{ijk}\eta^i\rho^j\chi^k +\mathrm{h.c.} \right) 
\crn& +\mu_4^2 h^+h^- + f_h \left( \rho^\dagger \eta h^+ +\mathrm{h.c.}\right) + (h^+h^-) \left( \lambda^h_1 \eta^{\dagger}\eta+ \lambda^h_2\rho^{\dagger}\rho +\lambda^h_3 \chi^{\dagger}\chi\right) + \lambda^h_4 \left(h^+h^-\right)^2,\label{eq_hpo1}	
\end{align}
where the last line includes all terms relating to the singly charged Higgs boson that does not appear in the previous versions~\cite{Diaz:2004fs, Buras:2012dp}.  The triple coupling $f_h$ is a very important parameter  controlling the mixing between the singly charged Higgs components of the two $SU(3)_L$ Higgs triplets and the Higgs singlet $h^\pm$.  It is emphasized that the existence of $f_h$ is a very interesting feature of the 331$\beta$ that did not mention previously,  because of the  nontrivial property that  the coupling $\rho^\dagger \eta h^+$ always respects  $U(1)_X$ symmetry for arbitrary $\beta$. 

 As we mentioned above, the VEV configuration considered in this work is the same as that chosen in Refs. \cite{Costantini:2020xrn, DeConto:2015eia}, which was  shown to be consistent with   the  unitarity, perturbativity and bounded-from-below (BFB) constraints. On the other hand, the  exact necessary and sufficient BFB constraints are still difficult to determine  \cite{Faro:2019vcd}. They relate to the copositive  (conditionally positive) conditions of the quartic term of the Higgs potential to guarantee the existence of local minima, as discussed in ref.~\cite{Kannike:2012pe}. Determining which local minimum is the global one defining  the stability of the Higgs potential corresponding to the  VEV structure chosen in this work  is more difficult.   A method introduced in ref. \cite{Maniatis:2006fs} can solve this problem, but it is still difficult to apply to  BSM models with complicated Higgs sectors such as the 3-3-1 models.  The discussions on the VEV structure  mentioned in Refs. \cite{Costantini:2020xrn,  DeConto:2015eia} were not addressed clearly to the vacuum stability issue.  In the model under consideration,  the global minimum corresponding to the VEV structure mentioned above requires more  relations between Higgs couplings.  We hope that the large number of Higgs couplings appearing in the Higgs potential \eqref{eq_hpo1} will allow the existence of these new relations consistent with the available constraints. They should be discussed in more detail when the Higgs phenomenology is focused. It is not our scope in this work, we therefore will not discuss more.

A detailed calculation to derive  masses  and mixing matrix  of the singly charged Higgs bosons is shown in appendix~\ref{app_cHiggs}. From this,  the relations between the mass and flavor eigenstates of singly charged Higgs bosons are 
\begin{align}
	\begin{pmatrix}
		\rho^{\pm}	&  \\
		\eta^{\pm}	&  \\
		h^{\pm}	& 
	\end{pmatrix} &= \begin{pmatrix}
		-s_{\beta}& c_{\alpha} c_{\beta} &  s_{\alpha} c_{\beta}\\
		c_{\beta}& c_{\alpha} s_{\beta} &  s_{\alpha} s_{\beta}\\
		0 & -s_{\alpha}   & c_{\alpha} 
	\end{pmatrix}
	\begin{pmatrix}
		\phi_W^{\pm}	&  \\
		H_1^{\pm}	&  \\
		H_2^{\pm}	& 
	\end{pmatrix}
	\label{scHigg},
\end{align}
where $\phi_W^\pm$ are the Goldstone bosons of $W^\pm$, $H^\pm_{1,2}$ are two physical states  with masses $m_{H^\pm_{1,2}}$, and $\alpha$ is a new mixing parameter defined in Eq. \eqref{eq_alpha}. Normally, $m_{H^\pm_{1,2}}$ and $\alpha$ are functions of the potential couplings. For convenience,   we will consider  $\alpha$, and $m_{H^+_{1,2}}$ as free parameters, while  $\mu_4$, $f$, and $f_h$ are chosen as functions of  all  free ones, namely:
\begin{align}
\label{eq_Higgscouplings}
\mu^2_{4}&= c^2_{\alpha} m^2_{H^\pm_2} + s^2_{\alpha} m^2_{H^\pm_1} -\frac{1}{2}\left( \lambda^h_{1} s^2_{\beta}v^2 + \lambda^h_{2} c^2_{\beta}v^2  +\lambda^h_{3} u^2\right), 
\crn f &=-\frac{c_{\beta } s_{\beta } \left(-\tilde{\lambda}_{12} v^2+2 c_{\alpha }^2 m_{H_1^{\pm}}^2+2 s_{\alpha }^2 m_{H_2^{\pm }}^2\right)}{2 u},
\crn f_h &=- \frac{\sqrt{2} s_{\alpha}c_{\alpha}\left(  m^2_{H^\pm_1} -  m^2_{H^\pm_2} \right)}{v}. 
\end{align}
The case of $s_{\alpha}=0$ or $c_{\alpha}=0$ will return to the decouple limit between $h^{\pm}$ and the $SU(3)_L$ Higgs triplets mentioned in Ref.~\cite{Hue:2021xap}, where this limit is allowed. In contrast,  we will see that $s_{2\alpha}=2s_{\alpha}c_{\alpha} \neq 0$, equivalently, the triple Higgs coupling $f_h\neq0$, is one of the necessary condition to give large one-loop contributions to AMM in the $331\beta$.

The relations between the mass and flavor eigenstates of other charged Higgs bosons are: 
\begin{align}
\left(
\begin{array}{c}
	\eta^{\pm A} \\
	\chi^{\pm A} \\
\end{array}
\right) &= \left(
\begin{array}{cc}
	s_{1u} & c_{1u} \\
-c_{1u} & s_{1u} \\
\end{array}
\right) \left(
\begin{array}{c}
	\phi^{ \pm A}_Y \\
	H^{\pm A} \\
\end{array}
\right), \; 
%\label{qacHigg}\\
%
\left(
\begin{array}{c}
	\rho^{\pm B} \\
	\chi^{\pm B} \\
\end{array}
\right) =\left(
\begin{array}{cc}
	s_{2u} & -c_{2u} \\
	c_{2u} & s_{2u} \\
\end{array}
\right) \left(
\begin{array}{c}
	\phi^{\pm B}_V \\
	H^{\pm B} \\
\end{array}
\right),
\label{qbcHigg}   
\end{align}
   where 
   $\phi_W^\pm$, $\phi^{ \pm A}_Y$ and $\phi^{\pm B}_V$  are the Goldstone bosons of $W^\pm$, $Y^{\pm A}$ and $V^{\pm B}$, respectively. The masses of the charged Higgs bosons $H^{\pm A}, H^{\pm B}$ are
   \bea
     &&m^2_{H^{ A}}=(u^2+v_1^{2})\left(\frac{-fv_2}{v_1u}+\frac{1}{2}\tilde{\lambda}_{13}\right),
     %\crn&&
  \;  m^2_{H^{B}}=(u^2+v^2_2)\left(\frac{-fv_1}{uv_2}+\frac{1}{2}\tilde{\lambda}_{23}\right).\label{charge_Higgs_mass}
   \eea
Because the neutral Higgs bosons couple to charged lepton through the Yukawa couplings of the form $S^0 \overline{e_a} e_a$, which is the same form as that of the  SM-like Higgs boson predicted by the SM, the corresponding one-loop contributions to AMM is very small.  Hence, they will be ignored in our calculation from now on. The discussion on the identification of the SM-like Higgs boson can be found in Ref. \cite{Hung:2019jue}. In total, there are six charged Higgs bosons, one neutral pseudoscalar Higgs and three neutral scalar Higgs bosons. 
%Bosonic particles with electric charges of $\pm B$ do not involve in the present calculation.  

\section{ \label{sec_SS} The minimal seesaw and inverse seesaw mechanisms in the neutral lepton sector}

In this section, we will collect important properties of the MSS and ISS  mechanisms used in our calculation. In the GSS framework,  the neutrino mixing matrix is parameterized in the following form:
\begin{equation} 
U^{\nu}= \left(
\begin{array}{cc}
	\left(	I_3-\frac{1}{2}RR^{\dagger} \right) U_{\mathrm{PMNS}} & RV \\
	-R^\dagger U_{\mathrm{PMNS}} & \left(I_K -\frac{1}{2}R^{\dagger} R\right)V \\
\end{array}
\right)  +\mathcal{O}(R^3), 
\label{eq_Unu0}	
\end{equation}
where    $V$ is a $K\times K$ unitary matrix;   $R$ is a $3\times K$ matrix satisfying $|R_{aI}|<1$ for all $a=1,2,3$, and $I=1,2,...,K$. The $3\times 3$ unitary matrix  $U_{\mathrm{PMNS}}$ is the Pontecorvo-Maki-Nakagawa-Sakata (PMNS) matrix \cite{ParticleDataGroup:2020ssz}.    The GSS relations are 
\begin{align}
	R&\simeq M^{\dagger}_D {M^*_N}^{-1} ,  	\; m_{\nu}\simeq -M^T_DM_N^{-1}M_D =U^*_{\mathrm{PMNS}}\hat{m}_{\nu} U^{\dagger}_{\mathrm{PMNS}} , \crn 
	V^*\hat{M}_NV &\simeq M_{N} + \frac{1}{2}R^TR^*M_N +\frac{1}{2}M_N R^{\dagger}R, 
	\label{eq_RSS}
\end{align}
where $\hat{m}_{\nu}=\mathrm{diag}(m_{n_1},m_{n_1},m_{n_3})$ and  $\hat{M}_N=\mathrm{diag}(m_{n_{4}}, m_{n_5},..., m_{n_{(K+3)}})$  consist  of three active neutrino and  $K$ new heavy neutrino masses, respectively. In many formulas discussed below, we will use the equality that $\left(\hat{m}_{\nu}\right)_{cc}=m_{n_c}$ with $c=1,2,3$ are active neutrino masses. 

The general parameterization of $M_D$ was introduced in Ref. \cite{Casas:2001sr}.  In the limit of the MSS mechanism with $K=3$, we will use the
simplest forms of  $M_N $ and  $M_D\equiv m_D$  as follows  \cite{Arganda:2014dta, Ibarra:2010xw}, 
\begin{equation}
M_N =M_0 I_3, \;	M_D\equiv m_D=i  \sqrt{M_0\hat{m}_\nu}  U^{\dagger}_{\mathrm{PMNS}}.\label{eq_mDss}
\end{equation}
The relations in \eqref{eq_RSS} reduce to the following simple form:
\begin{equation}
	 R = -iU_{\mathrm{PMNS}}  \left( \frac{\hat{m}_{\nu}}{M_0}\right)^{1/2}, \quad  
	V\simeq I_3, \quad \hat{M}_N\simeq  M_N, \; m_{n_{4,5,6}}\simeq M_0. \label{masafla1}
\end{equation}
%
%and $U^{\nu}_{ai}\simeq R_{a(i-3)}$ for $i=4,5,6$. 

In the ISS mechanism with $K=6$,  the   total  neutrino mixing matrix $U^\nu$ in \eqref{eq_Unu0} is $9\times9$.  In the 331$\beta$ model, the well-known ISS form of the total neutrino matrix  can be derived  from the requirement that the model respects a global $U(1)_{\mathcal{L}}$ symmetry called the generalized  lepton number, which is  defined from the following formula: $\mathcal{L}$: $ L\equiv -\frac{4}{\sqrt{3}}T^8 +\mathcal{L}$, where $L$ is the normal lepton number
 defined in the SM that $L(\ell)=1$ for all SM leptons $\ell=e,\mu,\tau,\nu_{eL},\nu_{\mu L}, \nu_{\tau L}$ and zero for all other  SM particles including quarks, gauge, and Higgs bosons \cite{Tully:2000kk,Chang:2006aa, CarcamoHernandez:2017cwi}.  The specific $\mathcal{L}$ assignments for all Higgs bosons and fermions in  two 3-3-1 models with right-handed neutrinos and the minimal ones  in Refs. \cite{Tully:2000kk, Chang:2006aa,CarcamoHernandez:2017cwi} are the same and independent with $\beta$, therefore they are valid for the 331$\beta$ model with $\mathcal{L}(L'_{aL})=1/3$, $\mathcal{L}(e'_{aR})=1$.  In addition, introducing  $\mathcal{L}(\rho)=\mathcal{L}(\eta)=2/3$,  $\mathcal{L}(\chi)=-4/3$, $\mathcal{L}(h^+)=0$, and $\mathcal{L}(E'_{aR})=-1$ will result in the Lagrangian \eqref{eq_ylepton1}  conserving $\mathcal{L}$, except  the mass term $\overline{X_{IR}}(X_{JR})^c$, which includes  soft-breaking  terms. Namely,  choosing that $\mathcal{L}(X_{IR})=1$ with $I\leq3$ and $\mathcal{L}(X_{IR})=-1$ with $I>3$,  the conserved Lagrangian \eqref{eq_ylepton1} implies that   $Y^h=\left(\mathcal{O}_{3\times3},\; Y^h_2\right)^T$ and $Y^X=\left(Y^X_1,\;\mathcal{O}_{3\times3}\right)^T$, where $Y^h_2$, and $Y^X_1$ are two $3\times3$ matrices, and  $\mathcal{O}_{3\times3}$ is  the $3\times3$ null matrix.   This leads to the ISS form of $M_D=(m_D,\; \mathcal{O}_{3\times3})^T$. The $6\times6$ Majorojana mass matrix $M_N$ consists of three parts denoted as  $M_R$, $ \mu_{X}$, and $ \mu'_{X}$. The conserved mass term $\left( M_R\right)_{ab}\equiv M_{N,a(b+3)}$ with $a,b\le3$ can be arbitrary large, while the soft-breaking term $ \left( \mu'_{X}\right)_{ab}\equiv M_{N,ab}$ and  $ \left( \mu_{X}\right)_{ab}\equiv M_{N,(a+3)(b+3)}$  should be small. Inserting the ISS form of $M_D$ into the GSS relations to derive the active neutrino mass term $m_{\nu}$, we find that small $\mu'_X$ does not affect significantly the final result.  Hence,  we assume the simplest case of $\mu'_X= \mathcal{O}_{3\times3}$ without loss of generality. 
 
 Now, the Dirac and Majorana mass matrices  have   well-known ISS forms  as follows  \cite{Arganda:2014dta, Ibarra:2010xw}
 \begin{align}
 	M^T_D= (m^T_D,\hs \mathcal{O}_{3\times3}), \hs M_N=\left(
 	\begin{array}{cc}
 		\mathcal{O}_{3\times3}& M_R \\
 		M^T_R & \mu_X \\
 	\end{array}
 	\right). 
 	\label{repara}
 \end{align}
  Defining    $M=M_R\mu_X^{-1}M_R^T$, 
 the ISS relations now are  
  \begin{align}
  	R&= M^{\dagger}_D{M^*_N}^{-1}  = \left(-m_D^{\dagger}M^{*-1},\hs m^\dagger_D\left(M^\dagger_R\right)^{-1} \right),
  	 \crn m_{\nu}&=-M^T_DM_N^{-1}M_D =  m_D^T \left( M_R^T\right)^{-1}\mu_XM_R^{-1}m_D,
  	\crn V^*\hat{M}_NV^{\dagger} &\simeq M_{N} + \frac{1}{2}R^TR^*M_N +\frac{1}{2}M_N R^{\dagger}R.  
  	\label{eq_RISS}
  \end{align}

  In the ISS framework,  $m_D$ is parameterized in terms of many free parameters, hence it is enough to choose  $\mu_X=\mu_X I_3$. The parameter $\mu_X$ is a new scale making the most important difference  between  the neutrino mixing  matrices in the ISS and MSS. We also assume that   $M_R=\hat{M}_R= M_0I_3$.  A simple parameterization of  $m_D$  is $	m_D= \mathrm{diag}(\sqrt{M_{11}},\; \sqrt{M_{22}},\; \sqrt{M_{33}})\sqrt{\hat{m}_\nu}  U^{\dagger}_{\mathrm{PMNS}}$ \cite{Arganda:2014dta}, which is completely different from the total antisymmetric  $m_D$ given in Ref. \cite{Hue:2021xap}.  
 The ISS condition $|\hat{m}_{\nu}|\ll |\mu_X|\ll |m_D|\ll M_0$  gives $\frac{\sqrt{\mu_X \hat{m}_{\nu}}}{M_0}\simeq0$. Then we have  
 \begin{align}
 \hat{M}_N= \left(\begin{matrix}
 	\hat{M}_R	& \mathcal{O}_{3\times3} \\ 
 	\mathcal{O}_{3\times3}	& \hat{M}_R
 \end{matrix} \right)\simeq  M_0I_6 , 
 \;  V\simeq \dfrac{1}{\sqrt{2}}
 \left(\begin{matrix}
 	-iI_3 	& I_3   \\ 
 	iI_3 	& I_3 
 \end{matrix} \right). \label{eq_UNiss}	
 \end{align}
 The important results for the ISS mechanism are:
 \begin{align}
 	\label{eq_mDRISS}
 	 m_D&= M_0\hat{x}_\nu^{1/2}  U^\dagger_{\mathrm{PMNS}},
 	\crn R &=\left( -U_{\mathrm{PMNS}}\frac{\sqrt{\mu_X \hat{m}_{\nu}}}{M_0},\;  U_{\mathrm{PMNS}}\hat{x}_\nu^{1/2} \right) \simeq \left(\mathcal{O}_{3\times3},\;  U_{\mathrm{PMNS}}\hat{x}_\nu^{1/2} \right),
 \end{align}
where $\; \hat{x}_\nu\equiv \frac{\hat{m}_\nu}{\mu_X}$ satisfying max$[\left( |\hat{x}_\nu|\right)_{ab}]\ll1$ for all $a,b=1,2,3$.

In numerical discussion, we will use the best-fit values  of the neutrino oscillation data~\cite{ParticleDataGroup:2020ssz} corresponding to  the normal order (NO) scheme  with $m_{n_1}<m_{n_2}<m_{n_3}$, namely 
\begin{align}
	\label{eq_d2mijNO}
	&s^2_{12}=0.32,\;   s^2_{23}= 0.547,\; s^2_{13}= 0.0216 ,\;  \delta= 218 \;[\mathrm{Deg}] , 
	\crn &\Delta m^2_{21}=7.55\times 10^{-5} [\mathrm{eV}^2], \;
 \Delta m^2_{32}=2.424\times 10^{-3} [\mathrm{eV}^2].
\end{align}
In numerical calculation,   we  will use the following formulas 
\begin{align}
\label{eq_NO1}
\hat{m}_{\nu}&=  \left( \hat{m}^2_{\nu}\right)^{1/2}= \mathrm{diag} \left( m_{n_1}, \; \sqrt{ m_{n_1}^2 +\Delta m^2_{21}},\; \sqrt{m_{n_1}^2 +\Delta m^2_{21} +\Delta m^2_{32}} \right),
%Corrected the UPMNS 04Feb2022
\crn U_{\mathrm{PMNS}} &=\left(
\begin{array}{ccc}
	c_{12} c_{13} & c_{13} s_{12} & s_{13} e^{-i \delta } \\
	-c_{23} s_{12}-c_{12} s_{13} s_{23} e^{i \delta } & c_{13} c_{23}-s_{12} s_{13} s_{23} e^{i \delta } & c_{13} s_{23} \\
	s_{12} s_{23}-c_{12} c_{23} s_{13} e^{i \delta } & -c_{23} s_{12} e^{i \delta } s_{13}-c_{13} s_{23} & c_{13} c_{23} \\
\end{array}
\right)
\crn&\simeq \left(
\begin{array}{ccc}
	0.816 & 0.560 & 0.147 e^{-i \delta } \\
	-0.381-0.09 e^{ i \delta } & 0.555-0.062 e^{ i \delta } & 0.732 \\
	0.418-0.082 e^{ i \delta } & -0.61-0.056 e^{ i \delta } & 0.666 \\
\end{array}
\right).
\end{align}
These neutrino masses satisfy the constraint  from Plank 2018 \cite{Planck:2018vyg} that   $\sum_{i=1}^{3}m_{n_a}\leq 0.12\; \mathrm{eV}$. With the best-fit values of $\Delta m^2_{ij}$ we have $ m_{n_1} \leq 0.028$ eV.

The other well-known numerical parameters are given in Ref.~\cite{ParticleDataGroup:2020ssz}, namely 
\begin{align}
	\label{eq_ex}
	g &=0.652,\; \alpha_e=\frac{1}{137}= \frac{e^2}{4\pi} ,\; s^2_{W}=0.231,\crn
	m_e&=5\times 10^{-4} \;\mathrm{GeV},\; m_{\mu}=0.105 \;\mathrm{GeV}, \; m_W=80.385 \; \mathrm{GeV}. 
\end{align}
Also, the inverted order (IO)  scheme with $m_{n_3}<m_{n_1}<m_{n_2}$ can be considered a similar way, but the qualitative results are the same as those from the NO scheme, so we will not present here.

The non-unitary part of the active neutrino mixing matrix $\left(	I_3-\frac{1}{2}RR^{\dagger} \right) U_{\mathrm{PMNS}}$ is constrained by other phenomenology such as electroweak precision, lepton flavor violating  decays of charged leptons (cLFV)~\cite{Fernandez-Martinez:2016lgt,  Pinheiro:2021mps, Agostinho:2017wfs},  namely 
\begin{align} \label{eq_maxRRd}
\eta\equiv	\frac{1}{2}\left| RR^{\dagger}\right|<
	\begin{pmatrix}
	2\times 10^{-3}	& 3.5\times 10^{-5}  &  8.\times 10^{-3}\\
		3.5\times 10^{-5}&8\times 10^{-4}  & 5.1\times 10^{-3} \\
		8\times 10^{-3}& 5.1\times 10^{-3} & 2.7\times 10^{-3} 
	\end{pmatrix}.
\end{align}
 This constraint is consistent with the data used popularly in recent works  \cite{Dao:2021vqp,Mondal:2021vou}.  The constraint on $\eta$ may be more strict, depending on  particular models. For example in the type III general  and inverse seesaw models, $|\eta_{aa}|\leq \mathrm{O}(10^{-4})$~\cite{Biggio:2019eeo, Escribano:2021css}.  We will choose the values that  $|\eta_{33}|\leq 10^{-3}$ in our numerical discussion.

 In the next section, we will consider the one-loop contributions to $\Delta a_{\mu,e}$. 

 \section{ \label{eq_discussion} Analytical formulas for AMM  and numerical discussion}
\label{analytical_results}
From the above information we  obtain all vertices giving one-loop contributions to  $e_b \to e_a \gamma$ decay rates and $a_{e_a}$.  They are collected from Lagrangian \eqref{eq_ylepton1}. All relevant couplings  are listed in  the following Lagrangian
\begin{align}
	\label{eq_g2Lagrangian}
	%22 Aug: change sign H^\pm_k
	\mathcal{L}&= \frac{g}{\sqrt{2} m_W}\sum_{k=1}^2\sum_{a=1}^3 \sum_{i=1}^{K+3} \overline{n_i} \left[ 	\lambda^{L,k}_{ia}  P_L +\lambda^{R,k}_{ia}P_R \right] e_a H^+_k   
	\crn &-\frac{g}{\sqrt{2} m_Y} \sum_{a,c=1}^3 V^{L*}_{ac}\overline{E_c} \left[ \frac{m_{e_a}}{t_{1u}} P_L+m_{E_c}t_{1u}P_R \right] e_a H^{A}\crn
	&+\sum_{a=1}^3\sum_{i=1}^{K+3}\frac{g}{\sqrt{2}} U^{\nu*}_{ai} \overline{n_{i}}\gamma^{\mu}P_L e_aW^+_{\mu}   +\sum_{a,c=1}^3 \frac{g}{\sqrt{2}} V^{L*}_{ac} \overline{E_c}\gamma^{\mu}P_L e_aY^{A}_{\mu} +\mathrm{h.c.},
\end{align}
where
\begin{align}
	\label{eq_lakLR}
	\lambda^{L,1}_{ia}&= \sum_{I=1}^KM_{D,Ia}t^{-1}_{\beta}c_{\alpha}U^{\nu}_{(I+3)i}\simeq t^{-1}_{\beta}c_{\alpha}\times \left[\begin{array}{cc}
		-\left( M^T_{D}R^{\dagger}U_{\mathrm{PMNS}}\right) _{ai},	& \quad i\leq 3 \\
	\left( M^T_{D} \left(I_K- \frac{1}{2}R^{\dagger}R\right)V\right) _{a(i-3)},	& \quad i > 3 
	\end{array}\right.
, 
\crn	\lambda^{L,2}_{ia} &\simeq	\lambda^{L,1}_{ia}t_{\alpha},
	\crn	\lambda^{R,1}_{ia}&= m_{e_a}t_{\beta} c_{\alpha} U^{\nu*}_{ai} + \sum_{I=1}^K \frac{v}{\sqrt{2}} Y^{h}_{Ia} s_{\alpha}U^{\nu*}_{(I+3)i}
	\crn&\simeq  \left[\begin{array}{cc}
		m_{e_a}t_{\beta} c_{\alpha} \left(\left( I_3 -\frac{1}{2}R^*R^T \right) U^*_{\mathrm{PMNS}}\right)_{ai} -\frac{vs_{\alpha}}{\sqrt{2}}\left( Y^{hT}R^{T}U^*_{\mathrm{PMNS}}\right) _{ai},	& \quad i\leq 3 \\
	m_{e_a}t_{\beta} c_{\alpha} \left( RV\right)^*_{a(i-3)}+	\frac{vs_{\alpha}}{\sqrt{2}}\left( Y^{hT} \left(I_K- \frac{1}{2}R^{T}R^*\right)V^*\right) _{a(i-3)}	& \quad i > 3 
	\end{array}\right.,
	%3,Sep,2021 needed checked 
	\crn	\lambda^{R,2}_{ia}&= m_{e_a}t_{\beta} s_{\alpha} U^{\nu*}_{ai} - \sum_{I=1}^K \frac{v}{\sqrt{2}} Y^{h}_{Ia} c_{\alpha}U^{\nu*}_{(I+3)i}
	\crn&\simeq   \left[\begin{array}{cc}
		m_{e_a}t_{\beta} s_{\alpha} \left(\left( I_3 -\frac{1}{2}R^*R^T \right) U^*_{\mathrm{PMNS}}\right)_{ai} 	-\frac{vc_{\alpha}}{\sqrt{2}}\left( Y^{hT}R^{T}U^*_{\mathrm{PMNS}}\right) _{ai},	& \quad i\leq 3 \\
		m_{e_a}t_{\beta} s_{\alpha} \left( RV\right)^*_{a(i-3)}-	\frac{vc_{\alpha}}{\sqrt{2}}\left( Y^{hT} \left(I_K- \frac{1}{2}R^{T}R^*\right)V^*\right) _{a(i-3)}	& \quad i > 3 
	\end{array}\right.. 
\end{align}
We can see that  $\lambda^{L,k}_{ia}$ with $k=1,2$ contains a factor $t_{\beta}^{-1}$, which is the inverse value included in $\lambda^{L,1}_{ia}$ introduced in Refs. \cite{Hue:2020wnn, Hue:2021xap}, where the regions predicting large $(g-2)_{\mu}$  require large  $t_{\beta}>40$, consistent with the  perturbative constraint $t_{\beta}\geq 0.3$. In contrast, the 331$\beta$ model may support small $t_{\beta}$ for large $(g-2)_{e,\mu}$, which may be excluded if $t_{\beta}<0.3$ is required.  Hence,  the valid regions satisfying the experimental AMM data must be determined through detailed numerical investigation. 

We do not list here the couplings of neutral gauge and Higgs bosons because they give suppressed contributions to  $a^{\mathrm{NP}}_{e_a}$. In particularly, the relevant couplings are only with usual charged leptons $s^0\overline{e_a}e_a$ and $V^0_{\mu}\overline{e_a}\gamma^{\mu}e_a$. The  one-loop contribution from $V_0=Z$ is the same as that predicted by the SM.  Another one  from heavy neutral gauge boson $V_0=Z'$  is  suppressed by a factor of $m^2_{Z}/m^2_{Z'}$. The contributions from neutral Higgs bosons are not larger than the one from the SM-like Higgs boson with a suppressed order of $\mathcal{O}(10^{-14})$. 

 The form factors $c^X_{(ab)R}$ relating to new one-loop contributions from  exchanging $X$ boson to the $\Delta a_{e_a}$ and cLFV decays were introduced in Ref.~\cite{Crivellin:2018qmi}, see appendix \ref{app_CLR}.  Formulas of   $c^X_{(ab)R}$  from  $X=H^A,W^\pm, Y^{\pm A}$ are:
\begin{align}
	c^{H^A}_{(ab)R}&= \frac{e g^2 m_{e_a}}{32\pi^2  m_Y^2 m^2_{H^A}} \sum_{c=1}^3 V^{L}_{ac} V^{L*}_{bc} \left\{ \frac{}{} m^2_{E_c} \left[  f_{\Phi} \left( t_{H,c}\right) + B  g_\Phi  \left( t_{H,c}\right) \right] \right.  
	\crn  &+  \left.  \left[ m^2_{e_b}   t^{-2}_{1u} +  m^2_{E_c}t^2_{1u} \right] \left[  \tilde{f}_\Phi \left( t_{H,c}\right) + B \tilde{g}_\Phi \left( t_{H,c}\right) \right] \right\}, 	\label{eq_cbaHA} \\ 
	c^W_{(ab)R}&\equiv \frac{e g^2m_{e_b} }{32 \pi^2m^2_{W} }\sum_{i=1}^{K+3} U^{\nu}_{ai}U^{\nu*}_{bi}   \tilde{f}_V \left( t_{W,i}\right) ,  	\label{eq_cbaW}\\
	c^Y_{(ab)R}&\equiv \frac{e g^2m_{e_b} }{32 \pi^2 m^2_{Y}} \sum_{c=1}^3 V^{L}_{ac}V^{L*}_{bc}  \left[ \tilde{f}_V \left( t_{Y,c}\right) +B \tilde{g}_V \left( t_{Y,c}\right) \right]  ,  	\label{eq_cbaY}
\end{align} 
where $t_{H,c}\equiv m^2_{E_c}/m^2_{H^A}$, $t_{W,i}\equiv m^2_{n_i}/m^2_{W}$, and  $t_{Y,c}\equiv m^2_{E_c}/m^2_{Y}$. 
 
The particular parameterisations of the MSS and ISS used in this work give   the limit $m_{n_i}=0$ with $i=1,2,3$; $m_{n_i}=M_0$ with all $i>3$;  $c^X_{(ab)R}=0$ with $a\neq b$  and  $X=H^\pm_{1,2},H^A,W,Y$.  To avoid large cLFV rates, we also consider the simple limit that $M_{ab}=M_0\delta_{ab}$, $m_{E_1}=m_{E_2}=m_{E_3}\equiv m_E$, and $V^L=I_3$, so that $c^{H^A}_{(ab)R}=0$ and $	c^Y_{(ab)R}=0$ for $a\neq b$.  Therefore, the cLFV decay rates are much smaller than the current experimental constraints~\cite{MEG:2016leq, BaBar:2009hkt}. We will not discuss them from now on. 

The one-loop contribution of  $X=H^\pm_{1,2},H^A,W,Y$ to AMM of a charged lepton $e_a$ is 
\begin{align}
	a_{e_a}(X)=- \frac{4m_{e_a}}{e}\mathrm{Re} \left[ 	c^X_{(aa)R}\right]. 
\end{align}
And the deviation from the SM is defined as follows:
\begin{align}
	\Delta a_{e_a}= \sum_{X}  a_{e_a}(X) +  \Delta a_{e_a}(W), \;  \Delta a_{e_a}(W)\equiv a_{e_a}(W)-a^{(1)\mathrm{SM}}_{e_a}(W), 
\end{align}
where $X=H^\pm_{1,2},H^A,Y$, and $a^{(1)\mathrm{SM}}_{\mu}(W) \simeq 3.83 \times 10^{-9}$~\cite{Jegerlehner:2009ry}.   In the $331\beta$ model, the SM-like  Higgs and gauge bosons have the same couplings with   usual  charged lepton $e_a$ as those predicted by the SM, hence they do not contribute to $\Delta a_{e_a}$. Also, the heavy neutral Higgs and gauge  bosons will give  one-loop  contributions   smaller than the ones of the SM-like gauge and Higgs bosons by  suppressed factors of $m^2_h/m^2_{H^0}<10^{-1}$ and $m^2_Z/m^2_{Z'}< 6.\times 10^{-4}$. We have used heavy neutral  Higgs mass $m_{H^0}>1$ TeV, and $m_{Z'}>3.7$ TeV from the constraints concerned for 3-3-1 models from LHC \cite{Coutinho:2013lta, Salazar:2015gxa, Nepomuceno:2019eaz} and the combination of weak charge data of Cesium and proton \cite{Long:2018fud}.  

One-loop contributions from heavy charged lepton $E_{a} $  exchanges  are 
\begin{align}
	\Delta a_{\mu}(H^A)& \simeq  - \frac{e g^2 m^2_{\mu}}{8\pi^2  m_W^2} \times  
	\left\{ \frac{m_W^2}{m^2_Y}\left[ t_{H^A}f_{\Phi} \left( t_{H^A}\right) + B  t_{H^A} g_\Phi  \left( t_{H^A}\right) \right]
	\right. \crn& \left. 
	+  \left( \frac{m^2_{\mu}}{m^2_{E} c^2_{\beta}}    +    \frac{m_W^4c^2_{\beta}}{m_Y^4}\right)  \left[  t_{H^A} \tilde{f}_\Phi \left(t_{H^A}\right) + B t_{H^A} \tilde{g}_\Phi \left( t_{H^A}\right) \right] \right\}, 	\label{eq_amuHA} \\
	\Delta a_{\mu}(Y) &\simeq  - \frac{e g^2 m^2_{\mu}}{8\pi^2  m_W^2}  \times  \frac{m_W^2 }{ m^2_{Y}} \left[ \tilde{f}_V \left( t_{Y}\right) + B \tilde{g}_V \left( t_{Y}\right)\right],
	 	\label{eq_anuY}
\end{align}
where $t_{Y}=m^2_E/m^2_Y$, $t_{H^A}=m^2_E/m^2_{H^A}$. 
The above formulas are independent from both MSS and ISS mechanisms affecting only the one-loop contributions from  singly charged Higgs bosons. $	\Delta a_{\mu}(H^A)$ has a chirally-enhanced term  but contains a suppressed factor $m^2_W/m^2_Y$.  

Firstly, we will show that the one-loop contribution from  $W^\pm $ is always close to the SM prediction.   Using the approximation that  $t_{W,i}=0$ with $i\le 3$ and $t_{W_i}=x_W=m^2_{n_i}/m_W^2$ with $i>3$,  we have  
\begin{align}
	\label{eq_cbaW1}
	c^W_{(aa)R}&= \frac{e g^2m_{e_a} }{32 \pi^2m^2_{W} } \left[  \tilde{f}_V \left(0\right) + \left( R^*R^T\right)_{aa} \times \left(\tilde{f}_V \left(x_W\right)-\tilde{f}_V \left(0\right)\right) \right],    
\end{align}
  leading to the  following contribution from $W$ to $ a_{e_a}$ with $\tilde{f}_V \left(0\right)= -5/12$:
\begin{align}
	\label{eq_Deltaaea}
	 a_{e_a}(W)&= -\frac{g^2 m^2_{e_a}}{8\pi^2 m^2_W}  \left[  -\frac{5}{12} +  \left( R^*R^T\right)_{aa} \times \left(\tilde{f}_V \left(x_W\right)+ \frac{5}{12}\right) \right].
\end{align}
Because $ |\tilde{f}_V \left(x_W\right)+ \frac{5}{12}|\leq \frac{5}{12}$ , see the bellow discussion, in the limit $\left( R^*R^T\right)_{aa}\leq 10^{-3}\ll1$ given in \eqref{eq_maxRRd},  $a_{\mu}(W)$ equals to the one-loop contribution predicted by the SM \cite{Jegerlehner:2009ry}: 
\begin{equation}\label{eq_amuSMW}
a^{(1)\mathrm{SM}}_{\mu}(W)  \simeq \frac{g^2 m^2_{\mu}}{8\pi^2 m^2_W}\times \frac{5}{12}\simeq 383\times 10^{-11},\; 	 \frac{g^2 m^2_{\mu}}{8\pi^2 m^2_W}\simeq 9.19\times 10^{-9}.
\end{equation}

Finally,  one-loop contributions from the two singly charged Higgs bosons will be shown precisely in the two  frameworks of MSS and ISS. The analytic formulas were collected in appendix \ref{app_calculation}. Before discussing the total  contributions, we just show here the most important part $a_{0,\mu}(H^\pm)$  which can be large  enough to reach the allowed ranges consistent with $\Delta a^{\mathrm{NP}}_\mu$:
\begin{align}
	\label{eq_aHpmLR}
	a_{\mu}(H^\pm)  =& a_{\mu}(H^\pm_1) + a_{\mu}(H^\pm_2) \equiv  a_{\mu,0}(H^\pm) + \dots,
	\crn a_{\mu,0}(H^\pm) &=-\frac{g^2m_{\mu}\;}{8 \pi^2 m^2_W}  \sum_{k=1}^{2}\sum_{i=3}^{K+3} \left[ \frac{\lambda^{L,k*}_{ia } \lambda^{R,k}_{ia }m_{n_i} f_{\Phi}(x_{i,k})}{m^2_{H^\pm_k} }\right] 
	\crn\simeq & -9.19\times 10^{-9}    \left[  \frac{vt^{-1}_{\beta}c_{\alpha}s_{\alpha}}{\sqrt{2} m_{\mu}}\left(\frac{M_D^{\dagger}V^*V^{\dagger}Y^{h}}{M_0} \right)_{22} \right] 
	%
	%\crn&\times 
	\times \left[ x_1f_{\Phi}(x_{1}) - x_2f_{\Phi}(x_{2}) \right],
	\end{align}
where $x_k\equiv M_0^2/m^2_{H^\pm_k}$.  Note that $a_{\mu}(H^\pm)\neq0$ requires $s_{2\alpha} =2s_{\alpha}c_{\alpha}\neq 0$ and $x_1\neq x_2$. 

\begin{figure}[ht]
	\centering\begin{tabular}{cc}
		\includegraphics[width=7.5cm]{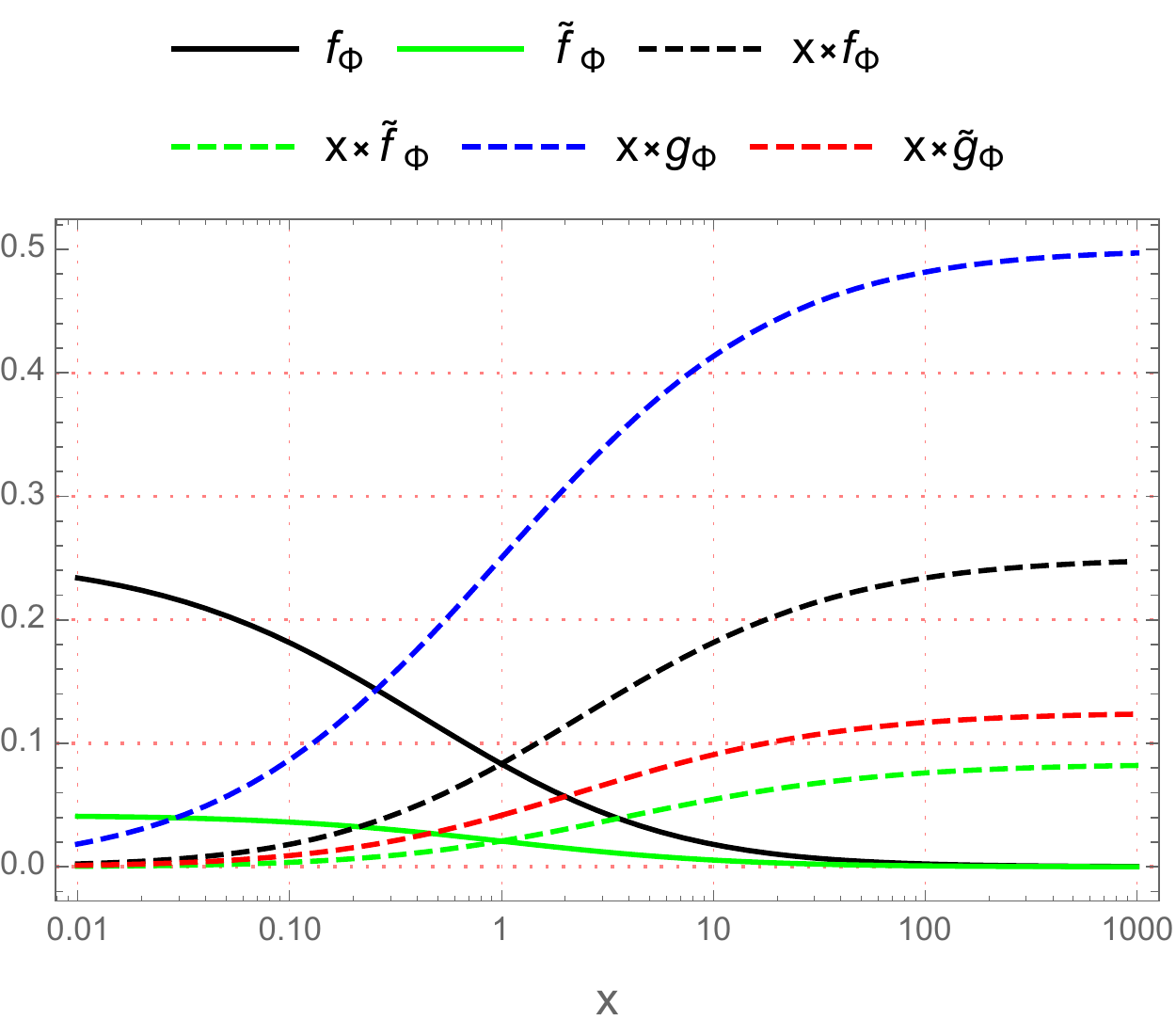}&	\includegraphics[width=7.5cm]{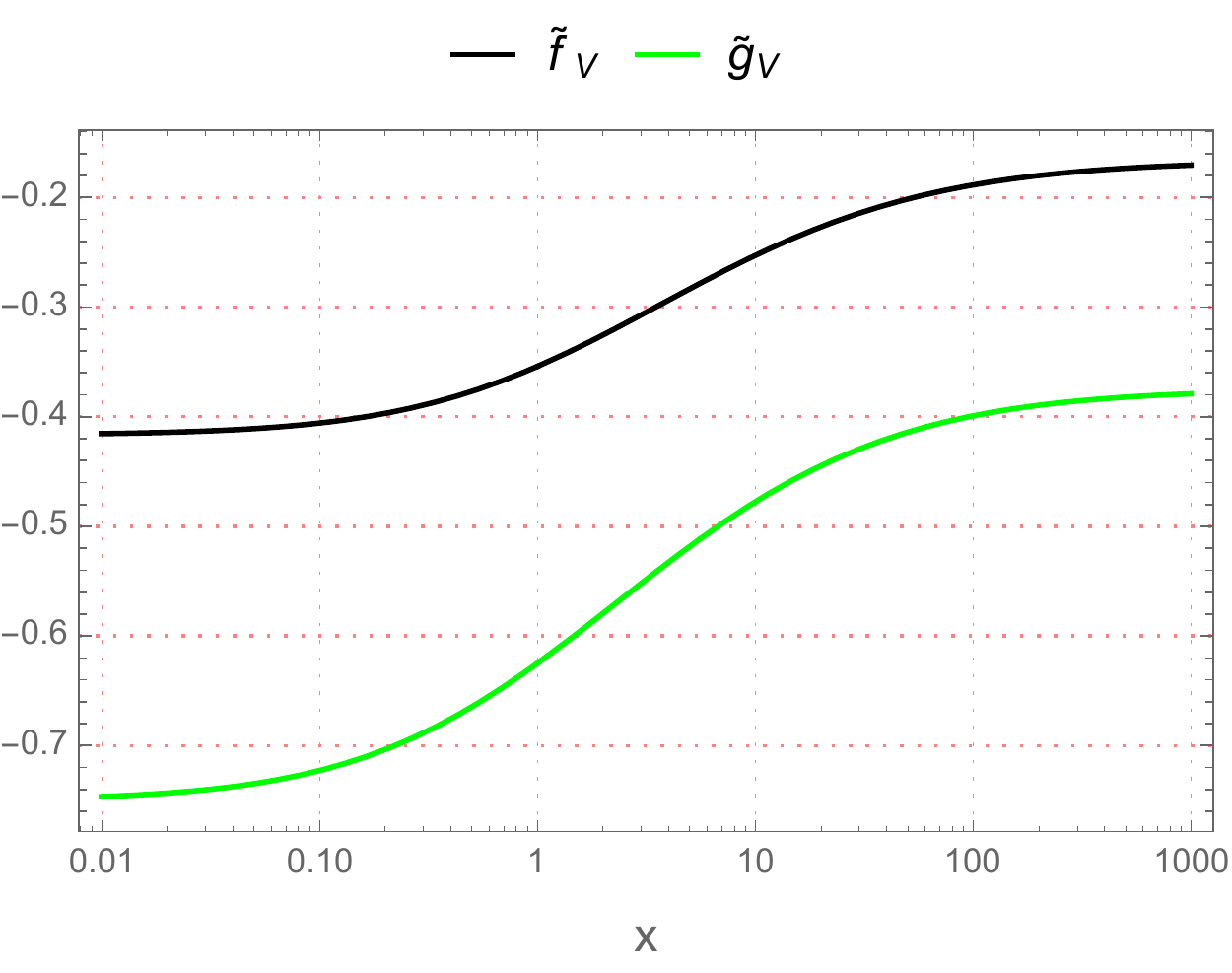}\\	
	\end{tabular}
	\caption{ The dependence of master formulas as functions of $x=m_E^2/m_X^2$ and $m^2_{n_i}/m_X^2$ with $X=W,Y,H^\pm_{1,2},H^A$. }\label{fig_fgx}
\end{figure}

In summary, general formulas for one-loop contributions to $a_{e_a}$ used in this work  were given in Ref.~\cite{Crivellin:2018qmi}. They are consistent with those calculated previously for the $331\beta$ models~\cite{Lavoura:2003xp,  Hue:2017lak}.   In the $331\beta$ model under consideration, all the relevant one-loop contributions will be derived in the forms depending on  the two classes of the  following master functions: $\left\{f_{\Phi}(x), \; \tilde{f}_{\Phi}(x),\; xf_{\Phi}(x),\; x\tilde{f}_{\Phi}(x),\; xg_{\Phi}(x),\; x\tilde{g}_{\Phi}(x)\right\}$ and $\{ \tilde{f}_{V}(x),\; \tilde{g}_{V}(x)\}$ for charged Higgs and gauge boson exchanges, respectively. The additional factor $x$ originates from the specific properties of the charged Higgs bosons in the $331\beta$ framework.  The dependence of these  functions on $x$ is shown in Fig. \ref{fig_fgx}, where all allowed ranges are shown precisely.

To estimate the one-loop contributions to AMM, it is useful to see the limits for the above master functions as follows:
\begin{align}
	\label{eq_limitfgvx}
&\lim_{x\to 0} f_{\Phi}(x)= f_{\Phi}(0)= \frac{1}{4}, 
\quad \lim_{x\to \infty} f_{\Phi}(x)= f_{\Phi}(\infty)=0,	
\crn 	&\lim_{x\to 0} \tilde{f}_{\Phi}(x)= \tilde{f}_{\Phi}(0)=\frac{1}{24},  
\quad \lim_{x\to \infty} \tilde{f}_{\Phi}(x)= \tilde{f}_{\Phi}(\infty)=0,
\crn &\lim_{x\to 0} \tilde{f}_{V}(x)=\tilde{f}_{V}(0)= -\frac{5}{12}, 
\quad \lim_{x\to \infty} \tilde{f}_{V}(x)= \tilde{f}_{V}(\infty)= -\frac{1}{6},
\crn &\lim_{x\to 0} \tilde{g}_{V}(x)= \tilde{g}_{V}(0)= -\frac{3}{4},
\quad \lim_{x\to \infty} \tilde{g}_{V}(x)= \tilde{g}_{V}(\infty)= -\frac{3}{8},
\crn 	&\lim_{x\to 0}\left[ x\times f_{\Phi}(x)\right]=0, 
\quad \quad \lim_{x\to \infty}\left[ x\times f_{\Phi}(x)\right]=\frac{1}{4},
	\crn 	&\lim_{x\to 0}\left[ x\times  \tilde{f}_{\Phi}(x)\right]=0,  
\quad \quad \lim_{x\to \infty}\left[ x\times \tilde{f}_{\Phi}(x)\right]=\frac{1}{12},
\crn 	&\lim_{x\to 0}\left[ x\times  g_{\Phi}(x)\right]=0, 
\quad \quad \lim_{x\to \infty}\left[ x\times g_{\Phi}(x)\right]= \frac{1}{2},
\crn 	&\lim_{x\to 0}\left[ x\times  \tilde{g}_{\Phi}(x)\right]=0,  
\quad \quad \lim_{x\to \infty}\left[ x\times \tilde{g}_{\Phi}(x)\right]=\frac{1}{8}. 
\end{align}
Because  $|\beta|\le \sqrt{3}$, we have $-1\le B\le2$. It is easily to show that:
\begin{align}
\label{eq_fgbound}
& \left| xf_{\Phi} \left( x\right) + B x g_\Phi  \left(x\right) \right| \leq  \mathcal{O}(1),
\; \left| x\tilde{f}_{\Phi} \left( x\right) + B x \tilde{g}_\Phi  \left(x\right) \right| \leq  \mathcal{O}(1),
\; \left|\tilde{f}_{V} \left( x\right) + B \tilde{g}_V  \left(x\right) \right| \leq \frac{7}{6},
\crn&0\leq  \tilde{f}_{V}\left( x\right)+\frac{5}{12} \leq \frac{5}{12} , \;   0\le  xf_{\Phi} \left( x\right) \le \frac{1}{4}.
\end{align}
First, we consider the one-loop contribution from the SM gauge $W^\pm$ where the deviation from the SM prediction derived from Eq.~\eqref{eq_Deltaaea} satisfies:
\begin{align}
\label{eq_SMdeviation}	
\left| \Delta a_{\mu}(W) \right|\simeq 9.19\times 10^{-9} \left|\left( R^*R^T\right)_{aa} \times \left(\tilde{f}_V \left(x_W\right)+ \frac{5}{12}\right)\right| < 2.5\times 10^{-11}<  a^{\mathrm{NP}}_{\mu},
\end{align}
where the constraint $|\left( R^*R^T\right)_{aa}|\leq2\times 10^{-3}$ consistent with non-unitary condition~\eqref{eq_maxRRd}. In Eq. \eqref{eq_SMdeviation},  $\Delta a_{\mu}$ is considered as the $1\sigma$ range of the discrepancy between the SM's prediction and experiments shown in Eq. \eqref{eq_damu}, namely $\left| \Delta a_{\mu}(W) \right|\ll \Delta a_{\mu}\in \left[ 1.92\times 10^{-9},\; 3.1\times 10^{-9}\right]$. Therefore,  we will use the following approximation for both frameworks MSS and ISS:  
\begin{align}
	\label{eq_amuWapp}	
\Delta a_{\mu}(W) \simeq 0. 
\end{align}

For the recent bound of the $SU(3)_L$ scale, we can use the lower bound $m_Y\ge 1$ TeV, consistent with the recent constraint concerned for 3-3-1 models \cite{Coutinho:2013lta, Salazar:2015gxa, Long:2018fud, Nepomuceno:2019eaz}. Now the one-loop contributions from  $H^A$ and  $Y^A$ can be estimated as follows:
\begin{align}
\label{eq_maxHAY}
&0<-  a_{\mu}(H^A)\leq 	9.19\times 10^{-9} \times \left[ \frac{m_W^2}{m^2_Y}+ \frac{m^2_{\mu}}{m^2_{E} c^2_{\beta}}    +    \frac{m_W^4c^2_{\beta}}{m_Y^4} \right]< 6.3 \times 10^{-11}\ll 	\Delta a^{\mathrm{NP}}_{\mu},
\crn &0<\Delta a_{\mu}(Y)\leq 9.19\times 10^{-9} \times  \frac{m_W^2}{m^2_Y}\times \frac{7}{6}<  7  \times 10^{-11}\ll 	\Delta a^{\mathrm{NP}}_{\mu}. 
\end{align}
 where  a crude lower bound $m_{E}c_{\beta}\ge5$ GeV was used.   We conclude that the two one-loop contributions originated from heavy Higgs $H^A$ and charged gauge boson $Y$ is much smaller than   $\Delta a^{\mathrm{NP}}_{\mu}\sim \mathcal{O}(10^{-9})$, which is considered as the $1\sigma$ range  given in Eq. \eqref{eq_damu} from now on. This agrees with all previous works, for example for the heavy charged gauge bosons~\cite{Pinheiro:2021mps}. We will ignore them from now on.
 
We now discuss on the dominant contributions of the two singly charged Higgs bosons given in Eq.~\eqref{eq_aHpmLR}, where small $t_{\beta}$ supports large values of these contributions. The reasonable values for a numerical estimation are  $t^{-1}_{\beta}s_{\alpha}c_{\alpha}\simeq 0.5$, and $v/(\sqrt{2}m_{\mu})=1.6\times 10^{3}$, max$[| x_1f_{\Phi} \left( x_1\right) - x_2f_{\Phi} \left( x_2\right) |]\simeq 0.25$, we have 
 \begin{align}
 	\label{eq_aHpmLR0}
 	\left|\sum_{k=1}^2\Delta a_{\mu}(H^\pm_1)\right|  &\leq   	1.9\times 10^{-9}\left[10^3  \left(M_0^{-1} M_D^{\dagger}Y^{h}\right)_{22} \right] \sim 	\Delta a^{\mathrm{NP}}_{\mu}  \left[10^3  \left(M_0^{-1} M_D^{\dagger}Y^{h}\right)_{22} \right]. 
 	 \end{align}
 In the next discussion for two specific frameworks of MSS and ISS, the allowed values of $\left| \left(M_0^{-1} M_D^{\dagger}Y^{h}\right)_{22}\right|$ will depend strictly on the characteristics of the two models. We will show that the  condition \eqref{eq_aHpmLR0} will satisfy for only the ISS mechanism, which allows this value to reach the $(g-2)_{\mu}$ data. 
 For convenience, we will use the following estimation,
 \begin{align}
 	\label{eq_defaultRanges}
 	\frac{v}{\sqrt{2}m_{\mu}}&\simeq 1.6\times 10^3; \; | x_1f_{\Phi} \left( x_1\right) - x_2f_{\Phi} \left( x_2\right) | \leq 0.25; \; 0.3\le t_{\beta}\leq10; 
 	\crn \left| s_{\alpha}c_{\alpha}\right|&= \left| \frac{\sin(2\alpha)}{2}\right|\le0.5; \; m_{H^\pm_1},\; m_{H^\pm_2} \ge 800 \;\mathrm{GeV};   M_0\ge 100 \; \mathrm{GeV}. 
 \end{align}
After that,  other  possible values of $m_{H^\pm_K}$,  $M_0$, and  $t_{\beta}$ will also be discussed.

Now we will derive the specific analytic formulas of one-loop contributions to $ \Delta a_{\mu}$  corresponding to the two mechanisms MSS and ISS. We note that all above discussions for $ \Delta a_{\mu}$  are applied in the same way to derive  to $ \Delta a_{e}$, therefore, we just 
mention to $ \Delta a_{\mu}$ in the numerical discussion. 

\subsection{ The MSS mechanism}
The MSS relations given in Eqs. \eqref{eq_mDss} and \eqref{masafla1}  result in that 
\begin{align}
	\label{eq_SS}
	M_D^{\dagger}R^{\dagger}= \hat{m}_{\nu}, \;  R = -iU_{\mathrm{PMNS}}  \left( \frac{\hat{m}_{\nu}}{M_0}\right)^{1/2}, \;\quad m_{n_{4,5,6}}\simeq M_0.
\end{align}
The detailed derivation of the one-loop contributions from singly charged Higgs bosons is given in appendix~\ref{app_calculation}. Using $\tilde{f}_{\Phi}(0)= \frac{1}{24}$, the one-loop  contribution from $H^\pm_1$ is 

%corrected 18, Nov,2021a-> 2
\begin{align}
	\label{eq_amuH1SS}
	\Delta a^{\mathrm{MSS}}_{\mu}(H^\pm_1) &= - 9.19\times 10^{-9}
	\crn&\times\mathrm{Re}\left\{ \left[ c^2_{\alpha} \left(\frac{m^2_{n_2} }{M_0^2}\right)^{1/2} + \frac{vt_{\beta}^{-1}c_{\alpha}s_{\alpha}}{\sqrt{2}m_{\mu}}\sum_{c=1}^{3}  \left(\frac{m^2_{n_c} }{M_0^2}\right)^{1/4} U_{\mathrm{PMNS},2c}\left(-iY^{h}\right)_{c2}\right]x_1 f_{\Phi}(x_1)
	%LL corrected 13 sep 2021
	\right.\crn&  \quad +t_{\beta}^{-2} c^2_{\alpha} \sum_{c=1}^3 \left|U_{\mathrm{PMNS},2c}\right|^2 \left[ \frac{m_{n_c}^2}{m^2_{H^\pm_1}} \left( \frac{1}{24} -\tilde{f}_{\Phi}(x_1)  \right) +  \frac{m_{n_c} }{M_0} x_1\tilde{f}_{\Phi}(x_1) \right]
	\crn & \quad+ \frac{m^2_{\mu} t_{\beta}^2c_{\alpha}^2}{m^2_{H^\pm_1}} \left[ \frac{1}{24} - \sum_{c=1}^{3} \left|U_{\mathrm{PMNS},2c}\right|^2 \frac{m_{n_c} }{M_0} \left( \frac{1}{24} -\tilde{f}_{\Phi}(x_1) \right) \right]
	\crn & \quad+ \frac{v^2s_{\alpha}^2}{2 m^2_{H^\pm_1}} \left[ \sum_{c=1}^{3} \left|Y^{h}_{c2}\right|^2 \frac{m_{n_c} }{M_0}  \left(  \frac{1}{24} -\tilde{f}_{\Phi}(x_1)  \right)  + \left(Y^{h\dagger}Y^{h}\right)_{22} \tilde{f}_{\Phi}(x_1)\right]
	\crn& \quad-\left.  \frac{vm_{\mu}t_{\beta} s_{2\alpha}}{\sqrt{2} m^2_{H^\pm_1}} \left[ \left(-iU_{\mathrm{PMNS}} \left(\frac{\hat{m}^2_{\nu} }{M^2_0}\right)^{1/4}Y^{h}\right)_{22}\left( \frac{1}{24}- \tilde{f}_{\Phi}(x_1)  \right)  \right] 	\right\}. 
\end{align}
It can be seen that  only two terms proportional to $\left(\frac{ m^2_{n_c}}{M^2_0}\right)^{1/4} \mathrm{Re}[(-iU_{\mathrm{PMNS},2c} Y^{h}_{c2})]$ can give   contributions having consistent sign with  $\Delta a^{\mathrm{NP}}_{\mu}$, and $\mathrm{Re}[(-iU_{\mathrm{PMNS},2c}Y^{h}_{c2})]$ must be negative (see the first and last lines in the real part of  Eq. \eqref{eq_amuH1SS}).  We just focus on these two contributions. The remaining terms  always give negative contributions to  $\Delta a^{\mathrm{NP}}_{\mu}$. The two mentioned terms can be estimated as follows:
\begin{align}
&0<	 \left(\frac{m^2_{n_c} }{M_0^2}\right)^{1/4} \times x_1 f_{\Phi}(x_1) \leq   \left(\frac{(0.12\; \mathrm{eV})^2 }{m_{H^{\pm}_1}^2}\right)^{1/4} \times x_1^{3/4} f_{\Phi}(x_1) <1.1\times 10^{-7},
\crn &0<	\left(\frac{m^2_{n_c} }{M_0^2}\right)^{1/4} \times \left( \frac{1}{24} -\tilde{f}_{\Phi}(x_1)\right)  \leq  \frac{1}{24} \left(\frac{(0.12\; \mathrm{eV})^2 }{m_{H^{\pm}_1}^2}\right)^{1/4}   <10^{-7},
\end{align}  
where we have used $m_{H^\pm_1}\geq 100$ GeV and max$[x_1^{3/4} f_{\Phi}(x_1)]<0.1 $.  But in this situation the  factor  $ \left|\frac{v}{\sqrt{2} m_{\mu}} Y^{h}_{c2}t^{-1}_{\beta} \right| \le  10^4$ is still not large enough so that the total can give any significant  contributions to $\Delta a^{\mathrm{MSS}}_{\mu}$. In conclusion,  the MSS mechanism still fails to explain the experimental AMM data of $\mu$. 

\subsection{The ISS mechanism}
The ISS mechanism will be considered instead of the MSS one.  The change is for only  singly charged Higgs bosons $H^\pm_k$.  Following Eqs. \eqref{eq_UNiss} and  \eqref{eq_mDRISS},   the results for $H^\pm_{1,2}$  are  
\begin{align}
	\label{eq_Hpm1}
	a^{\mathrm{ISS}}_{\mu}(H^\pm_1)& = - 9.19\times 10^{-9}
	%LR
\crn&	\times \mathrm{Re}\left\{ \sum_{c=1}^{3} \left[ c^2_{\alpha} \left|U_{\mathrm{PMNS},2c} \right|^2 \frac{m_{n_c}}{\mu_X} + \frac{vt_{\beta}^{-1}c_{\alpha}s_{\alpha}}{\sqrt{2}m_{\mu}}U_{\mathrm{PMNS},2c} \left(\frac{m_{n_c}}{\mu_X} \right)^{1/2} \left(Y^{h}_{2}\right)_{c2} \right]x_1f_{\Phi}(x_1)
	% LL corrected 13 sep 2021
	\right.\crn&  \quad +  \sum_{c=1}^{3}  \left[ \left|U_{\mathrm{PMNS},2c} \right|^2 \left(t_{\beta}^{-2} c^2_{\alpha} \frac{m_{n_c}}{\mu_X} \right) \right] x_1 \tilde{f}_{\Phi}(x_1)
	%RR
	\crn & \quad+ \frac{m^2_{\mu} t_{\beta}^2c_{\alpha}^2}{m^2_{H^\pm_1}} \left[ \frac{1}{24} - \sum_{c=1}^{3} \left[ \left|U_{\mathrm{PMNS},2c} \right|^2\frac{m_{n_c}}{\mu_X} \right] \left( \frac{1}{24} -\tilde{f}_{\Phi}(x_1) \right) \right]
	\crn & \quad+ \frac{v^2s_{\alpha}^2}{2 m^2_{H^\pm_1}} \left[ \sum_{c=1}^{3} \left[ \left| \left(Y^{h}_{2}\right)_{c2}\right|^2 \frac{m_{n_c}}{\mu_X} \right]  \left(  \frac{1}{24} -\tilde{f}_{\Phi}(x_1)  \right)  + \left(Y^{h\dagger}_2Y^{h}_2\right)_{22} \tilde{f}_{\Phi}(x_1)\right]
	\crn& \quad-\left.  \frac{vm_{\mu}t_{\beta} s_{2\alpha}}{ \sqrt{2} m^2_{H^\pm_1}} \left( \frac{1}{24}- \tilde{f}_{\Phi}(x_1)  \right) \sum_{c=1}^{3} \left[ U_{\mathrm{PMNS},2c}\left(Y^{h}_{2}\right)_{c2} \left(\frac{m_{n_c}}{\mu_X} \right)^{1/2}\right]  	\right\},
	\crn a^{\mathrm{ISS}}_{\mu}(H^\pm_2)&= a_{\mu}(H^\pm_1) \left[ x_1 \to\; x_2,\;s_{\alpha}\to - c_{\alpha}, \; c_{\alpha} \to s_{\alpha}\right],
\end{align}
where we have used the form $Y^h=(O_{3\times3},\; Y^h_2)^T$ for the ISS framework. 
Here,  the parameter $\mu_X$ appears in the ISS mechanism instead of  $M_0$ corresponding to the MSS. The second term in the first line of Eq.~\eqref{eq_Hpm1} is  from $H^\pm_1$ emphasized previously in Eq.~\eqref{eq_aHpmLR}. 

%Note change \hat{x}_{\nu} -2-> -3, 6, Sep, 2021
Using the constraint \eqref{eq_maxRRd} for $RR^{\dagger}=U_{\mathrm{PMNS}} \hat{x}_{\nu}U_{\mathrm{PMNS}}^{\dagger}$ we have $\hat{x}_{\nu}<\mathcal{O}(10^{-3})$.  Therefore, we will choose a safe upper bound for the NO scheme  as follows 
\be \label{eq_maxhxnu}
\mathrm{Max}[\left(\hat{x}_{\nu}\right)_{aa}]= \frac{m_{n_3}}{\mu_X}\simeq \left(  \frac{\Delta m^2_{32}}{\mu_X^2}\right)^{1/2} \leq 2\times 10^{-3} \Rightarrow \mu_X\geq 2.5\times 10^{-8}\; \mathrm{GeV}.
\ee 
The default value of $\mu_X$ is fixed by $\mu_X= 2.5\times 10^{-8}\; \mathrm{GeV}$.
 
With the allowed range given in Eq.~\eqref{eq_defaultRanges}, it can be proved that:
\begin{align}
&	0<\sum_{c=1}^3c^2_{\alpha}  \left|U_{\mathrm{PMNS},2c} \right|^2 \frac{m_{n_c}}{\mu_X}x_1 f_{\Phi}(x_1) < \left|U_{\mathrm{PMNS},23} \right|^2 \times 5\times 10^{-3}\times \frac{1}{12}\simeq 8.9\times 10^{-5}, 
\crn &0< 	a^{\mathrm{ISS}}_{\mu,1}(H_1^\pm)\sim  \sum_{c=1}^{3}  \left[ \left|U_{\mathrm{PMNS},2c} \right|^2 \left(t_{\beta}^{-2} c^2_{\alpha} \frac{m_{n_c}}{\mu_X} \right) \right] x_1 \tilde{f}_{\Phi}(x_1) <1.3\times 10^{-3},
\crn&0< \frac{m^2_{\mu} t_{\beta}^2c_{\alpha}^2}{m^2_{H^\pm_1}} \left[ \frac{1}{24} - \sum_{c=1}^{3} \left[ \left|U_{\mathrm{PMNS},2c} \right|^2\frac{m_{n_c}}{\mu_X} \right] \left( \frac{1}{24} -\tilde{f}_{\Phi}(x_1) \right) \right]<10^{-7}
\crn& 0<\frac{v^2s_{\alpha}^2}{2 m^2_{H^\pm_1}}\times  \sum_{c=1}^{3} \left[ \left|Y^{h}_{(c+3)2} \right|^2 \frac{m_{n_c}}{\mu_X} \right]  \left(  \frac{1}{24} -\tilde{f}_{\Phi}(x_1)  \right)  < 0.32\times 10^{-4},
\crn &0 <\frac{vm_{\mu}t_{\beta} s_{2\alpha}}{\sqrt{2} m^2_{H^\pm_1}} \left( \frac{1}{24}- \tilde{f}_{\Phi}(x_1)  \right) \sum_{c=1}^{3} \left[ \mathrm{Re}[U_{\mathrm{PMNS},2c}Y^{h}_{(c+3)2}] \left(\frac{m_{n_c}}{\mu_X} \right)^{1/2}\right]  < 10^{-6},
\crn
&0< a^{\mathrm{ISS}}_{\mu,2}(H_1^\pm)\sim \frac{v^2s_{\alpha}^2}{2 m^2_{H^\pm_1}} \left(Y^{h\dagger}Y^{h}\right)_{22} \tilde{f}_{\Phi}(x_1)= \frac{v^2s_{\alpha}^2}{2 m^2_{H^\pm_1}} \left(Y^{h\dagger}_2Y^{h}_2\right)_{22} \tilde{f}_{\Phi}(x_1)<1.6\times 10^{-2}. 
\end{align}
There are only two  contributions $a_{\mu,1(2)}(H_1^\pm)$  in the second and last lines  that may affect significantly  $ a_{\mu}(H^\pm_1)$ because of  the rather large upper bounds  $a_{\mu,1}(H_1^\pm)\leq 1.3c^2_{\alpha}\times 10^{-11}$ and   $a_{\mu,2}(H_1^\pm)\leq 23s^2_{\alpha}\times 10^{-11}$. But both of them give negative contributions to $\Delta a^{\mathrm{ISS}}_{\mu}$, hence should be small.  Ignoring all other contributions smaller  $10^{-4}\times  \Delta a^{\mathrm{NP}}_{\mu}$,  the remaining large contribution in Eq.~\eqref{eq_Hpm1} is the one  mentioned in Eq.~\eqref{eq_aHpmLR}. It has the following form in the ISS framework:
\begin{align}
	\label{eq_Hpm2}
	a^{\mathrm{ISS}}_{\mu,0}(H^\pm)& = - 9.19\times 10^{-9}
	%
%\crn &\times
 \mathrm{Re}\left\{\frac{vt_{\beta}^{-1}c_{\alpha}s_{\alpha}}{\sqrt{2}m_{\mu}} \left(\frac{m_{n_3}}{\mu_X} \right)^{1/2} Y^{d}_{2}  \left[ x_1f_{\Phi}(x_1) -x_2f_{\Phi}(x_2)\right]
	\right\}
	\crn&= -680.58 \times 10^{-9} \times t_{\beta}^{-1}c_{\alpha}s_{\alpha}\left(\frac{m_{n_3}}{\mu_X} \right)^{1/2} Y^{d}_{2} \left[ x_1f_{\Phi}(x_1) -x_2f_{\Phi}(x_2)\right],
\end{align}
where the part relating to $x_k$  is  the contribution from $H^{\pm}_k$ exchange, 
 and the new parameter $Y^d_2$ is  assumed to relate to Yukawa coupling matrix $Y^h_2$ through the following relation 
\begin{align}
	\label{eq_Yh31}
	U_{\mathrm{PMNS}} \left(\frac{\hat{m}_{\nu}}{m_{n_3}}\right)^{1/2} Y^{h}_{2}\equiv  \mathrm{diag} \left( Y^{d}_{1},\;Y^{d}_{2}, Y^{d}_{3}\right)=Y^d,
\end{align}
so that the two largest one-loop contributions from $H^{\pm}_{1,2}$ to AMM will allow  zero contributions  to the cLFV branching ratios Br$(e_b\to e_a\gamma)\sim \left( |c_{(ab)R}|^2+|c_{(ba)R}|^2\right)$ \cite{Crivellin:2018qmi},  because   $c_{(ab)R}(H^\pm), c_{(ba)R}(H^\pm) \sim 	\left(U_{\mathrm{PMNS}} \hat{m}_{\nu}^{1/2}Y^{h}_{2}\right)_{ab,ba} (H^\pm) =0$ for $a\neq b$. Now, the  matrix $Y^h_{2}$ is derived through the following relation:
\begin{equation}\label{eq_Yh2}
	Y^h_2=\left(\frac{\hat{m}_{\nu}}{m_{n_3}}\right)^{-1/2}U^{\dagger}_{\mathrm{PMNS}}Y^d,
\end{equation}
which will be used to check the perturbative limit of all  $\left| (Y^{h}_{2})_{ab}\right|<\sqrt{4\pi}\simeq 3.5$ while scanning values of $Y^d_{1,2,3}$. In this case,  the  factors $\left(\frac{(\hat{m}_{\nu})_{aa}}{m_{n_3}}\right)^{-1/2} \ge 1$ in the NO scheme and  all active neutrino masses lie in the denominators.  Therefore, these masses  must be non-zero and large enough to guarantee that  all entries of $Y^{h}_2$  satisfy the perturbative limits.  

 From now on, we will fix $m_{n_1}=0.01$ eV in our numerical discussion. Smaller $m_{n_1}$ will give smaller allowed $Y^d_1$ satisfying the perturbative limit of $Y^h_2$.  Using the best-fit points corresponding to the NO scheme of neutrino oscillation data given in Eq. \eqref{eq_d2mijNO},  $Y^h_2$ has the following form  
%check 23 April 2022: Ok
\begin{align}
	\label{eq_Yh2num}
	Y^{h}_2=\left(
	\begin{array}{ccc}
		1.84 Y^d_1 & (-0.7-0.125 i) Y^d_2 & (1.09-0.113 i) Y^d_3 \\
		1.1 Y^d_1 & (1.184-0.074 i) Y^d_2 & (-1.11-0.068 i) Y^d_3 \\
		(-0.116-0.09 i) Y^d_1 & 0.732 Y^d_2 & 0.666 Y^d_3 \\
	\end{array}
	\right).
\end{align}

For small $m_{n_1}$ including the case $m_{n_1}=0$, matrix $Y^h_2$ is  chosen as follows:
\begin{align}
	%	\label{eq_Yh20}
	%
	Y^h_2 = \left(
	\begin{array}{ccc}
		0 & 0 & 0 \\
		(2.9+0.3 i) Y^d_{1} & (0.50-0.32 i)Y^d_{2} & (0.61-0.68 i)Y^d_{3} \\
		(-1.23-0.21 i)Y^d_{1} & (1.15+0.12 i)Y^d_{2} & Y^d_{3} \\
	\end{array}
	\right),\nn 
\end{align}
which $Y^d$ defined from Eq. (65)  is not diagonal, but it always keeps $(Y^d)_{12}=(Y^d)_{21}=(Y^d)_{13}=0$. In addition,  $Y^d_3=0$ gives $(Y^d)_{32}=0$. This will avoid large contributions to the strict constraint of cLFV decay Br$(\mu \to e\gamma)$.  The numreical investigations show that the two choices of $Y^h_2$ mentioned above have the same qualitative results. Numerical illustration will be done with the $Y^h_2$ given in Eq. \eqref{eq_Yh2num}. Therefore,  we will fix $Y^d_3=0$  in our numerical investigation on $a^{\mathrm{ISS}}_{e,\mu}(H^\pm)$. We note that $Y^d_3$ also contributes to the AMM of the $\tau$ lepton,  which still has weak constraints from  recent experiments \cite{OPAL:1998dsa, L3:1998lhr, DELPHI:2003nah} and a combination derived from these experimental results \cite{Gonzalez-Sprinberg:2000lzf, Eidelman:2016aih},  see  discussions on this topic in Refs.  \cite{Tran:2020tsj, Crivellin:2021spu},  suggesting new experiments  to improve measurements.

 Similarly, the data of $\Delta a^{\mathrm{NP}}_{e}$ may be explained by the following contribution:
\begin{align}
	\label{eq_aeHpm2}
	a^{\mathrm{ISS}}_{e,0}(H^\pm)& = -9.19 \times 10^{-9} \times \frac{m_e^2}{m^2_{\mu}}
	%
	%\crn &\times
	\mathrm{Re}\left\{\frac{vt_{\beta}^{-1}c_{\alpha}s_{\alpha}}{\sqrt{2}m_{e}} \left(\frac{m_{n_3}}{\mu_X} \right)^{1/2} Y^{d}_{1}   \left[ x_1f_{\Phi}(x_1) -x_2f_{\Phi}(x_2)\right]
	\right\}
	\crn&=  -32409\times 10^{-13} \times t_{\beta}^{-1}c_{\alpha}s_{\alpha} \left(\frac{m_{n_3}}{\mu_X} \right)^{1/2} Y^{d}_{1} \left[ x_1f_{\Phi}(x_1) -x_2f_{\Phi}(x_2)\right].
\end{align}
In the simple forms of the matrices  $Y^{h}$ given in Eq. \eqref{eq_Yh2} and  $M_D$ we assumed here, the main difference between   $a^{\mathrm{ISS}}_{e,0}(H^\pm)$ and  $a^{\mathrm{ISS}}_{\mu,0}(H^\pm)$ is that they contain different free factors $Y^{d}_{1}$ and  $Y^{d}_{2}$, respectively. The numerical results show that this difference is enough to explain both AMM data of $e$ and $\mu$ at $1\sigma$ discrepancy  given in Eqs. \eqref{eq_damu} and \eqref{eq_dae}. The regions of the parameter space satisfying simultaneously these will be defined as the allowed regions from now on.

\begin{figure}[ht]
	\centering\begin{tabular}{cc}
	\includegraphics[width=9.cm]{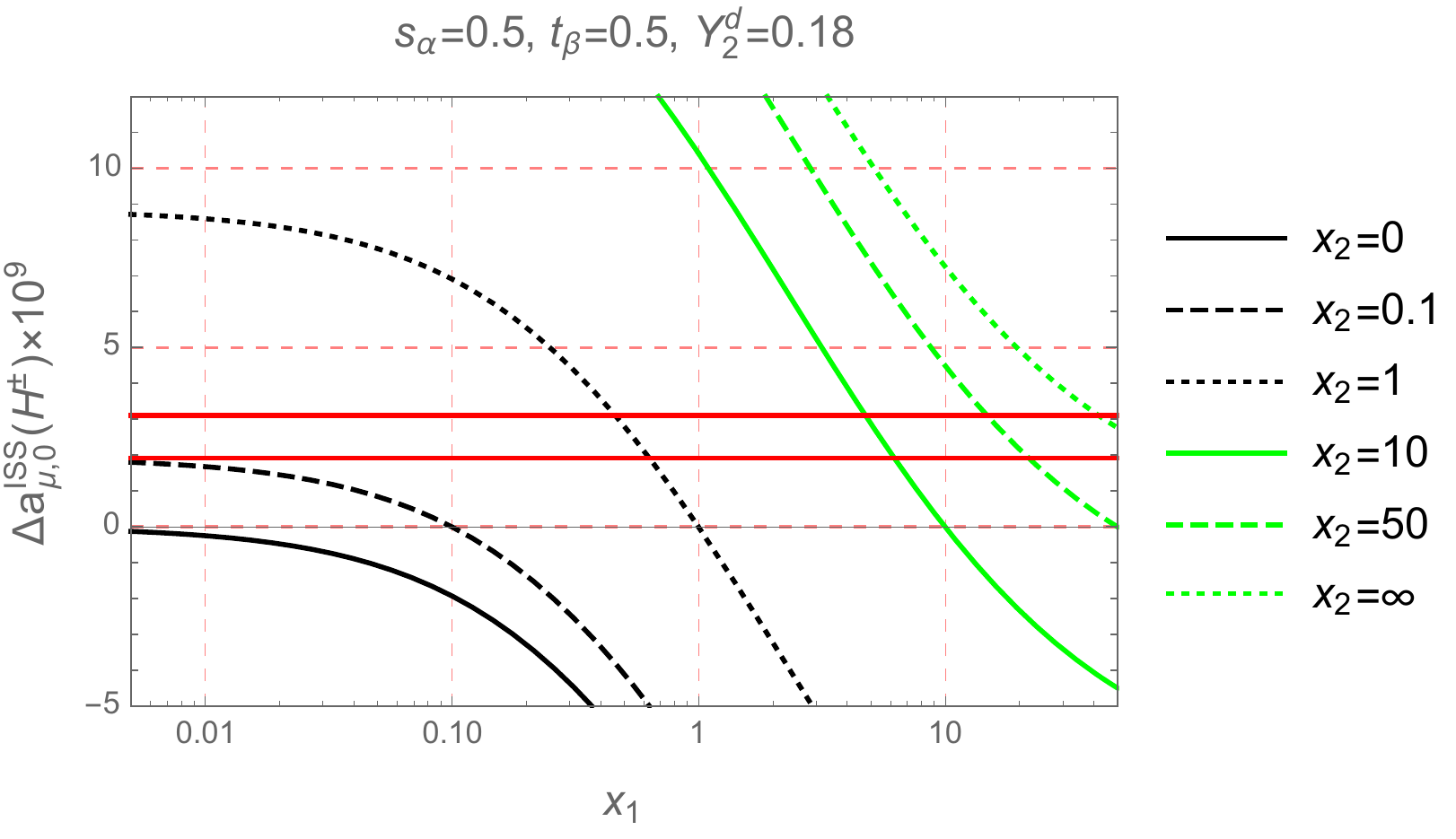}&	\includegraphics[width=7.cm]{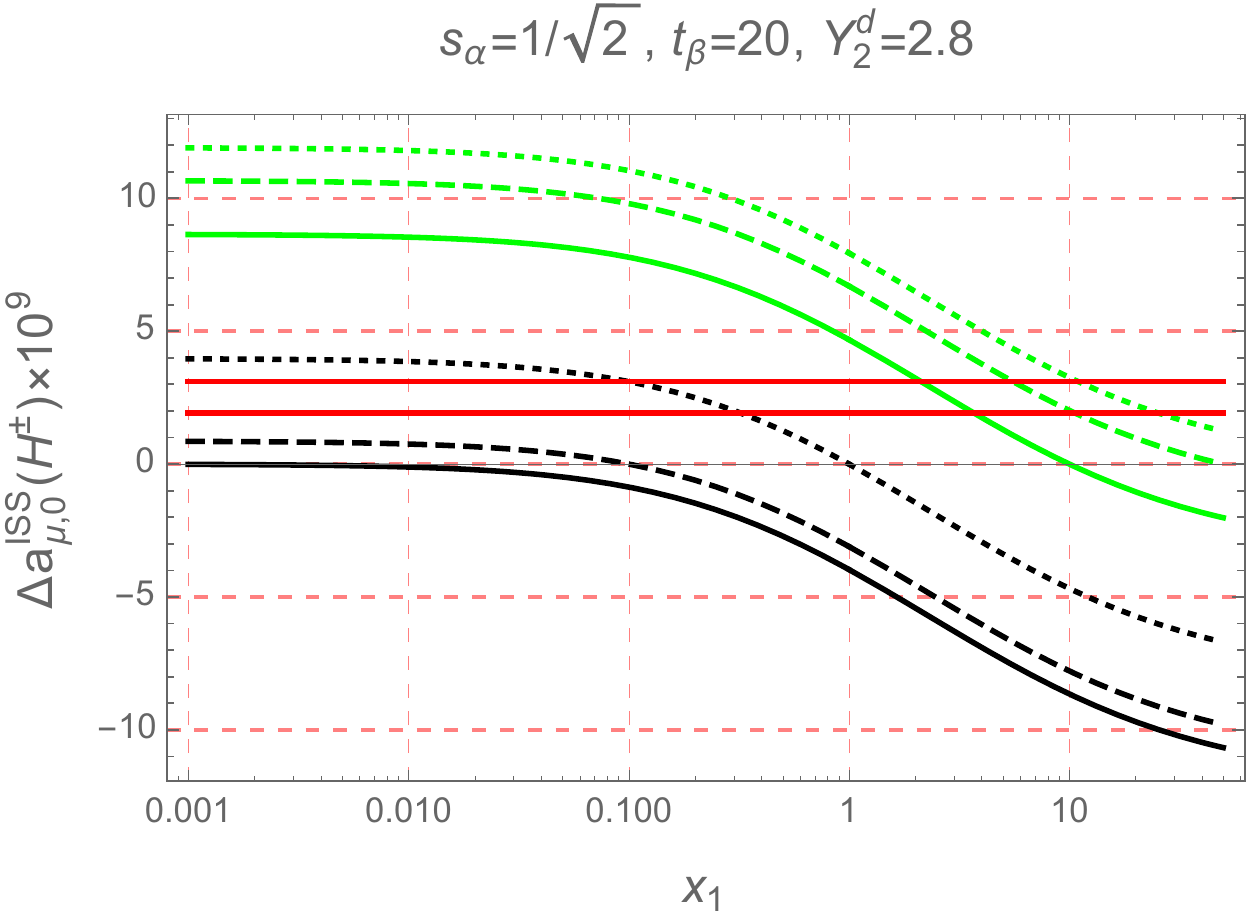}\\	
	\includegraphics[width=9.cm]{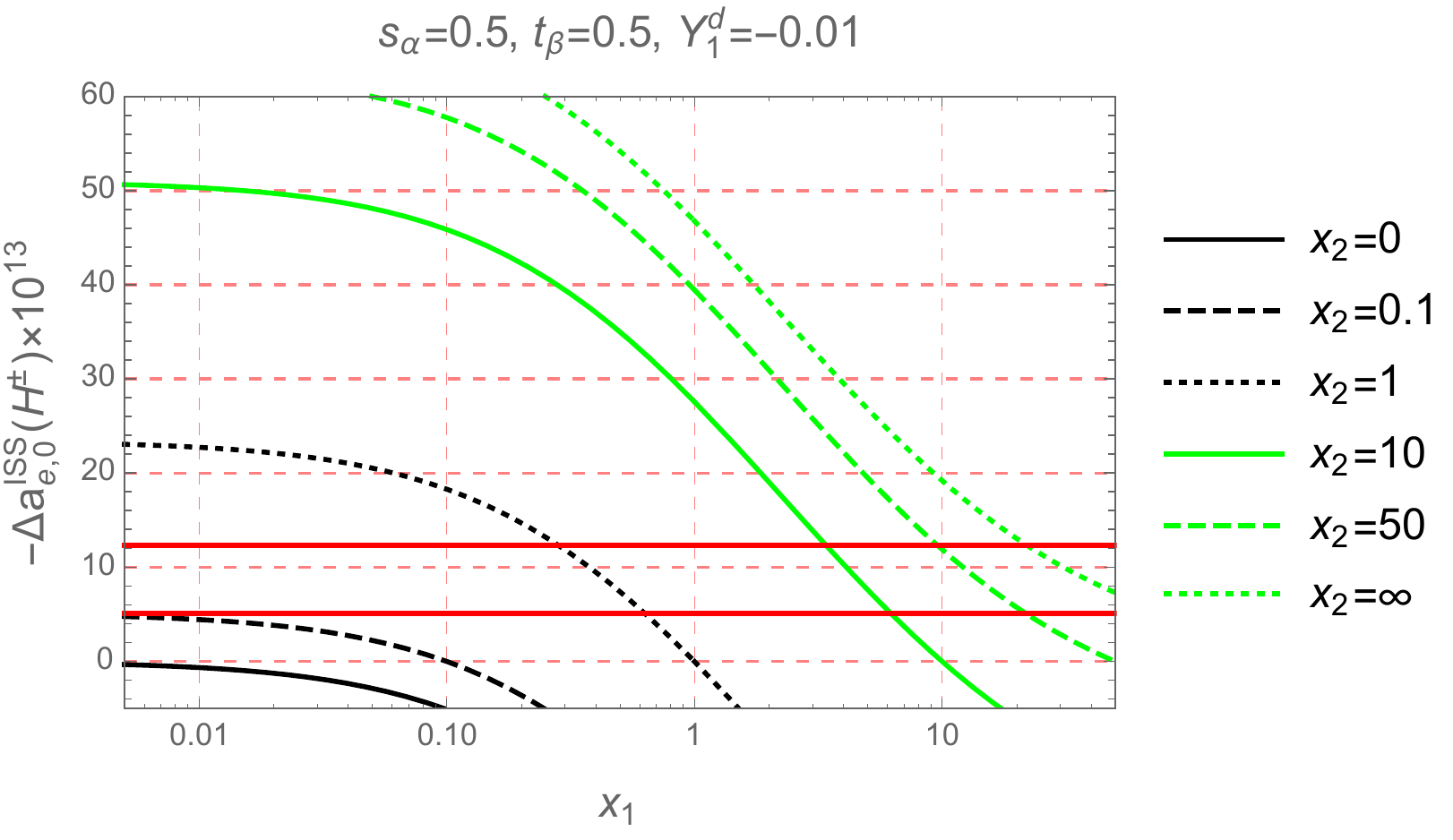}&	\includegraphics[width=7.cm]{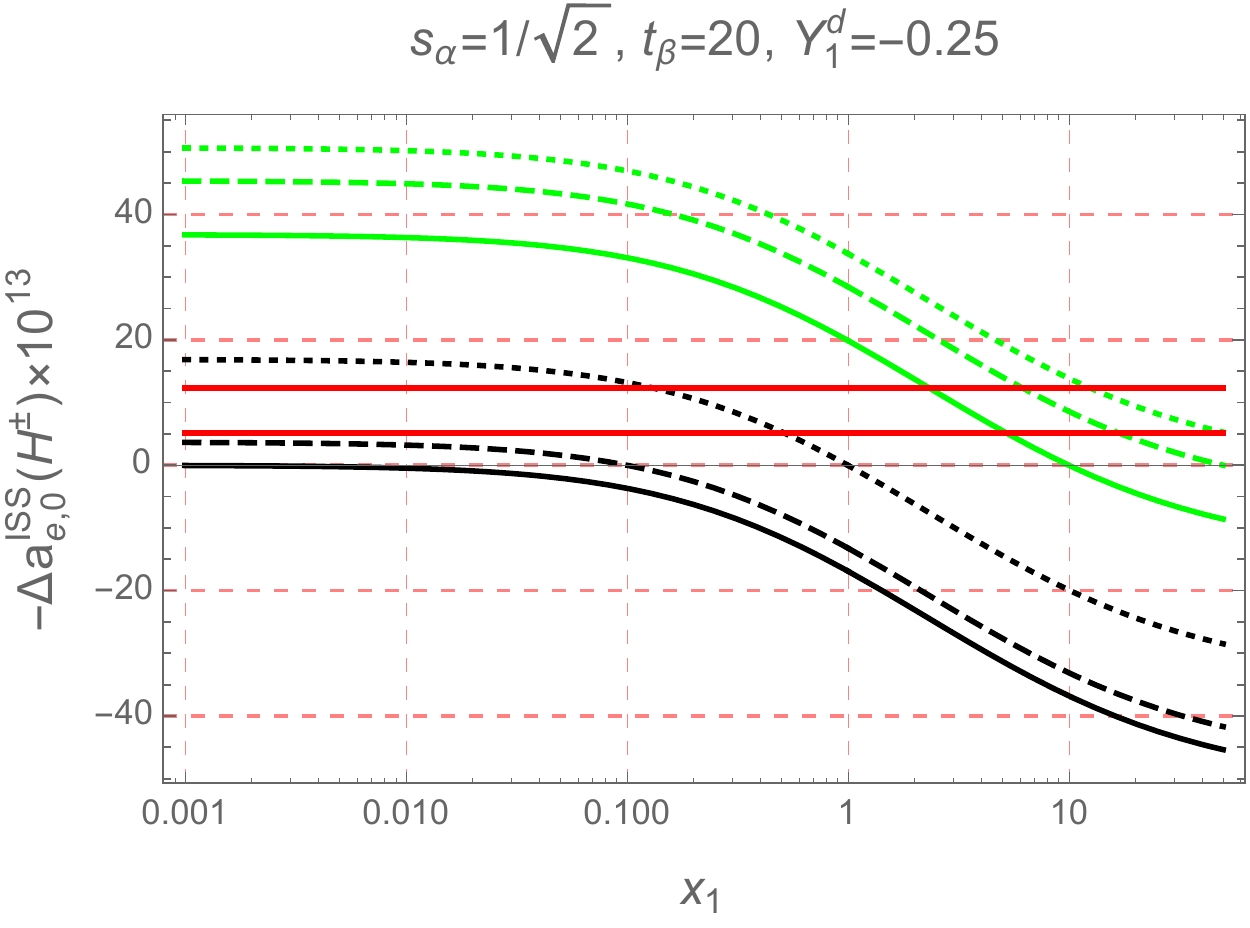}\\
	\end{tabular}
	\caption{ The dependence of $ \Delta a^{\mathrm{ISS}}_{\mu,0}(H^\pm)$ and $ \left[ -\Delta a^{\mathrm{ISS}}_{e,0}(H^\pm) \right]$  as functions of $x_1$ with different fixed $x_2$. The red   lines show the $1\sigma$ allowed ranges of $\Delta a^{\mathrm{NP}} _{\mu}$ and $\Delta a^{\mathrm{NP}} _{e}$   given in Eqs.~\eqref{eq_damu}  and \eqref{eq_dae}, respectively.  }\label{fig_amuHk}
\end{figure}

The first numerical illustrations are shown in Fig.~\ref{fig_amuHk}, where  free parameters are fixed in the ranges given in \eqref{eq_defaultRanges} and  predict valid regions  satisfying both the experimental AMM data of muon (two upper panels) and electron (two lower panels). 
In addition, in the upper left panel of Fig.~\ref{fig_amuHk}, the numerical values $s_{\alpha}=0.5$, $Y^{d}_{2}=0.21$, and $t_{\beta}=0.5$ safely  satisfy  perturbative limits of max$|\left(Y^h_2\right)_{ab}|<0.2$. On the other hand, numerical values of free parameters in the upper right panel are somewhat special: $s_{\alpha}=1/\sqrt{2}$ is maximal for $s_{2\alpha}=1$, large $Y^{d}_{2}=2.8$  close the perturbative limit max$|\left(Y^h_2\right)_{ab}|=3.32$, and $t_{\beta}=20 \gg1$ does not support large $ a^{\mathrm{ISS}}_{\mu}(H^\pm)$, which excludes the regions satisfying  $0<x_2<0.1$ and all $x_1>0$, for example. All values of $t_{\beta}>30$ are  excluded in this case.  We conclude that the AMM data will result in a upper bound of $t_{\beta}$.   The values of $Y^{d}_{1}$ are chosen so that there exist allowed values of $(x_1,x_2)$ satisfying simultaneously  both  $1\sigma$ experimental AMM data of  muon and electron.  Namely, the allowed values of $(x_1,\;x_2)$ in the two left panels are in the ranges $0<x_1\leq 20$ and $0.1< x_2$. Similarly, the allowed regions in the two right panels satisfy $0<x_1\leq10$ and $ x_2>0.1$. Large $t_{\beta}$ gives strong upper constraint on $M_0<750$ GeV derived from the perturbative limit of  max$|Y^X_{ab}|$=max$ \frac{\sqrt{2}\left(m_D\right)_{ab}}{v_2}=$ max$\frac{M_0\sqrt{2}\left(\hat{x}_{\nu}^{1/2}U^\dagger_{\mathrm{PMNS}}\right)_{ab}}{vc_{\beta}}<3.5$. Consequently, $x_k=M_0^2/m^2_{H_k}$ should not be too large so that $m_{H_k}$ are larger  than the lower bounds from experiments.

In general, the allowed regions of parameter space depend strongly on the $\hat{x}_{\nu}$, namely larger $\hat{x}_{\nu}$ will allow larger $t_{\beta}$, and smaller values of other parameters including $ s_{2\alpha}\equiv 2 s_{\alpha}c_{\alpha}$, $Y^{d}_{1}$, and $Y^{d}_{2}$. Defining that  $\hat{x}_{\nu}= \left(\hat{m}_{\nu}/m_{n_3}\right) \times \hat{x}_{\nu3}$ with $ \hat{x}_{\nu3}=\frac{m_{n_3}}{\mu_X}$ and fixed $m_{n_1}=0.01$ eV,  $a^{\mathrm{ISS}}_{e,\mu}(H^\pm)$ given in Eq. \eqref{eq_Hpm1} depends strongly on $\hat{x}_{\nu3}$. With large  $\hat{x}_{\nu3}  \in \left[10^{-3},\; 5\times 10^{-3} \right]$ the allowed ranges of free parameters are given in Table~\ref{t_allowedRhnu0a}.
%From file N2ppp,5000 point
\begin{table}[ht]
	\begin{tabular}{c|cccccccccc}
		&$t_{\beta }$ & $s_{\alpha } \left\{-,\;+ \right\}$ & $M_0$ [TeV] & $m_{H_1^{\pm }}$ [TeV] & $m_{H_2^{\pm }}$ [TeV] &  $Y_{1}^{d} \left\{-,\;+ \right\}$  &   $ Y_{2}^{d}\left\{-,\;+ \right\}$\\
		\hline 
		Min &$\;0.320$ &$ \{-0.987,\; 0.004\}$  & 0.194 & 0.806 & 0.801 &$ \{ -0.364,\; 0.019\}$ &$ \{ -2.95,\; 0.311\}$ \\
		\hline 
		Max & $\; 22.932$ & $ \{-0.039,\; 0.998 \}$& 4.998 & 49.67 & 49.03 &$ \{-0.018,\; 0.376\}$ & $ \{-0.388,\;  2.95\}$
	\end{tabular}
	\caption{ Allowed ranges of free parameters  with large $\;10^{-3}\leq  \hat{x}_{\nu3} = \frac{m_{n_3}}{\mu_X} \leq 5\times 10^{-3} $, the notations $-$ ($+$) denote the negative (positive) ranges of the allowed regions. } \label{t_allowedRhnu0a}
\end{table}
We note that $s_{\alpha}$ never vanishes, namely large $\hat{x}_{\nu3} \le 5\times  10^{-3} $ can allow rather small  $|s_{\alpha}|\ge 10^{-3}$,  provided that $t_{\beta}\to 0.3$.   This property distinguishes completely to the conclusion given in Ref. \cite{Hue:2021xap}, where the allowed regions with fixed $s_{\alpha}=0$  require  a necessary condition of large  $t_{\beta}>30$. 

In this last discussion we will focus on the allowed regions consisting  of light masses of  heavy neutrinos and singly charged Higgs bosons. Namely in the ISS realization, heavy neutrinos  can be detected by future searches at colliders such as  Large Hadron Collider (LHC) and the International Linear Collider (ILC), and  Large Hadron electron Collider (LHeC) \cite{LHeCStudyGroup:2012zhm}, 
where the heavy neutrinos mass range from $\mathcal{O}(10^2)$ GeV to few TeV were discussed \cite{Das:2012ze, Das:2014jxa, Das:2015toa, Das:2016hof, Das:2018usr}.  Namely,  because of  the not too small mixing $\sim \sqrt{x_{\nu,3}}$ between ISS  and active neutrinos $\nu_{aL}$, the main production channel of heavy neutrinos $n_I$ (I=4,...,9) with mass $M_0$  at LHC is $u\bar{d}\to n_Ie^+_a$ through the $s$ channel  exchanging $W$ boson.  Then  the decay channel of  $n_I$ may be  $n_I\to e_a^-W^+,n_{a} Z,\; n_a h$, where $h$ is the standard model-like Higgs boson. The ILC can produce heavy neutrino in the processes $e^+e^-\to \bar{n}_a n_I$ through $t$ and $s$-channels exchanging the $W$ and $Z$ bosons, respectively. The model under consideration also predicts   a channel producing two heavy neutrinos $e^+e^-\to \bar{n}_I n_I$ through exchanging $H^\pm_k$. In the following numerical discussion, the allowed regions  are  defined as they result in the two values of  $ \Delta a^{\mathrm{ISS}}_{\mu}$ and $ \Delta a^{\mathrm{ISS}}_{e}$  satisfying both AMM experimental data of $\mu$ and electron at 1 $\sigma$ discrepancy level, and all  Yukawa couplings satisfy perturbative limits, $|(Y^{h}_2)_{ab}|,|Y^X_{ab}|\leq \sqrt{4\pi}$ with $a,b\le3$. The region  of parameter space used to scan  is chosen as follows: 
\begin{align}
	%last change M_0 range, 4,sep,2012, included in standard math file.
	%correct ranges of x1,x2: 17, Sep, 2021
	\label{eq_scanRanges}
 	&\; m_{H^\pm_1},\; m_{H^\pm_2} \ge 800 \;\mathrm{GeV};   10 \; \mathrm{GeV}\le  M_0\le \; 5\times 10^3\;  \mathrm{GeV}; \; 0.01\le x_1,x_2\le 100,
 	 \crn &0.3 \le t_{\beta}\leq50;\;  \left| s_{\alpha}\right|\le 1.; \; |Y^{d}_1|,\;|Y^{d}_2|\le 4.5;\; 10^{-7} \leq \; \hat{x}_{\nu3}=\frac{m_{n_3}}{\mu_X} \leq 10^{-3}.
\end{align}
The scanning range of  $\hat{x}_{\nu3}$ satisfies the non-unitary constraint given in Eq.~\eqref{eq_maxRRd}. The numerical results confirm  that $\left|	a^{\mathrm{ISS}}_{\mu,1}(H^\pm)/	a^{\mathrm{ISS}}_{\mu}(H^\pm) \right|<4\%$, and  $\left|	a^{\mathrm{ISS}}_{\mu,2}(H^\pm)/	a^{\mathrm{ISS}}_{\mu}(H^\pm) \right|<10^{-5}$. Therefore, these suppressed values are not shown in detail.
The allowed regions are more strict than the scanned region given in~\eqref{eq_scanRanges},  see Table~\ref{t_allowedR}. 
\begin{table}[ht]
	\begin{tabular}{c|ccccccccccc}
		&$t_{\beta }$ & $s_{\alpha }$ & $\hat{x}_{\nu3}$& $M_0$ [TeV] & $m_{H_1^{\pm }}$ [TeV] & $m_{H_2^{\pm }}$ [TeV] & $Y^d_{1}$ & $Y^d_{2}$\\
		\hline 
		Min &0.3 & -0.99 &$3.9\times 10^{-7}$ & 0.318 & 0.8 & 0.8  & -0.361 & -2.95 \\
		\hline 
		Max &  21.42 & 0.996 & $10^{-3}$ &5. & 48 & 48. & 0.352& 2.949 \\
	\end{tabular}
	\caption{ Allowed ranges of free parameters corresponding to the scanning region  \eqref{eq_scanRanges}.} \label{t_allowedR}
\end{table}
In addition,  values  of $|s_{\alpha}|$, $|Y^{d}_{1}|$, and $|Y^{d}_{2}|$ are bounded from below:
\begin{align}
	\label{eq_allowedRanges}
	s_{\alpha} &\in [-0.99,\; -0.029] \cup [0.026,\; 0.996] \to s_{2\alpha} \in [-1,\; -0.058] \cup [0.051,\; 1] , 
	\crn \; Y^{d}_{1} &\in [-0.361,\; -0.024] \cup [0.015,\;0.352], \; Y^{d}_{2} \in [-2.95,\;-0.176] \cup [0.523,\; 2.95],
\end{align}
where we define $s_{2\alpha}=2s_{\alpha} c_{\alpha}$.  Here although the lower bound of $t_{\beta}$ is the perturbative limit chosen in the scanned range, the upper bound is more strict than the largest value of  the scanned range $t_{\beta}<21.42<50$.  In general, the allowed regions require all lower bounds for free parameters  $\hat{x}_{\nu3}\geq 3.9\times 10^{-7}$, $M_0 \geq 318.$ GeV, $|Y^{d}_{2}| \geq 0.176$, and $|s_{2\alpha}|>0.051$.  Values of $Y^{d}_{1}$ are bounded in a more strict range of $0.015<|Y^{d}_{1}|<0.352$.  The lower bound of $M_0$ supports  many promoting channels to search for heavy neutrinos at both LHC and ILC \cite{Das:2012ze, Das:2014jxa, Das:2015toa, Das:2016hof}. 

\begin{figure}[ht]
	\centering\begin{tabular}{cc}
		\includegraphics[width=7.5cm]{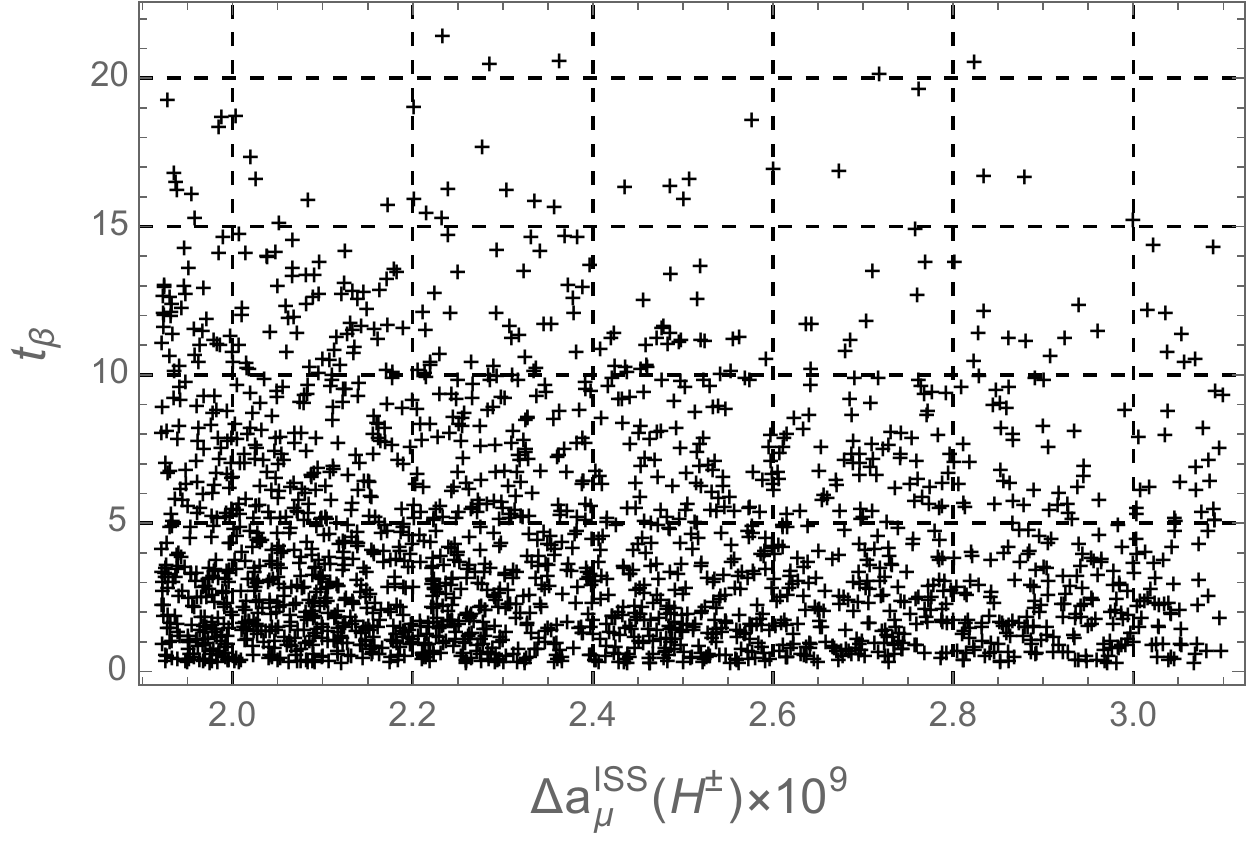}&	\includegraphics[width=7.8cm]{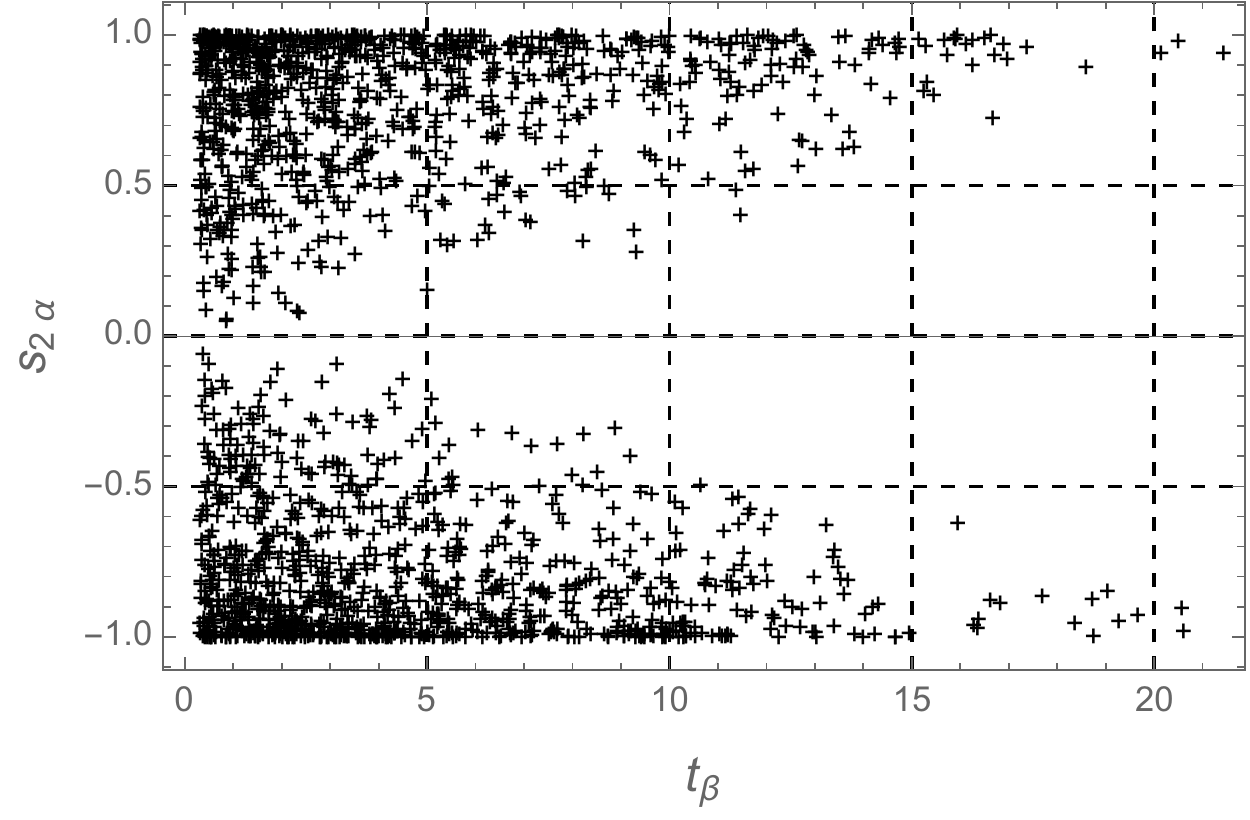}\\	
		\includegraphics[width=8.cm]{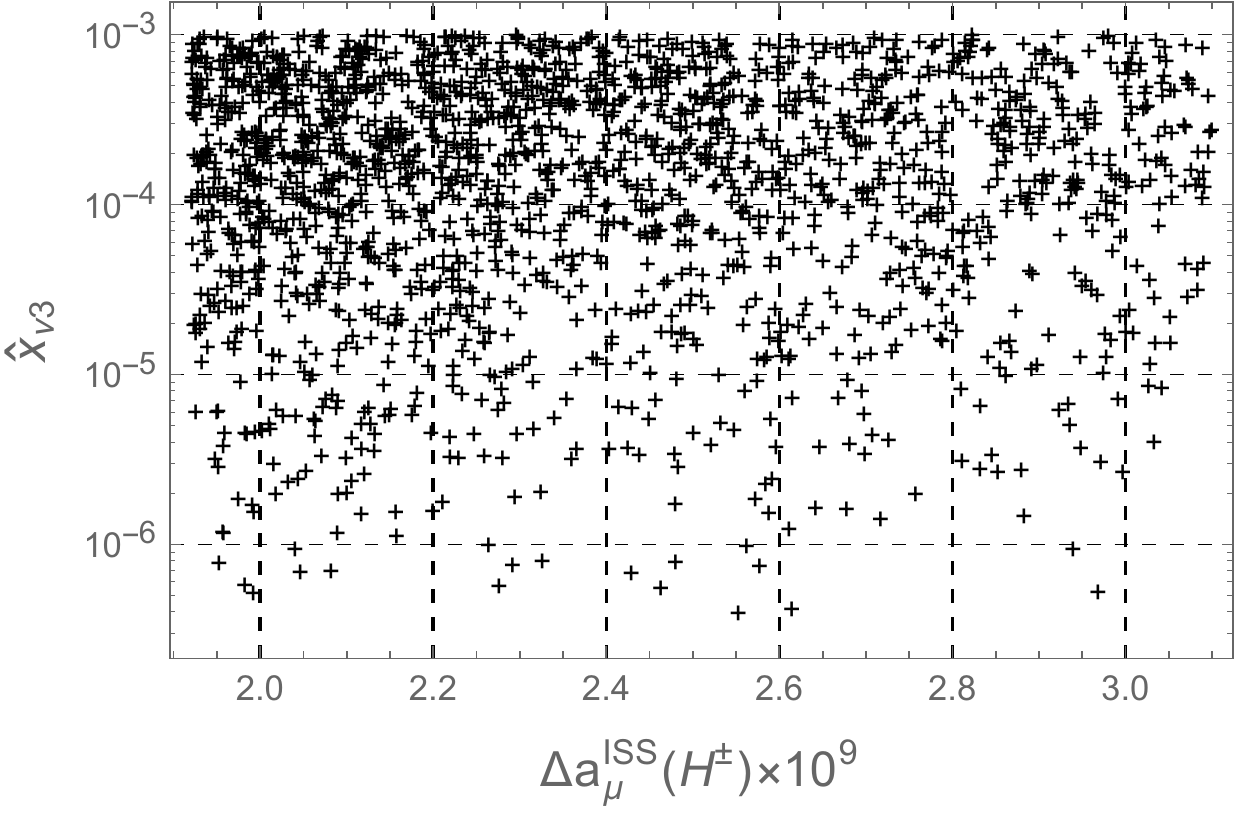}&	\includegraphics[width=7.5cm]{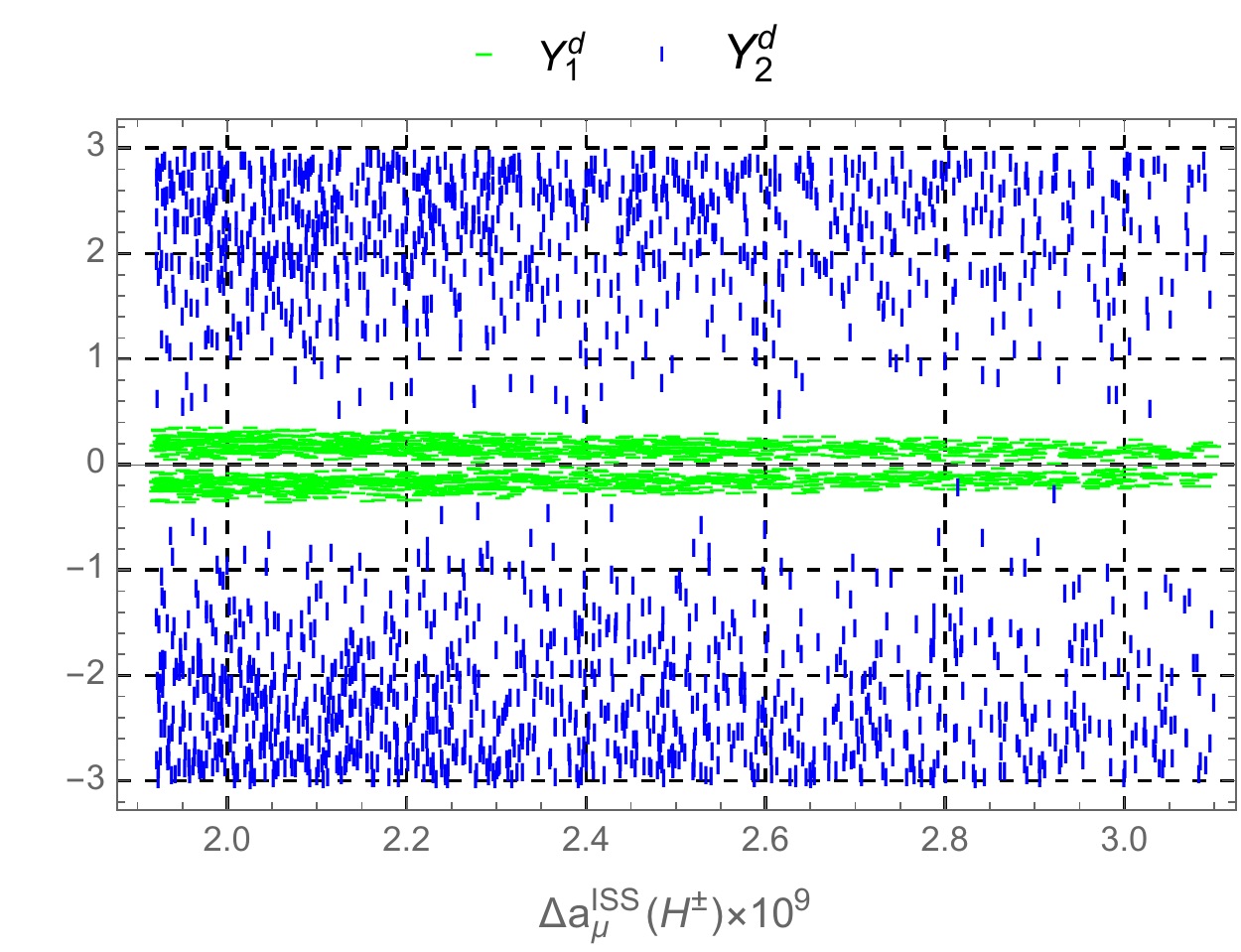}\\
	\end{tabular}
	\caption{ The correlations of free parameters  vs. $ \Delta a^{\mathrm{ISS}}_{\mu}(H^\pm)$ and $t_{\beta}$ in the allowed regions.  }\label{fig_amuX}
\end{figure}

\begin{figure}[ht]
	\centering\begin{tabular}{cc}
		\includegraphics[width=7.5cm]{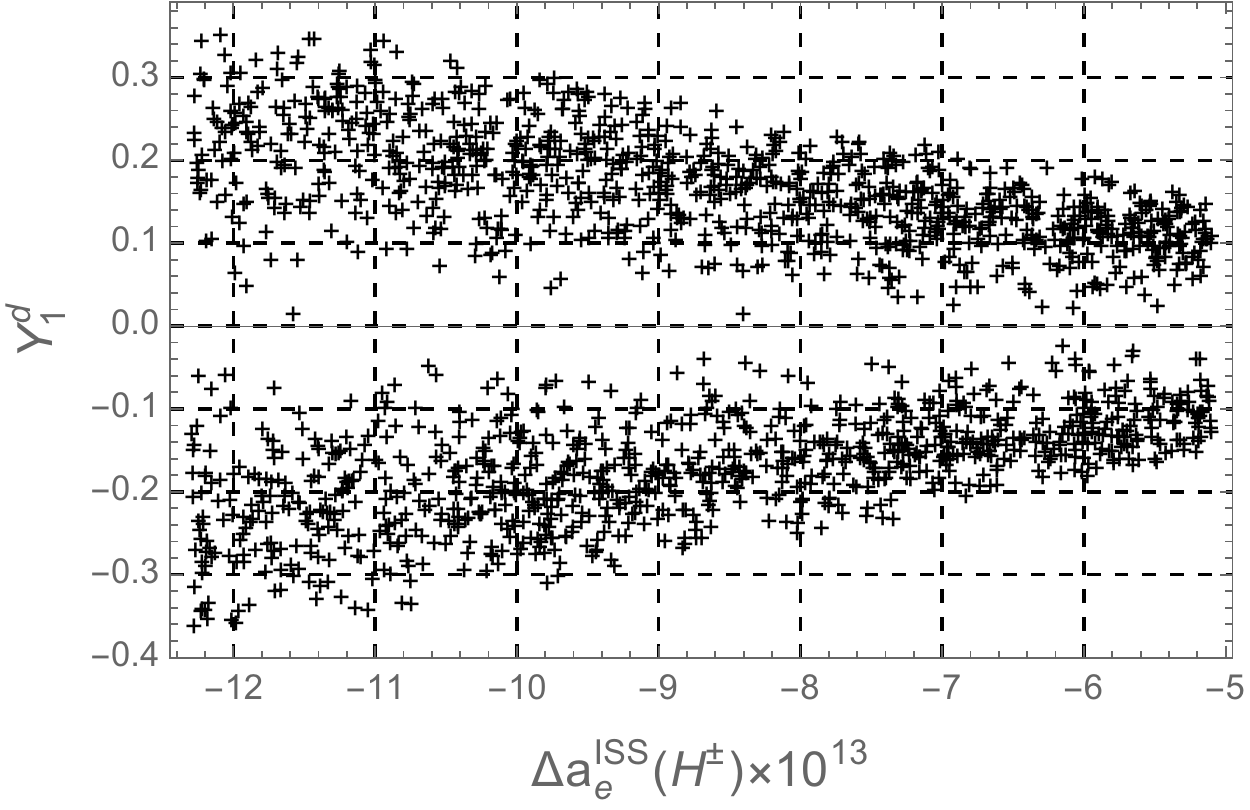}&
		\includegraphics[width=7.5cm]{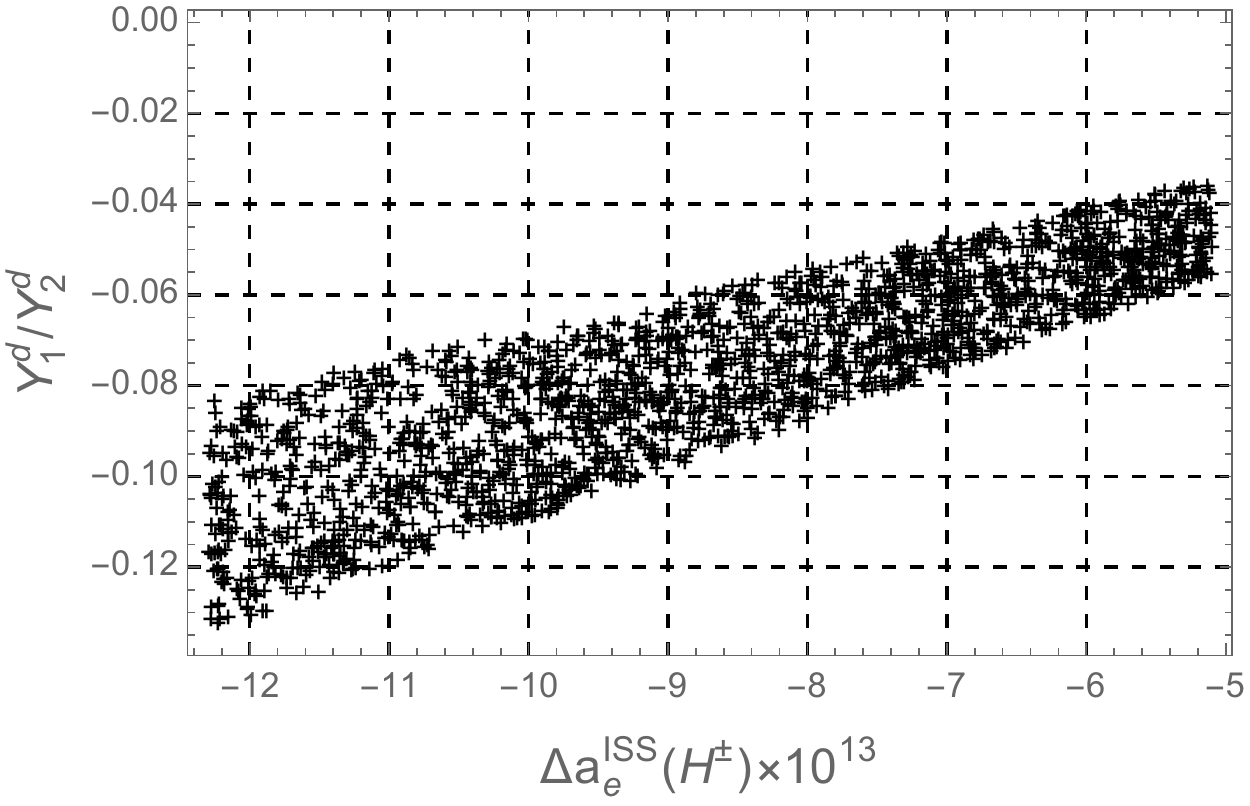}\\	
	\end{tabular}
	\caption{ The  correlations between $Y^d_1$ and   $Y^d_1/Y^d_2$ vs $\Delta a^{\mathrm{ISS}}_{e_a,0}$.  }\label{fig_aeYd}
\end{figure}

The correlations between  free parameters and $a^{\mathrm{ISS}}_{\mu}(H^\pm)$ in the allowed regions are illustrated in Fig.~\ref{fig_amuX} with 2000 allowed points collected. 
The correlations of  $ \Delta a_{e}(H^\pm)$ vs. $ \Delta a_{\mu}(H^\pm)$ can be seen from the correlations relating with  $Y^{d}_{1}$ shown in the lower right panel of Fig.~\ref{fig_amuX}. We can see that very large $t_{\beta}$ allows only small $ \Delta a_{\mu}(H^\pm)$.  The dependence of  $s_{\alpha}$, $Y^{d}_{1}$, and $Y^{d}_{2}$  on $ \Delta a_{\mu}(H^\pm)$ is rather weak. The Fig. 3 shows the consistent approximation we discussed above that $a^{\mathrm{ISS}}_{e_a}\simeq a^{\mathrm{ISS}}_{e_a,0}(H^\pm)\sim t^{-1}_{\beta} \hat{x}_{\nu_3}^{1/2} Y^d_as_{2\alpha}$.  Namely,  the two left panels prefer allowed points  with small $t_{\beta}$ and  large   $\hat{x}_{\nu_3}$. While large values of $ \Delta a_{\mu}$ near the  upper allowed bound exclude large $t_{\beta}$ and too small $\hat{x}_{\nu_3}$. In the upper right panel, large $t_{\beta}$ requires large $|s_{2\alpha}|$ so that the ratio $s_{2\alpha}/t_{\beta}$ is large enough to keep $ \Delta a_{\mu}$ in the allowed range.  In the lower right panel, we cannot realize  the linear dependence of $Y^d_2$ on $ \Delta a_{\mu}$  because many values of $Y^d_2$ are excluded by the pertubative limit of $Y^h_2$. On the other hand,   this property can be seen for $Y^d_1$ because of the relations $|Y^d_1| \sim |\Delta a_{\mu,0}/\Delta a_{e,0}|$.  In particularly, 
based on the two dominant contribution of AMM given in Eq. \eqref{eq_Hpm2} and \eqref{eq_aeHpm2}, $\Delta a^{\mathrm{ISS}}_{e_a,0}\simeq \Delta a^{\mathrm{ISS}}_{e_a}$,  it is easily to show that $|\Delta a^\mathrm{ISS}_{\mu}/\Delta a^{\mathrm{ISS}}_{e}|\simeq $$|\Delta a^\mathrm{ISS}_{\mu,0}/\Delta a^{\mathrm{ISS}}_{e,0}| =|m_{\mu}Y^d_2/(m_{e}Y^d_1)|$.  Identifying these with the  experimental data will lead to a consequence that $|Y^d_1|= |m_{\mu}Y^d_2\Delta a^{\mathrm{NP}}_e/(m_{e}\Delta a^{\mathrm{NP}}_\mu)| \simeq \mathcal{O}(10^{-2})|Y^d_2|$. Therefore, $|Y_1|$ can get small values of $\mathcal{O}(10^{-2})$.  Illustrations  are shown in  Fig. \ref{fig_aeYd},
where the left panel shows that  $|Y^{d_1}|\sim \Delta a^{\mathrm{ISS}}_{e_a,0}$ depends nearly linearly on $\Delta a^{\mathrm{ISS}}_{e}$. The right panel shows the valid of the relation $\Delta a^{\mathrm{ISS}}_{e}\simeq a^{\mathrm{ISS}}_{e,0} \sim Y^d_1/Y^d_2$ we mentioned above. The band widths  appear in the plots originate  from the $1\sigma$ ranges of AMM experimental data.   

It is also emphasized that $Y^d_{1,2}$ may give loop corrections to lepton masses $m_{e,\mu}$ \cite{Baker:2020vkh, Baker:2021yli}, where large  $|Y^d_{1,2}|$ may lead to the fine-tuning problem that loop corrections  $\delta m_{\mu,e}\gg m_{\mu,e}$.  Our model considered here has the same property with the models in class I with new neutral lepton having $Y_\psi=0$.  Discussions in Ref.  \cite{Baker:2021yli} suggest  that   the allowed regions we discussed above may consist of points with small $|Y^d_{1,2}|$ enough to avoid this fine-tuning.  Determining exactly these regions of the parameter space should be done in the future.

\begin{figure}[ht]
	\centering\begin{tabular}{cc}
		\includegraphics[width=7.5cm]{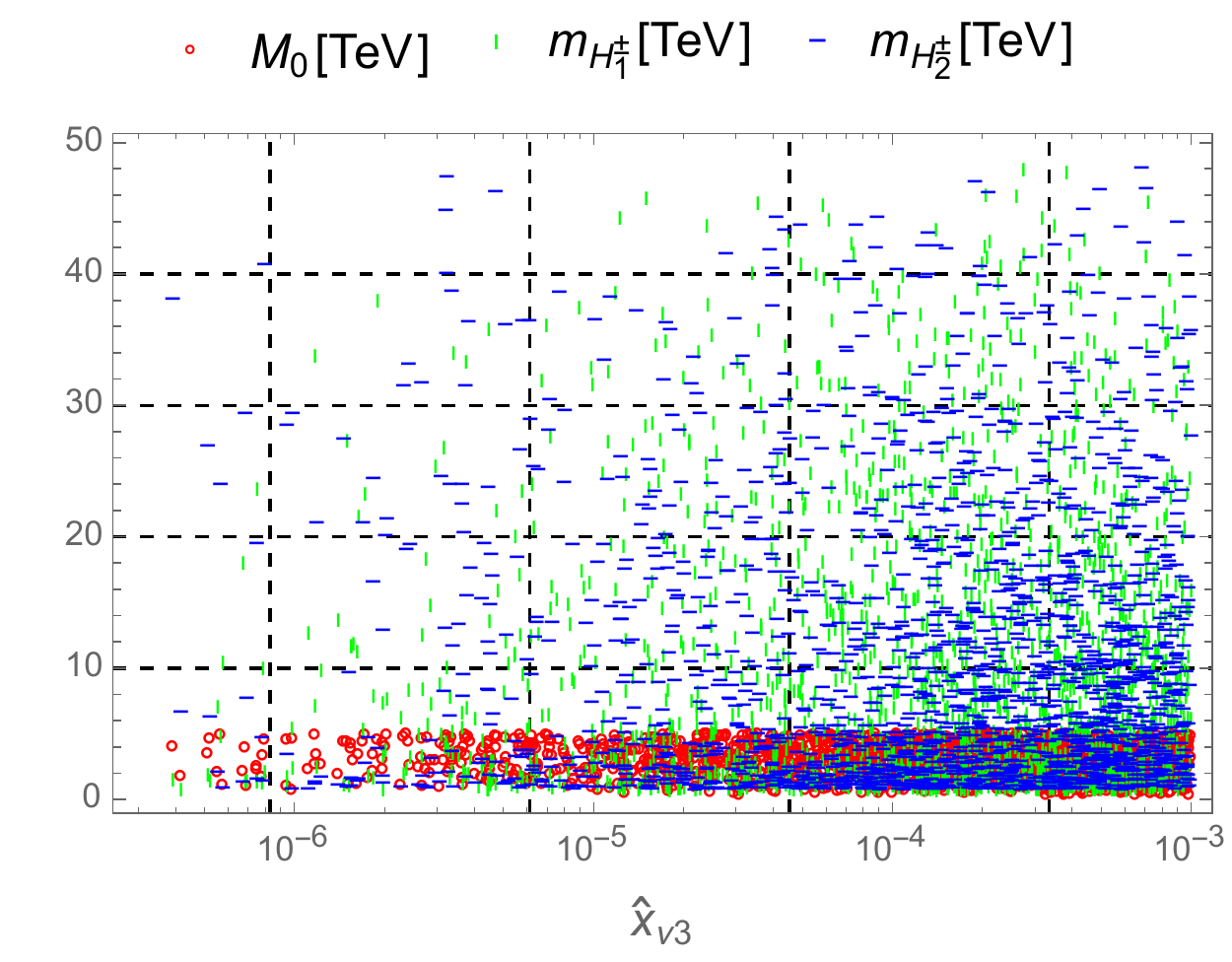}&	\includegraphics[width=7.5cm]{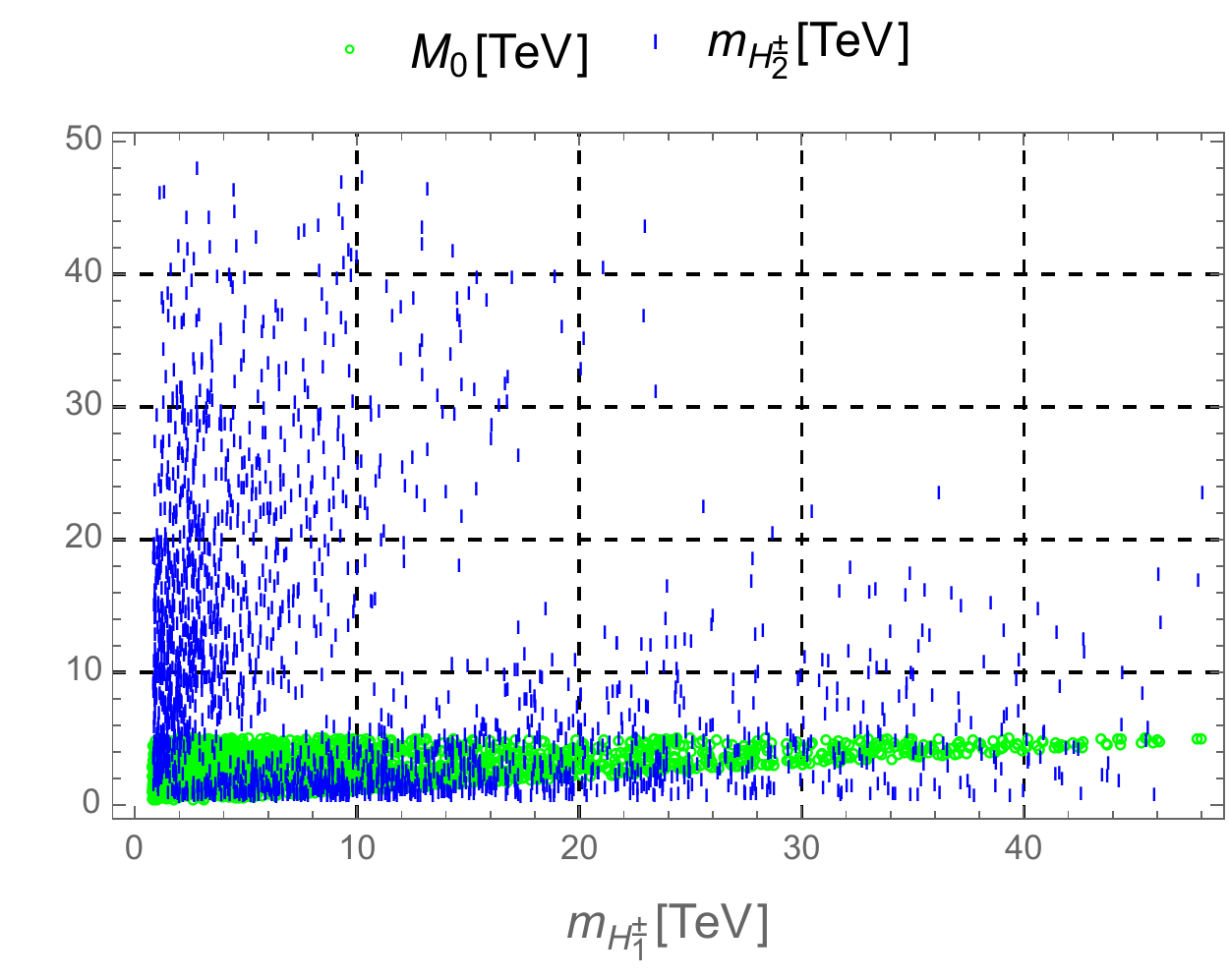}\\	
	\end{tabular}
	\caption{  The  correlations between different masses  vs $\hat{x}_{\nu3}$ (left panel) and $m_{H^\pm_1}$ (right panel).} \label{fig_mX}
\end{figure}

The mass parameters $M_{0}$ and $m_{H^\pm_{1,2}}$  are  independent with $ \Delta a_{\mu}(H^\pm)$ in the allowed regions.  It is more interesting  to see  the relations between two singly charged  Higgs boson masses and $M_0$, and between $\hat{x}_{\nu 3}$ and two   charged Higgs boson masses and $M_0$, see  Fig.  \ref{fig_mX}. 
In the left panel, small $\hat{x}_{\nu 3}$ is disfavored and allowed with only large $M_0$ up to the upper bound of the scanned range. In the right panel, the allowed region favors  both small values of  $m_{H^\pm_1}$ and $m_{H^\pm_2}$, but requires  $|m_{H^\pm_1}-m_{H^\pm_2}|\geq 252.4$ GeV.  

\begin{figure}[ht]
	\centering\begin{tabular}{ccc}
		\includegraphics[width=5.5cm]{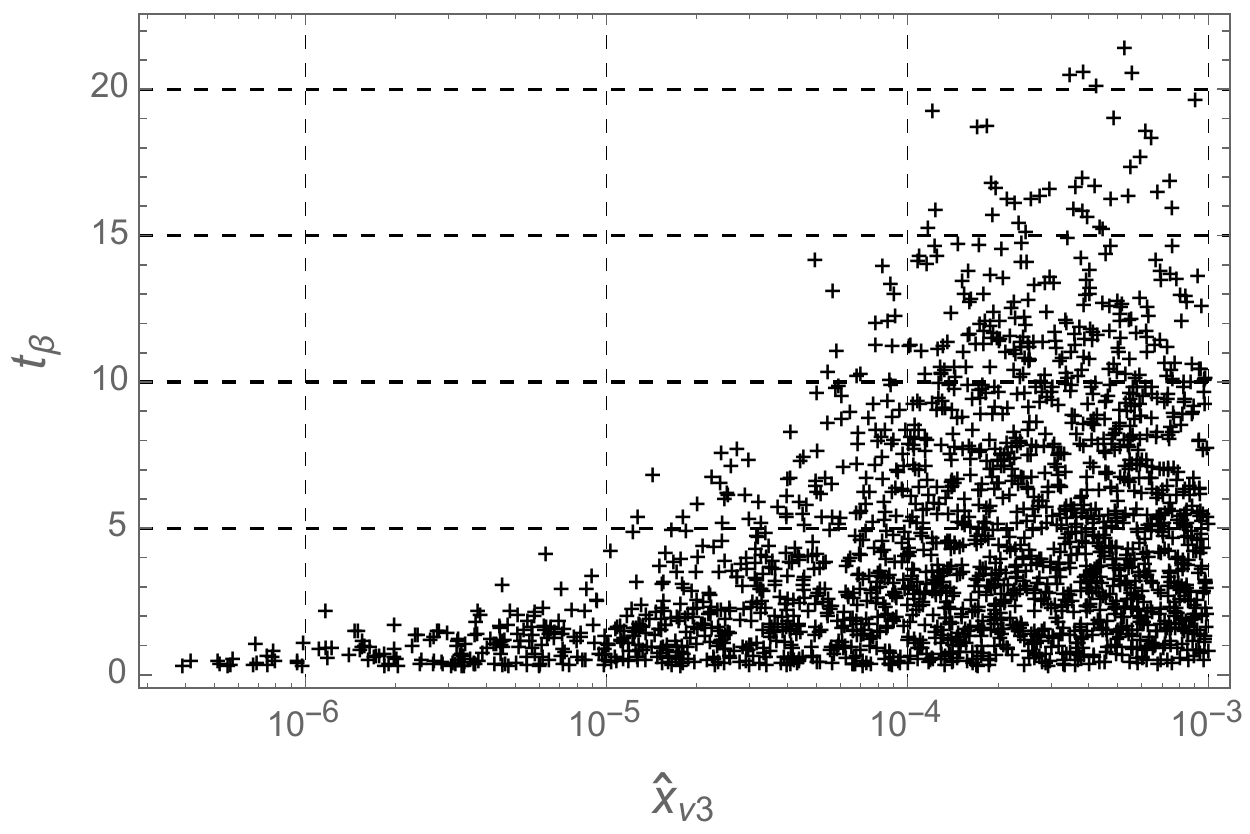}&
		\includegraphics[width=5.8cm]{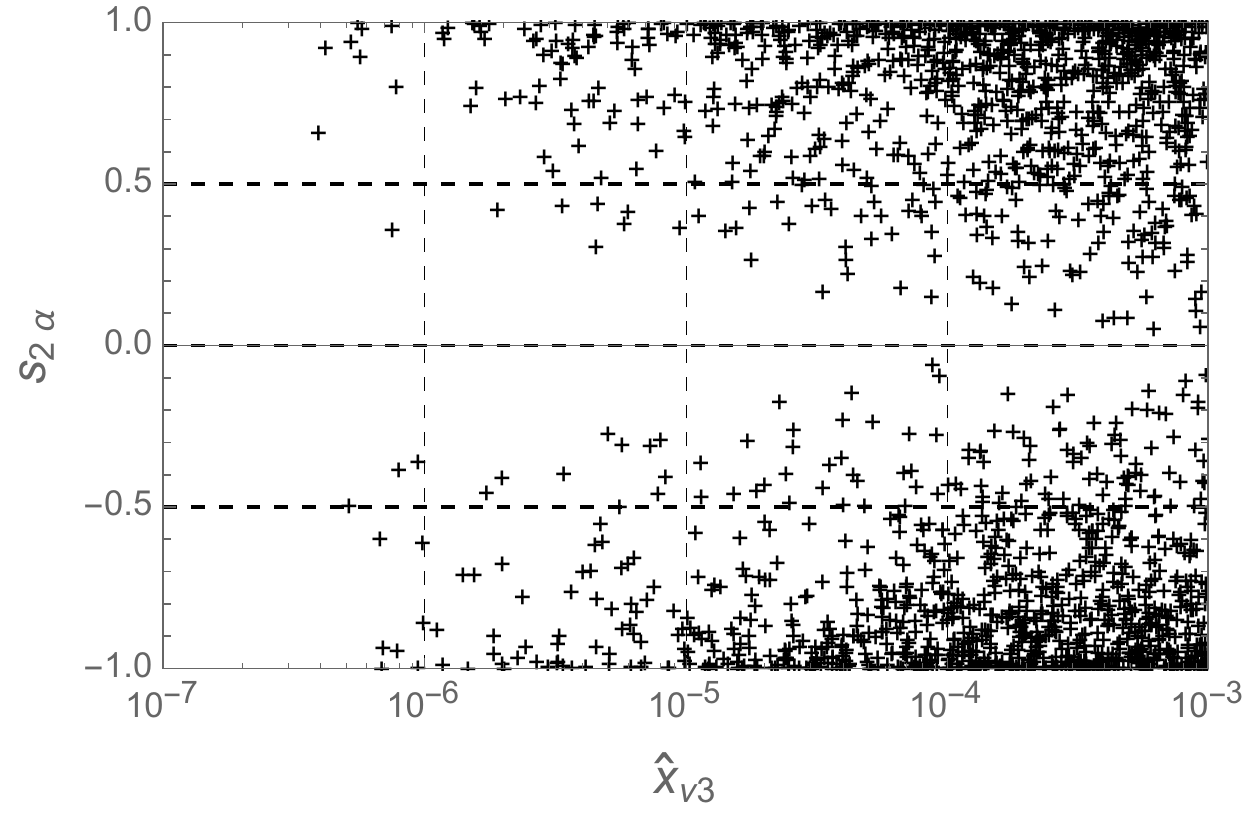}
		&	\includegraphics[width=5.5cm]{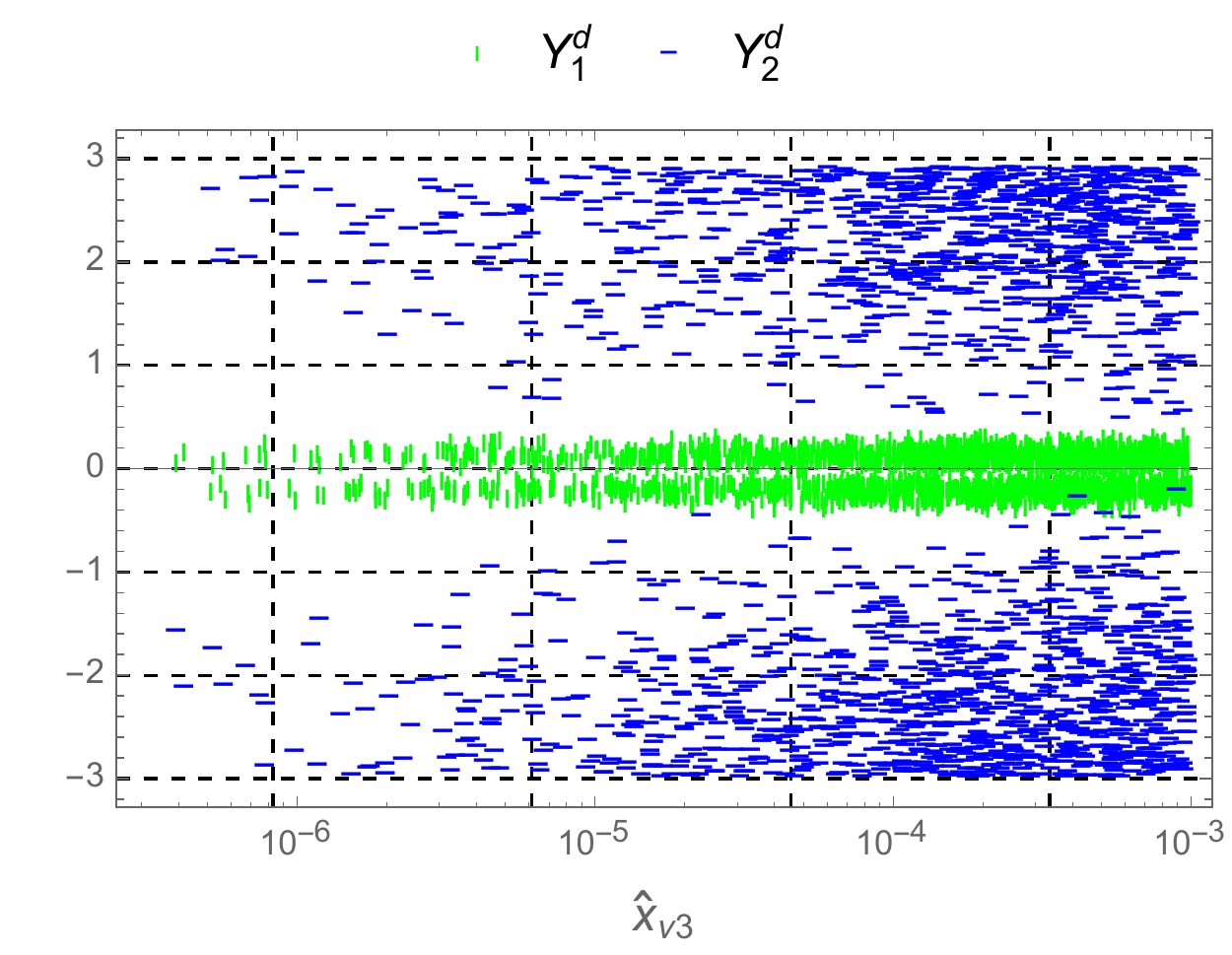}\\	
	\end{tabular}
	\caption{ The  correlations between different free parameters  vs $\hat{x}_{\nu3}$.  }\label{fig_fbetaX}
\end{figure}

Other interesting correlations  between different  free parameters versus   $\hat{x}_{\nu3}$ are shown in  Fig.~\ref{fig_fbetaX}. 
First, the allowed regions favor  large $\hat{x}_{\nu3}$, which supports small $s_{2\alpha}$ and large $t_{\beta}$.  In addition, careful numerical investigations show that the recent constraint on  $\hat{x}_{\nu3}$ given in Eq. \eqref{eq_maxRRd} does not allow  $s_{\alpha}=0$ or $t_{\beta}>30$. This conclusion excludes completely the allowed regions indicated in Ref. \cite{Hue:2021xap}, where large $t_{\beta}>30$ is one of the  necessary requirements to explain the experimental $(g-2)_{\mu}$ data. This important difference appears because of  the different Higgs triplets in the Yukawa term generating $M_D$ and Higgs couplings, depending on which models 331ISS or 331$\beta$. The future update on $ \hat{x}_{\nu3}$ will lead to a significant lower bound of $s_{2\alpha}$, for example $ \hat{x}_{\nu3}\le  10^{-4}$ will result in $|s_{2\alpha}|\ge0.07$, $t_{\beta}\le15$, and $|Y^d_2|\ge0.4$. On the other hand, small  $\hat{x}_{\nu3} <5. 10^{-7}$ requires simultaneously small $t_{\beta}>0.3$, large $|s_{2\alpha}|\to 1$, and large $Y^d_2$ corresponding to max$| \left(Y^{h}_{2}\right)_{ab}| \to 3.0$. This is the reason why $\hat{x}_{\nu3}$ must be bounded from below.

\section{ \label{eq_conclusion} Conclusion}

 The two models $331\beta$ under consideration and 331ISS given in Ref.~\cite{Hue:2021xap}  have two identical Yukawa couplings generating masses to charged leptons and top quarks, therefore  keep  the same lower bound $t_{\beta}>0.3$. But they predict two opposite  ranges of $t_{\beta}$ in the regions explaining successfully the experimental data of $(g-2)_{e,\mu}$. Namely, the regions predicted by the 331$\beta$ model   requires small $t_{\beta}<30$, in contrast to the requirement of  $t_{\beta}>30$ indicated for  the 331ISS model. These opposite  predictions of allowed $t_{\beta}$  depend on which Higgs triplets appear in the Yukawa terms needed to  generate Dirac neutrino mass matrix and couplings of singly charged Higgs bosons.

We have indicated that the $331\beta$ model adding heavy neutrinos and singly charged Higgs bosons $h^\pm$ as $SU(3)_L$ singlets  can explain both experimental data of $(g-2)_{\mu,e}$ in the ISS framework. Apart from the well-known property that this mechanism generates active neutrino masses and mixing consistent with neutrino oscillations data, it also allows both large values of non-unitary mixing parameters and heavy neutrino masses larger than order of $\mathcal{O}(10^2)$ GeV.  The new singly charged Higgs bosons as $SU(3)_L$ singlets will mix with the other  Higgs components predicted by the 331$\beta$ model, leading to new free couplings $Y^{h}$ of singly charged Higgs bosons with heavy ISS neutrinos and charged leptons. All of these features  result in the  chirally-enhanced one-loop contributions from heavy ISS neutrino exchanges to the AMM of electron and muon.  These contributions can be  large up to the order of $\Delta a^{\mathrm{NP}}_{e,\mu}$.  We have confirmed this conclusion from numerical illustrations in the limits of the simplest forms of the total neutrino mass matrix  and the Yukawa coupling matrix $Y^{h}$  needed  to avoid large Br$(e_b\to e_a\gamma)$. The phenomenology of the model $331\beta$ will be richer when these limits are relaxed, and should be studied in more detail.. 

\section*{Acknowledgments}
We thank    Prof. Arindam, Prof.  Kei Yagyu, Prof. Hidezumi Terazaw, Dr J. M. Yang, Dr. Marcin Badziak,   Dr. Bogdan Malaescu, and Dr. Lei Wang  for useful and  interesting comments and communications.  L. T. Hue is thankful to  Van Lang University. This research is funded by the Vietnam National Foundation for Science and Technology Development (NAFOSTED) under the grant number 103.01-2019.387.

\appendix
\section{\label{app_CLR} One loop contribution to the form factor $c_{(ba)R}$ for cLFV decays $e_b\rightarrow e_a\gamma$ and $\Delta a_{e_a}$}
We  collect here the results given in Ref.~\cite{Crivellin:2018qmi}, which were used directly to construct our analytic formulas corresponding to the particular properties of the 3-3-1 models. 
The general Lagrangian for needed interactions ($b\equiv i$, $a\equiv f$):
\begin{align}
	\label{eq_lagiangian}
	\mathcal{L}_{\Phi}&= \overline{\Psi} \left( \Gamma_{\Psi \Phi}^{aL}P_L + \Gamma_{\Psi \Phi}^{aR}P_R\right)e_a \Phi^*+\mathrm{h.c.}, \crn 
	\mathcal{L}_{V}&= \overline{\Psi} \left( \Gamma_{\Psi V}^{aL} \gamma^{\mu}P_L + \Gamma_{\Psi V}^{aR} \gamma^{\mu} P_R\right)e_a V_{\mu}^*+\mathrm{h.c.},  
\end{align}
The form factors $c_{(ab)R}$ corresponding to the one-loop contribution of a boson $X$ coupling with a fermion $\psi$ and usual charged $e_a$ are:
\begin{align}
	\label{eq_cbaX}
	c^X_{(ab)R}&\equiv \frac{e}{16\pi^2 m^2_{X}} \left\{ \Gamma^{aL*}_{\Psi X} \Gamma^{bR}_{\Psi X} m_{\Psi} \left[  f_X \left( t_X\right) +Q g_X \left( t_X\right) \right] \right.  
	\crn  &+  \left.  \left[ m_{e_b} \Gamma^{aL*}_{\Psi X} \Gamma^{bL}_{\Psi X} + m_{e_a} \Gamma^{aR*}_{\Psi X} \Gamma^{bR}_{\Psi X} \right] \left[  \tilde{f}_X \left( t_X\right) +Q \tilde{g}_X \left( t_X\right) \right] \right\},  
\end{align}
where $X=\Phi,V$, $t_X\equiv m^2_{\Psi}/m^2_X$, $Q\equiv Q_\Psi$ is the electric charge of the fermion $\Psi$, and the master functions are 
\begin{align}
	\label{eq_MasterFunc}
	f_\Phi (x)&= 2\tilde{g}_\Phi(x)=\frac{x^2-1 -2x\ln x}{4(x-1)^3},\crn 
	g_\Phi&=\frac{x-1 -\ln x}{2(x-1)^2}, \crn 
	\tilde{f}_\Phi(x)&= \frac{2x^3 +3x^2 -6x +1 -6x^2 \ln x}{24(x-1)^4}, \crn
	f_V(x)&= \frac{x^3 -12 x^2 +15 x -4 +6 x^2\ln x}{4(x-1)^3}, \crn
	g_V(x)&=\frac{x^2-5x +4+3 x \ln x}{2(x-1)^2}, \crn
	\tilde{f}_V(x)&= \frac{-4x^4 +49x^3 -78 x^2 +43x -10 -18x^3\ln x}{24(x-1)^4}, \crn
	\tilde{g}_V(x)&= \frac{-3(x^3 -6x^2 +7x -2 +2x^2\ln x)}{(x-1)^3}. 
\end{align}
We note that except $g_{\Phi}(x)$, all of the remaining master functions given in \eqref{eq_MasterFunc} are bounded in finite ranges, namely $0\leq f_\Phi (x), g_\Phi (x), \tilde{f}_\Phi (x), f_V(x), g_V(x), -\tilde{f}_V(x), -\tilde{g}_V(x)\leq  a \leq 2$. Regarding $g_{\Phi}(x)$, although $\lim_{x\to0}g_{\Phi}(x)= \infty$, the appearance of the factor $m_{\Psi}/m^2_{X}$ along with this function will result in the fact that the relevant contributions should be calculated by the modified function $ g(x)\to \sqrt{x}g_{\Phi}(x)$ that is always finite and have bound $0\leq\sqrt{x}g_{\Phi}(x)\leq \frac{1}{4} $. In the 3-3-1 models  discussed in this work, the modified function is $xg_{\Phi}(x)$ mentioned in Eq.~\eqref{eq_amuHA} is also finite for all $x$. Furthermore,  $\Gamma^{bR}_{\Psi V}=0$ for all charged gauge bosons $V=W,\; Y$ hence $f_V(x)$ and $g_V(x)$ do not appear in our calculation.

\section{\label{app_calculation} Detailed  steps of calculation}
The one-loop contributions of the singly charged Higgs bosons  to AMM is
\begin{align}
	\label{eq_aHpm}
	a_{e_a} (H^\pm_k)& = \frac{-f_a}{m^2_{H^\pm_k} }  \sum_{i=1}^{K+3} \left[ \lambda^{L,k*}_{ia } \lambda^{R,k}_{ia }m_{n_i} f_{\Phi}(x_{i,k}) + m_{e_a} \left(  \lambda^{L,k*}_{ia } \lambda^{L,k}_{ia } + \lambda^{R,k*}_{ia } \lambda^{R,k}_{ia }\right)  \tilde{f}_{\Phi}(x_{i,k}) \right]
	\crn&= \frac{-f_a}{m^2_{H^\pm_k} }  \left\{ \sum_{i=1}^{3} \left[ \lambda^{L,k*}_{ia } \lambda^{R,k}_{ia }m_{n_i} f_{\Phi}(0) + m_{e_a} \left(  \lambda^{L,k*}_{ia } \lambda^{L,k}_{ia } + \lambda^{R,k*}_{ia } \lambda^{R,k}_{ia }\right)  \tilde{f}_{\Phi}(0) \right] 
	\right.\crn&\left. +\sum_{i=4}^{K+3} \left[ \lambda^{L,k*}_{ia } \lambda^{R,k}_{ia }M_0 f_{\Phi}(x_{k}) + m_{e_a} \left(  \lambda^{L,k*}_{ia } \lambda^{L,k}_{ia } + \lambda^{R,k*}_{ia } \lambda^{R,k}_{ia }\right)  \tilde{f}_{\Phi}(x_{k}) \right]  \right\}, 
\end{align}
where $x_{i,k}\equiv m^2_{n_i}/m^2_{H^\pm_k}$, $f_a= \frac{g^2m_{e_a}\;}{8 \pi^2 m^2_W} >0$.  
 Using the approximations that $m^2_{n_i}/m^2_{H^\pm_k}\simeq0$ for $i\leq3$, otherwise   $m^2_{n_i}/m^2_{H^\pm_k}\simeq M_0^2/m^2_{H^\pm_k}=x_k$, we have $ f_{\Phi}(x_{i,k})\simeq f_{\Phi}(0)$ for $i\leq 3$ and $ f_{\Phi}(x_{i,k})\simeq f_{\Phi}(x_k)$ for $i>3$, leading to the precise analytic formulas  for different lef-right parts as follows
% 22 Aug 2021
\begin{align}
	\label{eq_aea0}
	%%%%%%LR: Ok, check 13, Sep, 2021
	&\sum_{i=1}^{K+3}\lambda^{L,1*}_{ia } \lambda^{R,1}_{ia }m_{n_i} f_{\Phi}(x_{i,1})
	\crn&=  \left\{ -m_{e_a} c^2_{\alpha} \left[ M_D^{\dagger}R^Tm_{\nu}\left( I_3-\frac{1}{2}RR^{\dagger}\right)\right]_{aa} +\frac{v}{\sqrt{2}}t_{\beta}^{-1} s_{\alpha}c_{\alpha} \left[ M_D^{\dagger}R^Tm_{\nu}R Y^{h}\right]_{aa} \right\} f_{\Phi}(0)
	%i>3
	\crn &+ \left\{ m_{e_a}c^2_{\alpha}\left[ M_D^{\dagger}\left( I_K-\frac{1}{2}R^TR^{*} \right) V^*V^{\dagger}R^{\dagger}\right]_{aa}
	\right.\crn&\quad\;\left.+ \frac{v}{\sqrt{2}}t^{-1}_{\beta}c_{\alpha}s_{\alpha}\left[ M_D^{\dagger}\left(I_K-\frac{1}{2} R^TR^*\right)V^*V^{\dagger}\left(I_K-\frac{1}{2} R^{\dagger}R\right)Y^{h}\right]_{aa} \right\}M_0f_{\Phi}(x_{1}),
	%%%%%LL: LL, Ok, check 13, Sep, 2021
	\crn &	\sum_{i=1}^{K+3} m_{e_a}\lambda^{L,1*}_{ia } \lambda^{L,1}_{ia } \tilde{f}_{\Phi}(x_{i,k})
	\crn&= m_{e_a}t_{\beta}^{-2}c_{\alpha}^2 \left\{ \left( M_D^{\dagger}R^TR^*M_D\right)_{aa} \tilde{f}_{\Phi}(0) 
	+ \left[ M_D^{\dagger}\left(I_K- \frac{1}{2}R^TR^*\right)^2M_D\right]_{aa} \tilde{f}_{\Phi}(x_1)\right\}, 
	%%%%23Aug2021 lan3
	\crn &	\sum_{i=1}^{K+3} \lambda^{R,1*}_{ia } \lambda^{R,1}_{ia } \tilde{f}_{\Phi}(x_{i,k})= m^2_{e_a}t^2_{\beta}c^2_{\alpha}\left[ \left( I_3-\frac{1}{2}RR^{\dagger}\right)^2_{aa} \tilde{f}_{\Phi}(0)  +  \left(RR^{\dagger}\right)_{aa}\tilde{f}_{\Phi}(x_1)  \right]
	\crn&+\frac{v^2s^2_{\alpha}}{2} \left\{  \left(Y^{h\dagger}R^{\dagger} R Y^{h}\right)_{aa}\tilde{f}_{\Phi}(0)  + \left[Y^{h\dagger} \left( I_K-\frac{1}{2}R^{\dagger}R\right)^2 Y^{h}\right]_{aa} \tilde{f}_{\Phi}(x_1) \right\}
	\crn&+\frac{v m_{e_a}t_{\beta} s_{2\alpha}}{\sqrt{2}} \mathrm{Re}\left\{ -\left[\left( I_3-\frac{RR^{\dagger}}{2}\right) RY^{h} \right]_{aa}\tilde{f}_{\Phi}(0) 
	%
%	\right.\crn& +\left.
+  \left[ R\left( I_K-\frac{R^{\dagger}R}{2}\right)Y^{h} \right]_{aa}\tilde{f}_{\Phi}(x_1) \right\},
\end{align}
where $s_{2\alpha}=2s_{\alpha}c_{\alpha}$, $U\equiv U_{\mathrm{PMNS}}$,  and $m_{\nu}\equiv U^*\hat{m}_{\nu}U^{\dagger} $. Ignoring suppressed term proportional to $\mathcal{O}(R^3)$ and setting $\tilde{f}_{\Phi}(0) =\frac{1}{24}$, we have 
\begin{align}
\label{eq_amu1}
	a_{e_a} (H^\pm_1) =- \frac{g^2 m_{e_a}^2}{8\pi^2 m^2_{W} }&\mathrm{Re}\left\{ \left[ c^2_{\alpha} \left( M_D^{\dagger} V^*V^{\dagger}R^{\dagger}\right)_{aa} + \frac{vt_{\beta}^{-1}c_{\alpha}s_{\alpha}}{\sqrt{2}m_{e_a}} \left(M_D^{\dagger}V^*V^{\dagger} Y^{h}\right)_{aa}\right]\frac{M_0 f_{\Phi}(x_1)}{m^2_{H^\pm_1}}
	\right.\crn&  \quad +t_{\beta}^{-2} c^2_{\alpha} \left[ \frac{\left( M_D^{\dagger}R^TR^*M_D \right)_{aa}}{m^2_{H^\pm_1}} \left( \frac{1}{24} -\tilde{f}_{\Phi}(x_1)  \right) + \frac{\left( M_D^{\dagger}M_D \right)_{aa}}{m^2_{H^\pm_1}}\tilde{f}_{\Phi}(x_1) \right]
	\crn & \quad+ \frac{m^2_{e_a} t_{\beta}^2c_{\alpha}^2}{m^2_{H^\pm_1}} \left[ \frac{1}{24} - (RR^{\dagger})_{aa}\left( \frac{1}{24} -\tilde{f}_{\Phi}(x_1) \right) \right]
	\crn & \quad+ \frac{v^2s_{\alpha}^2}{2 m^2_{H^\pm_1}} \left[ \left(Y^{h\dagger}R^{\dagger}RY^{h}\right)_{aa}\left(  \frac{1}{24} -\tilde{f}_{\Phi}(x_1)  \right)  + \left(Y^{h\dagger}Y^{h}\right)_{aa} \tilde{f}_{\Phi}(x_1)\right]
	\crn& \quad-\left.  \frac{v m_{e_a} t_{\beta} s_{2\alpha}}{\sqrt{2} m^2_{H^\pm_1}} \left[ \left(RY^{h}\right)_{aa}\left( \frac{1}{24}- \tilde{f}_{\Phi}(x_1)  \right)  \right] +\mathcal{O}(R^3)	\right\}, 
\end{align}
and 
\begin{equation}
	\label{eq_amu2}
	%----correct 16, Sep 2021
	a_{e_a} (H^\pm_2) =a_{e_a} (H^\pm_1)\left[ x_1 \to\; x_2,\;s_{\alpha}\to -c_{\alpha}, c_{\alpha} \to s_{\alpha}\right]. 
\end{equation}
 The total mixing matrices of neutrino corresponding to the MSS and ISS  frameworks are 
\begin{equation} 
	U^{\nu}= \left(
	\begin{array}{cc}
		U_{\mathrm{PMNS}}\left(	1 -\frac{\hat{m}_{\nu}}{2M_0} \right)  & 	-i U_{\mathrm{PMNS}}\left(	\frac{\hat{m}_{\nu}}{M_0} \right)^{1/2}  \\
		-i\left(	\frac{\hat{m}_{\nu}}{M_0} \right)^{1/2}  & 1 -\frac{\hat{m}_{\nu}}{2M_0}\\
	\end{array}
	\right) 
	\label{eq_UnuMSS}	
\end{equation}
and 
\begin{align}
	\label{eq_UnuISS}
	U^{\nu}=\begin{pmatrix}
		U_{\mathrm{PMNS}} \left( I_3-\frac{1}{2} \hat{x}_\nu\right)& \quad  i U_{\mathrm{PMNS}} \frac{\hat{x}_\nu^{1/2}}{\sqrt{2}} &  \quad U_{\mathrm{PMNS}}\frac{\hat{x}_\nu^{1/2}}{\sqrt{2}}\\
		0_{3\times 3}	& -\frac{i I_3}{\sqrt{2}} & \frac{I_3}{\sqrt{2}} \\
		-\hat{x}_\nu^{1/2}	&\frac{ i}{\sqrt{2}}\left( I_3 - \frac{\hat{x}_\nu}{2}  \right)& \frac{ 1}{\sqrt{2}}\left( I_3 - \frac{\hat{x}_\nu}{2} \right)
	\end{pmatrix}, 
\end{align}
respectively.  They satisfy the  unitary condition: $U^{\nu\dagger} U^{\nu}= U^{\nu} U^{\nu\dagger}=I_3  +\mathcal{O}\left(\left[\frac{\hat{m}_{\nu}}{M_0}\right]^{2} \right)$ and $U^{\nu\dagger} U^{\nu}= U^{\nu} U^{\nu\dagger}=I_9 +\mathcal{O}\left(\hat{x}_{\nu}^{2} \right)$. 
%
%\bibliographystyle{h-physrev}
%\bibliography{main} 

\begin{thebibliography}{99}
	%Phan trich-dan
	
\bibitem{Aoyama:2020ynm}
T.~Aoyama, N.~Asmussen, M.~Benayoun, J.~Bijnens, T.~Blum, M.~Bruno, I.~Caprini, C.~M.~Carloni Calame, M.~C\`e and G.~Colangelo, \textit{et al.}
%``The anomalous magnetic moment of the muon in the Standard Model,''
Phys. Rept. \textbf{887}, 1 (2020)
%doi:10.1016/j.physrep.2020.07.006
[arXiv:2006.04822 [hep-ph]].

%%%% 19 Sep 2021
\bibitem{Keshavarzi:2018mgv}
A.~Keshavarzi, D.~Nomura and T.~Teubner,
%``Muon $g-2$ and $\alpha(M_Z^2)$: a new data-based analysis,''
Phys. Rev. D \textbf{97}, no.11, 114025 (2018)
%doi:10.1103/PhysRevD.97.114025
[arXiv:1802.02995 [hep-ph]].

\bibitem{Colangelo:2018mtw}
G.~Colangelo, M.~Hoferichter and P.~Stoffer,
%``Two-pion contribution to hadronic vacuum polarization,''
JHEP \textbf{02}, 006 (2019)
%doi:10.1007/JHEP02(2019)006
[arXiv:1810.00007 [hep-ph]].

\bibitem{Hoferichter:2019mqg}
M.~Hoferichter, B.~L.~Hoid and B.~Kubis,
%``Three-pion contribution to hadronic vacuum polarization,''
JHEP \textbf{08}, 137 (2019)
%doi:10.1007/JHEP08(2019)137
[arXiv:1907.01556 [hep-ph]].

\bibitem{Davier:2019can}
M.~Davier, A.~Hoecker, B.~Malaescu and Z.~Zhang,
%``A new evaluation of the hadronic vacuum polarisation contributions to the muon anomalous magnetic moment and to $\mathbf{\boldsymbol\alpha(m_Z^2)}$,''
Eur. Phys. J. C \textbf{80}, no.3, 241 (2020)
[erratum: Eur. Phys. J. C \textbf{80}, no.5, 410 (2020)]
%doi:10.1140/epjc/s10052-020-7792-2
[arXiv:1908.00921 [hep-ph]].
%
\bibitem{Keshavarzi:2019abf}
A.~Keshavarzi, D.~Nomura and T.~Teubner,
%``$g-2$ of charged leptons, $\alpha (M^2_Z)$ , and the hyperfine splitting of muonium,''
Phys. Rev. D \textbf{101}, no.1, 014029 (2020)
%doi:10.1103/PhysRevD.101.014029
[arXiv:1911.00367 [hep-ph]].

\bibitem{Kurz:2014wya}
A.~Kurz, T.~Liu, P.~Marquard and M.~Steinhauser,
%``Hadronic contribution to the muon anomalous magnetic moment to next-to-next-to-leading order,''
Phys. Lett. B \textbf{734}, 144-147 (2014)
%doi:10.1016/j.physletb.2014.05.043
[arXiv:1403.6400 [hep-ph]].


\bibitem{Melnikov:2003xd}
K.~Melnikov and A.~Vainshtein,
%``Hadronic light-by-light scattering contribution to the muon anomalous magnetic moment revisited,''
Phys. Rev. D \textbf{70}, 113006 (2004)
%doi:10.1103/PhysRevD.70.113006
[arXiv:hep-ph/0312226 [hep-ph]].

%[19] P. Masjuan and P. Sanchez-Puertas, ´ Phys. Rev. D95, 054026 (2017), arXiv:1701.05829 [hep-ph].
\bibitem{Masjuan:2017tvw}
P.~Masjuan and P.~Sanchez-Puertas,
%``Pseudoscalar-pole contribution to the $(g_{\mu}-2)$: a rational approach,''
Phys. Rev. D \textbf{95}, no.5, 054026 (2017)
%doi:10.1103/PhysRevD.95.054026
[arXiv:1701.05829 [hep-ph]].

%[20] G. Colangelo, M. Hoferichter, M. Procura, and P. Stoffer, JHEP 04, 161 (2017), arXiv:1702.07347 [hep-ph]
\bibitem{Colangelo:2017fiz}
G.~Colangelo, M.~Hoferichter, M.~Procura and P.~Stoffer,
%``Dispersion relation for hadronic light-by-light scattering: two-pion contributions,''
JHEP \textbf{04}, 161 (2017)
%doi:10.1007/JHEP04(2017)161
[arXiv:1702.07347 [hep-ph]].

%[21] M. Hoferichter, B.-L. Hoid, B. Kubis, S. Leupold, and S. P. Schneider, JHEP 10, 141 (2018), arXiv:1808.04823 [hep-ph].

\bibitem{Hoferichter:2018kwz}
M.~Hoferichter, B.~L.~Hoid, B.~Kubis, S.~Leupold and S.~P.~Schneider,
%``Dispersion relation for hadronic light-by-light scattering: pion pole,''
JHEP \textbf{10}, 141 (2018)
%doi:10.1007/JHEP10(2018)141
[arXiv:1808.04823 [hep-ph]].

%[22] A. Gerardin, H. B. Meyer, and A. Ny ´ ffeler, Phys. Rev. D100, 034520 (2019), arXiv:1903.09471 [hep-lat].  
\bibitem{Gerardin:2019vio}
A.~G\'erardin, H.~B.~Meyer and A.~Nyffeler,
%``Lattice calculation of the pion transition form factor with $N_f=2+1$ Wilson quarks,''
Phys. Rev. D \textbf{100}, no.3, 034520 (2019)
%doi:10.1103/PhysRevD.100.034520
[arXiv:1903.09471 [hep-lat]].

%[23] J. Bijnens, N. Hermansson-Truedsson, and A. Rodr´ıguez-Sanchez, ´ Phys. Lett. B798, 134994 (2019), arXiv:1908.03331 [hep-ph].

\bibitem{Bijnens:2019ghy}
J.~Bijnens, N.~Hermansson-Truedsson and A.~Rodr\'\i{}guez-S\'anchez,
%``Short-distance constraints for the HLbL contribution to the muon anomalous magnetic moment,''
Phys. Lett. B \textbf{798}, 134994 (2019)
%doi:10.1016/j.physletb.2019.134994
[arXiv:1908.03331 [hep-ph]].

%[24] G. Colangelo, F. Hagelstein, M. Hoferichter, L. Laub, and P. Stoffer, JHEP 03, 101 (2020), arXiv:1910.13432 [hep-ph].
\bibitem{Colangelo:2019uex}
G.~Colangelo, F.~Hagelstein, M.~Hoferichter, L.~Laub and P.~Stoffer,
%``Longitudinal short-distance constraints for the hadronic light-by-light contribution to $(g-2)_\mu$ with large-$N_c$ Regge models,''
JHEP \textbf{03}, 101 (2020)
%doi:10.1007/JHEP03(2020)101
[arXiv:1910.13432 [hep-ph]].

%[31] G. Colangelo, M. Hoferichter, A. Nyffeler, M. Passera, and P. Stoffer, Phys. Lett. B735, 90 (2014), arXiv:1403.7512 [hep-ph]
\bibitem{Colangelo:2014qya}
G.~Colangelo, M.~Hoferichter, A.~Nyffeler, M.~Passera and P.~Stoffer,
%``Remarks on higher-order hadronic corrections to the muon g\ensuremath{-}2,''
Phys. Lett. B \textbf{735}, 90-91 (2014)
%doi:10.1016/j.physletb.2014.06.012
[arXiv:1403.7512 [hep-ph]].

%[32] T. Blum, N. Christ, M. Hayakawa, T. Izubuchi, L. Jin, C. Jung, and C. Lehner, Phys. Rev. Lett. 124, 132002 (2020), arXiv:1911.08123
\bibitem{Blum:2019ugy}
T.~Blum, N.~Christ, M.~Hayakawa, T.~Izubuchi, L.~Jin, C.~Jung and C.~Lehner,
%``Hadronic Light-by-Light Scattering Contribution to the Muon Anomalous Magnetic Moment from Lattice QCD,''
Phys. Rev. Lett. \textbf{124}, no.13, 132002 (2020)
%doi:10.1103/PhysRevLett.124.132002
[arXiv:1911.08123 [hep-lat]].

%[33] T. Aoyama, M. Hayakawa, T. Kinoshita, and M. Nio, Phys. Rev. Lett. 109, 111808 (2012), arXiv:1205.5370 [hep-ph].
\bibitem{Aoyama:2012wk}
T.~Aoyama, M.~Hayakawa, T.~Kinoshita and M.~Nio,
%``Complete Tenth-Order QED Contribution to the Muon g-2,''
Phys. Rev. Lett. \textbf{109}, 111808 (2012)
%doi:10.1103/PhysRevLett.109.111808
[arXiv:1205.5370 [hep-ph]].

%[34] T. Aoyama, T. Kinoshita, and M. Nio, Atoms 7, 28 (2019).
\bibitem{Aoyama:2019ryr}
T.~Aoyama, T.~Kinoshita and M.~Nio,
%``Theory of the Anomalous Magnetic Moment of the Electron,''
Atoms \textbf{7}, no.1, 28 (2019)
%doi:10.3390/atoms7010028

%[35] A. Czarnecki, W. J. Marciano, and A. Vainshtein, Phys. Rev. D67, 073006 (2003), [Erratum: Phys. Rev. D73, 119901 (2006)], arXiv:hepph/0212229 [hep-ph].
\bibitem{Czarnecki:2002nt}
A.~Czarnecki, W.~J.~Marciano and A.~Vainshtein,
%``Refinements in electroweak contributions to the muon anomalous magnetic moment,''
Phys. Rev. D \textbf{67}, 073006 (2003)
[erratum: Phys. Rev. D \textbf{73}, 119901 (2006)]
%doi:10.1103/PhysRevD.67.073006
[arXiv:hep-ph/0212229 [hep-ph]].

%[36] C. Gnendiger, D. Stockinger, and H. St ¨ ockinger-Kim, ¨ Phys. Rev. D88, 053005 (2013), arXiv:1306.5546 [hep-ph]
\bibitem{Gnendiger:2013pva}
C.~Gnendiger, D.~St\"ockinger and H.~St\"ockinger-Kim,
%``The electroweak contributions to $(g-2)_\mu$ after the Higgs boson mass measurement,''
Phys. Rev. D \textbf{88}, 053005 (2013)
%doi:10.1103/PhysRevD.88.053005
[arXiv:1306.5546 [hep-ph]].

%%%new references from communications, updated 19, Sep, 2021
%------Bogdan Malaescu,

%\bibitem{Davier:2019can}

\bibitem{Davier:2017zfy}
M.~Davier, A.~Hoecker, B.~Malaescu and Z.~Zhang,
%``Reevaluation of the hadronic vacuum polarisation contributions to the Standard Model predictions of the muon $g-2$ and ${\alpha (m_Z^2)}$ using newest hadronic cross-section data,''
Eur. Phys. J. C \textbf{77}, no.12, 827 (2017)
%doi:10.1140/epjc/s10052-017-5161-6
[arXiv:1706.09436 [hep-ph]].



\bibitem{Davier:2010nc}
M.~Davier, A.~Hoecker, B.~Malaescu and Z.~Zhang,
%``Reevaluation of the Hadronic Contributions to the Muon g-2 and to alpha(MZ),''
Eur. Phys. J. C \textbf{71}, 1515 (2011)
[erratum: Eur. Phys. J. C \textbf{72}, 1874 (2012)]
%doi:10.1140/epjc/s10052-012-1874-8
[arXiv:1010.4180 [hep-ph]].
%--------------end 19 Sep 2021

\bibitem{Borsanyi:2020mff}
S.~Borsanyi, Z.~Fodor, J.~N.~Guenther, C.~Hoelbling, S.~D.~Katz, L.~Lellouch, T.~Lippert, K.~Miura, L.~Parato and K.~K.~Szabo, \textit{et al.}
%``Leading hadronic contribution to the muon magnetic moment from lattice QCD,''
Nature \textbf{593}, no.7857, 51-55 (2021)
%doi:10.1038/s41586-021-03418-1
[arXiv:2002.12347 [hep-lat]].



\bibitem{Abi:2021gix}
B.~Abi \textit{et al.} [Muon g-2],
%``Measurement of the Positive Muon Anomalous Magnetic Moment to 0.46~ppm,''
Phys. Rev. Lett. \textbf{126},  141801 (2021)
%doi:10.1103/PhysRevLett.126.141801
[arXiv:2104.03281 [hep-ex]].

\bibitem{Muong-2:2006rrc}
G.~W.~Bennett \textit{et al.} [Muon g-2],
%``Final Report of the Muon E821 Anomalous Magnetic Moment Measurement at BNL,''
Phys. Rev. D \textbf{73}, 072003 (2006)
%doi:10.1103/PhysRevD.73.072003
[arXiv:hep-ex/0602035 [hep-ex]].



\bibitem{Hanneke:2008tm}
D.~Hanneke, S.~Fogwell and G.~Gabrielse,
%``New Measurement of the Electron Magnetic Moment and the Fine Structure Constant,''
Phys. Rev. Lett. \textbf{100}, 120801 (2008)
%doi:10.1103/PhysRevLett.100.120801
[arXiv:0801.1134 [physics.atom-ph]].

\bibitem{Parker:2018vye}
R.~H.~Parker, C.~Yu, W.~Zhong, B.~Estey and H.~M\"uller,
%``Measurement of the fine-structure constant as a test of the Standard Model,''
Science \textbf{360}, 191 (2018)
%doi:10.1126/science.aap7706
[arXiv:1812.04130 [physics.atom-ph]].

\bibitem{Morel:2020dww}
L.~Morel, Z.~Yao, P.~Clad\'e and S.~Guellati-Kh\'elifa,
%``Determination of the fine-structure constant with an accuracy of 81 parts per trillion,''
Nature \textbf{588}, no.7836, 61-65 (2020)
%doi:10.1038/s41586-020-2964-7

\bibitem{Gerardin:2020gpp}
A.~G\'erardin,
%``The anomalous magnetic moment of the muon: status of Lattice QCD calculations,''
Eur. Phys. J. A \textbf{57}, no.4, 116 (2021)
%doi:10.1140/epja/s10050-021-00426-7
[arXiv:2012.03931 [hep-lat]].

%%%%-ae in SM
\bibitem{Aoyama:2012wj}
T.~Aoyama, M.~Hayakawa, T.~Kinoshita and M.~Nio,
%``Tenth-Order QED Contribution to the Electron g-2 and an Improved Value of the Fine Structure Constant,''
Phys. Rev. Lett. \textbf{109}, 111807 (2012)
%doi:10.1103/PhysRevLett.109.111807
[arXiv:1205.5368 [hep-ph]].

%\bibitem{Aoyama:2012wk}
%T.~Aoyama, M.~Hayakawa, T.~Kinoshita and M.~Nio,
%``Complete Tenth-Order QED Contribution to the Muon g-2,''
%Phys. Rev. Lett. \textbf{109}, 111808 (2012)
%doi:10.1103/PhysRevLett.109.111808
%[arXiv:1205.5370 [hep-ph]].

\bibitem{Laporta:2017okg}
S.~Laporta,
%``High-precision calculation of the 4-loop contribution to the electron g-2 in QED,''
Phys. Lett. B \textbf{772}, 232-238 (2017)
%doi:10.1016/j.physletb.2017.06.056
[arXiv:1704.06996 [hep-ph]].


\bibitem{Aoyama:2017uqe}
T.~Aoyama, T.~Kinoshita and M.~Nio,
%``Revised and Improved Value of the QED Tenth-Order Electron Anomalous Magnetic Moment,''
Phys. Rev. D \textbf{97}, no.3, 036001 (2018)
%doi:10.1103/PhysRevD.97.036001
[arXiv:1712.06060 [hep-ph]].

%---------H.Terazawa
\bibitem{Terazawa:2018pdc}
H.~Terazawa,
%``Convergence of Perturbative Expansion Series in QED and the Muon g-2: One of the Oldest Problems in Quantum Field Theory and of the Latest Problems in the Standard Model,''
Nonlin. Phenom. Complex Syst. \textbf{21}, no.3, 268-272 (2018)

\bibitem{Volkov:2019phy}
S.~Volkov,
%``Calculating the five-loop QED contribution to the electron anomalous magnetic moment: Graphs without lepton loops,''
Phys. Rev. D \textbf{100}, no.9, 096004 (2019)
%doi:10.1103/PhysRevD.100.096004
[arXiv:1909.08015 [hep-ph]].

%g2-vector-like
\bibitem{Dermisek:2013gta}
R.~Dermisek and A.~Raval,
%``Explanation of the Muon g-2 Anomaly with Vectorlike Leptons and its Implications for Higgs Decays,''
Phys. Rev. D \textbf{88}, 013017 (2013)
%doi:10.1103/PhysRevD.88.013017
[arXiv:1305.3522 [hep-ph]].

\bibitem{Crivellin:2018qmi}
A.~Crivellin, M.~Hoferichter and P.~Schmidt-Wellenburg,
%``Combined explanations of $(g-2)_{\mu,e}$ and implications for a large muon EDM,''
Phys. Rev. D \textbf{98}, no.11, 113002 (2018)
%doi:10.1103/PhysRevD.98.113002
[arXiv:1807.11484 [hep-ph]].

\bibitem{Escribano:2021css} 
P.~Escribano, J.~Terol-Calvo and A.~Vicente,
%``$\boldsymbol{(g-2)_{e,\mu}}$ in an extended inverse type-III seesaw model,''
Phys. Rev. D \textbf{103}, no.11, 115018 (2021)
%doi:10.1103/PhysRevD.103.115018
[arXiv:2104.03705 [hep-ph]].

\bibitem{Hernandez:2021tii}
A.~E.~C.~Hern\'andez, S.~F.~King and H.~Lee,
%``Fermion mass hierarchies from vectorlike families with an extended 2HDM and a possible explanation for the electron and muon anomalous magnetic moments,''
Phys. Rev. D \textbf{103}, no.11, 115024 (2021)
%doi:10.1103/PhysRevD.103.115024
[arXiv:2101.05819 [hep-ph]].

\bibitem{Crivellin:2021rbq}
A.~Crivellin and M.~Hoferichter,
%``Consequences of chirally enhanced explanations of $(g-2)_\mu$ for $h\to \mu\mu$ and $Z\to \mu\mu$,''
JHEP \textbf{07}, 135 (2021)
%doi:10.1007/JHEP07(2021)135
[arXiv:2104.03202 [hep-ph]].

\bibitem{Dermisek:2021ajd}
R.~Dermisek, K.~Hermanek and N.~McGinnis,
%``Muon g-2 in two-Higgs-doublet models with vectorlike leptons,''
Phys. Rev. D \textbf{104}, no.5, 055033 (2021)
%doi:10.1103/PhysRevD.104.055033
[arXiv:2103.05645 [hep-ph]].

\bibitem{Chun:2020uzw}
E.~J.~Chun and T.~Mondal,
%``Explaining $g-2$ anomalies in two Higgs doublet model with vector-like leptons,''
JHEP \textbf{11}, 077 (2020)
%doi:10.1007/JHEP11(2020)077
[arXiv:2009.08314 [hep-ph]].

\bibitem{Frank:2020smf}
M.~Frank and I.~Saha,
%``Muon anomalous magnetic moment in two-Higgs-doublet models with vectorlike leptons,''
Phys. Rev. D \textbf{102}, no.11, 115034 (2020)
%doi:10.1103/PhysRevD.102.115034
[arXiv:2008.11909 [hep-ph]].

\bibitem{Endo:2020tkb}
M.~Endo and S.~Mishima,
%``Muon $g − 2$ and CKM unitarity in extra lepton models,''
JHEP \textbf{08}, no.08, 004 (2020)
%doi:10.1007/JHEP08(2020)004
[arXiv:2005.03933 [hep-ph]].

\bibitem{Cogollo:2020nrc}
D.~Cogollo, Y.~M.~Oviedo-Torres and Y.~S.~Villamizar,
%``Are 3-4-1 models able to explain the upcoming results of the muon anomalous magnetic moment?,''
Int. J. Mod. Phys. A \textbf{35}, no.23, 2050126 (2020)
%doi:10.1142/S0217751X20501262
[arXiv:2004.14792 [hep-ph]].

%---- K.~Yagyu added 19, Sep, 2021
\bibitem{Chen:2020tfr}
K.~F.~Chen, C.~W.~Chiang and K.~Yagyu,
%``An explanation for the muon and electron $g − 2$ anomalies and dark matter,''
JHEP \textbf{09}, 119 (2020)
%doi:10.1007/JHEP09(2020)119
[arXiv:2006.07929 [hep-ph]].

\bibitem{Bharadwaj:2021tgp}
H.~Bharadwaj, S.~Dutta and A.~Goyal,
%``Leptonic g \ensuremath{-} 2 anomaly in an extended Higgs sector with vector-like leptons,''
JHEP \textbf{11}, 056 (2021)
%doi:10.1007/JHEP11(2021)056
[arXiv:2109.02586 [hep-ph]].


%\letoquark
\bibitem{Crivellin:2020tsz}
A.~Crivellin, D.~Mueller and F.~Saturnino,
%``Correlating h\textrightarrow{}\ensuremath{\mu}+\ensuremath{\mu}- to the Anomalous Magnetic Moment of the Muon via Leptoquarks,''
Phys. Rev. Lett. \textbf{127}, no.2, 021801 (2021)
%doi:10.1103/PhysRevLett.127.021801
[arXiv:2008.02643 [hep-ph]].

\bibitem{Mondal:2021vou}
T.~Mondal and H.~Okada,
%``Inverse seesaw and (g \ensuremath{-} 2) anomalies in B \ensuremath{-} L extended two Higgs doublet model,''
Nucl. Phys. B \textbf{976}, 115716 (2022)
%doi:10.1016/j.nuclphysb.2022.115716
[arXiv:2103.13149 [hep-ph]].


\bibitem{Arbelaez:2020rbq}
C.~Arbel\'aez, R.~Cepedello, R.~M.~Fonseca and M.~Hirsch,
%``$(g-2)$ anomalies and neutrino mass,''
Phys. Rev. D \textbf{102}, no.7, 075005 (2020)
%doi:10.1103/PhysRevD.102.075005
[arXiv:2007.11007 [hep-ph]].

%%%--------2. Marcin Badziak

\bibitem{Badziak:2019gaf}
M.~Badziak and K.~Sakurai,
%``Explanation of electron and muon g \ensuremath{-} 2 anomalies in the MSSM,''
JHEP \textbf{10} (2019), 024
%doi:10.1007/JHEP10(2019)024
[arXiv:1908.03607 [hep-ph]].

%---J.~M.~Yang
\bibitem{Li:2021koa}S.~Li, Y.~Xiao and J.~M.~Yang,
%``Can electron and muon $g-2$ anomalies be jointly explained in SUSY?,''
Eur. Phys. J. C \textbf{82}, no.3, 276 (2022)
%doi:10.1140/epjc/s10052-022-10242-y
[arXiv:2107.04962 [hep-ph]].


\bibitem{Li:2020dbg}
S.~P.~Li, X.~Q.~Li, Y.~Y.~Li, Y.~D.~Yang and X.~Zhang,
%``Power-aligned 2HDM: a correlative perspective on $(g-2)_{e,\mu}$,''
JHEP \textbf{01}, 034 (2021)
%doi:10.1007/JHEP01(2021)034
[arXiv:2010.02799 [hep-ph]].

\bibitem{DelleRose:2020oaa}
L.~Delle Rose, S.~Khalil and S.~Moretti,
%``Explaining electron and muon $g$ \ensuremath{-} 2 anomalies in an Aligned 2-Higgs Doublet Model with right-handed neutrinos,''
Phys. Lett. B \textbf{816}, 136216 (2021)
%doi:10.1016/j.physletb.2021.136216
[arXiv:2012.06911 [hep-ph]].

\bibitem{Botella:2020xzf}
F.~J.~Botella, F.~Cornet-Gomez and M.~Nebot,
%``Electron and muon $g-2$ anomalies in general flavour conserving two Higgs doublets models,''
Phys. Rev. D \textbf{102}, no.3, 035023 (2020)
%doi:10.1103/PhysRevD.102.035023
[arXiv:2006.01934 [hep-ph]].




%%%----Lei Wang
\bibitem{Han:2018znu}
X.~F.~Han, T.~Li, L.~Wang and Y.~Zhang,
%``Simple interpretations of lepton anomalies in the lepton-specific inert two-Higgs-doublet model,''
Phys. Rev. D \textbf{99}, no.9, 095034 (2019)
%doi:10.1103/PhysRevD.99.095034
[arXiv:1812.02449 [hep-ph]].

\bibitem{Han:2021gfu}
X.~F.~Han, T.~Li, H.~X.~Wang, L.~Wang and Y.~Zhang,
%``Lepton-specific inert two-Higgs-doublet model confronted with the new results for muon and electron g-2 anomalies and multi-lepton searches at the LHC,''
Phys. Rev. D \textbf{104}, no.11, 115001 (2021)
%doi:10.1103/PhysRevD.104.115001
[arXiv:2104.03227 [hep-ph]].
%%%%%%%%%%%%%%%g-2 Higgs 


%%%%%%%%%%%%%%%%%331 old
\bibitem{Singer:1980sw}
M.~Singer, J.~W.~F.~Valle and J.~Schechter,
%``Canonical Neutral Current Predictions From the Weak Electromagnetic Gauge Group SU(3) X $u$(1),''
Phys. Rev. D \textbf{22}, 738 (1980)
%doi:10.1103/PhysRevD.22.738

\bibitem{Pisano:1992bxx}
F.~Pisano and V.~Pleitez,
%``An SU(3) x U(1) model for electroweak interactions,''
Phys. Rev. D \textbf{46}, 410-417 (1992)
%doi:10.1103/PhysRevD.46.410
[arXiv:hep-ph/9206242 [hep-ph]].

\bibitem{Frampton:1992wt}
P.~H.~Frampton,
%``Chiral dilepton model and the flavor question,''
Phys. Rev. Lett. \textbf{69}, 2889-2891 (1992)
%doi:10.1103/PhysRevLett.69.2889

\bibitem{Foot:1992rh}
R.~Foot, O.~F.~Hernandez, F.~Pisano and V.~Pleitez,
%``Lepton masses in an SU(3)-L x U(1)-N gauge model,''
Phys. Rev. D \textbf{47}, 4158-4161 (1993)
%doi:10.1103/PhysRevD.47.4158
[arXiv:hep-ph/9207264 [hep-ph]].

\bibitem{Pleitez:1992xh}
V.~Pleitez and M.~D.~Tonasse,
%``Heavy charged leptons in an SU(3)-L x U(1)-N model,''
Phys. Rev. D \textbf{48}, 2353-2355 (1993)
%doi:10.1103/PhysRevD.48.2353
[arXiv:hep-ph/9301232 [hep-ph]].

\bibitem{Foot:1994ym}
R.~Foot, H.~N.~Long and T.~A.~Tran,
%``$SU(3)_L \otimes U(1)_N$ and $SU(4)_L \otimes U(1)_N$ gauge models with right-handed neutrinos,''
Phys. Rev. D \textbf{50}, no.1, R34-R38 (1994)
%doi:10.1103/PhysRevD.50.R34
[arXiv:hep-ph/9402243 [hep-ph]].

\bibitem{Ozer:1995xi}
M.~Ozer,
%``SU(3)-L x U(1)-x model of the electroweak interactions without exotic quarks,''
Phys. Rev. D \textbf{54}, 1143-1149 (1996)
%doi:10.1103/PhysRevD.54.1143

\bibitem{Diaz:2004fs}
R.~A.~Diaz, R.~Martinez and F.~Ochoa,
%``SU(3)(c) x SU(3)(L) x U(1)(X) models for beta arbitrary and families with mirror fermions,''
Phys. Rev. D \textbf{72}, 035018 (2005)
%doi:10.1103/PhysRevD.72.035018
[arXiv:hep-ph/0411263 [hep-ph]].

\bibitem{Hue:2015mna}
L. T. Hue and L. D. Ninh,
%``The simplest 3-3-1 model,''
Mod. Phys. Lett. A \textbf{31},  1650062 (2016)
%doi:10.1142/S0217732316500620
[arXiv:1510.00302 [hep-ph]].

\bibitem{Fonseca:2016tbn}
R.~M.~Fonseca and M.~Hirsch,
%``A flipped 331 model,''
JHEP \textbf{08}, 003 (2016)
%doi:10.1007/JHEP08(2016)003
[arXiv:1606.01109 [hep-ph]].


 \bibitem{Buras:2012dp}
A.~J.~Buras, F.~De Fazio, J.~Girrbach and M.~V.~Carlucci,
%``The Anatomy of Quark Flavour Observables in 331 Models in the Flavour Precision Era,''
JHEP \textbf{02}, 023 (2013)
%doi:10.1007/JHEP02(2013)023
[arXiv:1211.1237 [hep-ph]].


%%%%%%%%%%%%%%%%%%%%%%%%%%%%%%g2-old 331
\bibitem{Ky:2000ku}
N.~A.~Ky, H.~N.~Long and D.~V. Soa,
%``Anomalous magnetic moment of muon in 3 3 1 models,''
Phys. Lett. B \textbf{486}, 140 (2000),
%doi:10.1016/S0370-2693(00)00696-1,
arXiv:hep-ph/0007010 [hep-ph].

\bibitem{Kelso:2014qka}
C.~Kelso, H.~N.~Long, R.~Martinez and F.~S.~Queiroz,
%``Connection of $g-2_{\mu}$, electroweak, dark matter, and collider constraints on 331 models,''
Phys. Rev. D \textbf{90}, no.11, 113011 (2014)
%doi:10.1103/PhysRevD.90.113011
arXiv:1408.6203 [hep-ph].

\bibitem{Binh:2015jfz}
D.~T.~Binh, D.~Huong, L. T.~Hue and H.~N.~Long,
%``Anomalous Magnetic Moment of Muon in Economical 3-3-1 Model,''
Commun. in Phys. \textbf{25}, no.1, 29-43 (2015)
%doi:10.15625/0868-3166/25/1/4582

\bibitem{DeConto:2016ith}
G.~De Conto and V.~Pleitez,
%``Electron and muon anomalous magnetic dipole moment in a 3\textendash{}3\textendash{}1 model,''
JHEP \textbf{05}, 104 (2017)
%doi:10.1007/JHEP05(2017)104
[arXiv:1603.09691 [hep-ph]].

\bibitem{deJesus:2020upp}
A.~S.~De Jesus, S.~Kovalenko, F.~S.~Queiroz, C.~Siqueira and K.~Sinha,
%``Vectorlike leptons and inert scalar triplet: Lepton flavor violation, $g-2$, and collider searches,''
Phys. Rev. D \textbf{102}, no.3, 035004 (2020)
%doi:10.1103/PhysRevD.102.035004
arXiv:2004.01200 [hep-ph].

\bibitem{deJesus:2020ngn}
\'A.~S.~de Jesus, S.~Kovalenko, C.~A.~de S.~Pires, F.~S.~Queiroz and Y.~S.~Villamizar,
%``Dead or alive? Implications of the muon anomalous magnetic moment for 3-3-1 models,''
Phys. Lett. B \textbf{809}, 135689 (2020)
%doi:10.1016/j.physletb.2020.135689,
arXiv:2003.06440 [hep-ph].

\bibitem{Lindner:2016bgg}
M.~Lindner, M.~Platscher and F.~S.~Queiroz,
%``A Call for New Physics : The Muon Anomalous Magnetic Moment and Lepton Flavor Violation,''
Phys. Rept. \textbf{731}, 1 (2018)
%doi:10.1016/j.physrep.2017.12.001
[arXiv:1610.06587 [hep-ph]].

\bibitem{Hue:2020wnn}
L.~Hue, P.~N.~Thanh and T.~D.~Tham,
%``Anomalous Magnetic Dipole Moment \((\mathrm{g}-2)\mu\) in 3-3-1 Model with Inverse Seesaw Neutrinos,''
Commun. in Phys. \textbf{30}, no.3, 221-230 (2020)
%doi:10.15625/0868-3166/30/3/14963

\bibitem{Hue:2021xap}
L.~T.~Hue, H.~T.~Hung, N.~T.~Tham, H.~N.~Long and T.~P.~Nguyen,
%``Large (g-2)\ensuremath{\mu} and signals of decays eb\textrightarrow{}ea\ensuremath{\gamma} in a 3-3-1 model with inverse seesaw neutrinos,''
Phys. Rev. D \textbf{104},  033007 (2021)
%doi:10.1103/PhysRevD.104.033007
[arXiv:2104.01840 [hep-ph]].

%--331 discrete
\bibitem{CarcamoHernandez:2019lhv}
A.~E.~C\'arcamo Hern\'andez, D.~T.~Huong and H.~N.~Long,
%``Minimal model for the fermion flavor structure, mass hierarchy, dark matter, leptogenesis, and the electron and muon anomalous magnetic moments,''
Phys. Rev. D \textbf{102},  055002 (2020)
%doi:10.1103/PhysRevD.102.055002
[arXiv:1910.12877 [hep-ph]].

\bibitem{CarcamoHernandez:2020pxw}
A.~E.~C\'arcamo Hern\'andez, Y.~Hidalgo Vel\'asquez, S.~Kovalenko, H.~N.~Long, N.~A.~P\'erez-Julve and V.~V.~Vien,
%``Fermion spectrum and $g-2$ anomalies in a low scale 3-3-1 model,''
Eur. Phys. J. C \textbf{81}, no.2, 191 (2021)
%doi:10.1140/epjc/s10052-021-08974-4
[arXiv:2002.07347 [hep-ph]].


\bibitem{Lavoura:2003xp}
L.~Lavoura,
%``General formulae for f(1) ---\ensuremath{>} f(2) gamma,''
Eur. Phys. J. C \textbf{29}, 191-195 (2003)
%doi:10.1140/epjc/s2003-01212-7
[arXiv:hep-ph/0302221 [hep-ph]].

% Lavoura:2003xp, Hue:2017lak
\bibitem{Hue:2017lak}
L.~T.~Hue, L.~D.~Ninh, T.~T.~Thuc and N.~T.~T.~Dat,
%``Exact one-loop results for $l_i \to l_j\gamma$ in 3-3-1 models,''
Eur. Phys. J. C \textbf{78}, no.2, 128 (2018)
%doi:10.1140/epjc/s10052-018-5589-3
[arXiv:1708.09723 [hep-ph]].

\bibitem{Long:2018fud}
H.~N.~Long, N.~V.~Hop, L.~T.~Hue and N.~T.~T.~Van,
%``Constraining heavy neutral gauge boson $Z'$ in the 3 - 3 - 1 models by weak charge data of Cesium and proton,''
Nucl. Phys. B \textbf{943}, 114629 (2019)
%doi:10.1016/j.nuclphysb.2019.114629
[arXiv:1812.08669 [hep-ph]].

\bibitem{Buras:2014yna}
A.~J.~Buras, F.~De Fazio and J.~Girrbach-Noe,
%``$Z$-$Z'$ mixing and $Z$-mediated FCNCs in $SU(3)_{C}  \times  SU(3)_{L}  \times U(1)_{X}$ models,''
JHEP \textbf{08}, 039 (2014)
%doi:10.1007/JHEP08(2014)039
[arXiv:1405.3850 [hep-ph]].

\bibitem{Buras:2016dxz}
A.~J.~Buras and F.~De Fazio,
%``331 Models Facing the Tensions in $\Delta F=2$ Processes with the Impact on $\varepsilon^\prime/\varepsilon$, $B_s\to\mu^+\mu^-$ and $B\to K^*\mu^+\mu^-$,''
JHEP \textbf{08}, 115 (2016)
%doi:10.1007/JHEP08(2016)115
[arXiv:1604.02344 [hep-ph]].

\bibitem{Buras:2015kwd}
A.~J.~Buras and F.~De Fazio,
%``$\varepsilon'/\varepsilon$ in 331 Models,''
JHEP \textbf{03}, 010 (2016)
%doi:10.1007/JHEP03(2016)010
[arXiv:1512.02869 [hep-ph]].

\bibitem{CarcamoHernandez:2005ka}
A.~E.~Carcamo Hernandez, R.~Martinez and F.~Ochoa,
%``Z and Z' decays with and without FCNC in 331 models,''
Phys. Rev. D \textbf{73}, 035007 (2006)
%doi:10.1103/PhysRevD.73.035007
[arXiv:hep-ph/0510421 [hep-ph]].



\bibitem{Descotes-Genon:2017ptp}
S.~Descotes-Genon, M.~Moscati and G.~Ricciardi,
%``Nonminimal 331 model for lepton flavor universality violation in $b{\rightarrow}s{\ell}{\ell}$ decays,''
Phys. Rev. D \textbf{98}, no.11, 115030 (2018)
%doi:10.1103/PhysRevD.98.115030
[arXiv:1711.03101 [hep-ph]].

\bibitem{Hue:2018dqf}
L. T. Hue and L. D. Ninh,
%``On the triplet anti-triplet symmetry in 3-3-1 models,''
Eur. Phys. J. C \textbf{79}, no.3, 221 (2019)
%doi:10.1140/epjc/s10052-019-6735-2
[arXiv:1812.07225 [hep-ph]].


 \bibitem{Hung:2019jue}
H.~T.~Hung, T.~T.~Hong, H.~H.~Phuong, H.~L.~T.~Mai and L.~T.~Hue,
%``Neutral Higgs decays $H \rightarrow Z \gamma,\gamma\gamma$ in 3-3-1 models,''
Phys. Rev. D \textbf{100}, no.7, 075014 (2019)
%doi:10.1103/PhysRevD.100.075014
[arXiv:1907.06735 [hep-ph]].


\bibitem{Allwicher:2021rtd}
L.~Allwicher, P.~Arnan, D.~Barducci and M.~Nardecchia,
%``Perturbative unitarity constraints on generic Yukawa interactions,''
JHEP \textbf{10}, 129 (2021)
%doi:10.1007/JHEP10(2021)129
[arXiv:2108.00013 [hep-ph]].


\bibitem{ParticleDataGroup:2020ssz}
P.~A.~Zyla \textit{et al.} [Particle Data Group],
%``Review of Particle Physics,''
PTEP \textbf{2020},  083C01 (2020)
%doi:10.1093/ptep/ptaa104

\bibitem{Casas:2001sr}
J.~A.~Casas and A.~Ibarra,
%``Oscillating neutrinos and $\mu \to e, \gamma$,''
Nucl. Phys. B \textbf{618}, 171 (2001)
%doi:10.1016/S0550-3213(01)00475-8
[arXiv:hep-ph/0103065 [hep-ph]].

\bibitem{Ibarra:2010xw}
A.~Ibarra, E.~Molinaro and S.~T.~Petcov,
%``TeV Scale See-Saw Mechanisms of Neutrino Mass Generation, the Majorana Nature of the Heavy Singlet Neutrinos and $(\beta\beta)_{0\nu}$-Decay,''
JHEP \textbf{09}, 108 (2010)
%doi:10.1007/JHEP09(2010)108
[arXiv:1007.2378 [hep-ph]].

\bibitem{Arganda:2014dta}
E.~Arganda, M.~J.~Herrero, X.~Marcano and C.~Weiland,
%``Imprints of massive inverse seesaw model neutrinos in lepton flavor violating Higgs boson decays,''
Phys. Rev. D \textbf{91},  015001 (2015)
%doi:10.1103/PhysRevD.91.015001
[arXiv:1405.4300 [hep-ph]].

\bibitem{Planck:2018vyg}
N.~Aghanim \textit{et al.} [Planck],
%``Planck 2018 results. VI. Cosmological parameters,''
Astron. Astrophys. \textbf{641}, A6 (2020)
[erratum: Astron. Astrophys. \textbf{652}, C4 (2021)]
%doi:10.1051/0004-6361/201833910
[arXiv:1807.06209 [astro-ph.CO]]

\bibitem{Pinheiro:2021mps}
J.~P.~Pinheiro, C.~A.~de S.~Pires, F.~S.~Queiroz and Y.~S.~Villamizar,
%``Confronting the inverse seesaw mechanism with the recent muon g-2 result,''
Phys. Lett. B \textbf{823}, 136764 (2021)
%doi:10.1016/j.physletb.2021.136764
[arXiv:2107.01315 [hep-ph]].

\bibitem{Fernandez-Martinez:2016lgt} 
E.~Fernandez-Martinez, J.~Hernandez-Garcia and J.~Lopez-Pavon,
%``Global constraints on heavy neutrino mixing,''
JHEP \textbf{08}, 033 (2016)
%doi:10.1007/JHEP08(2016)033
[arXiv:1605.08774 [hep-ph]].

\bibitem{Agostinho:2017wfs}
N.~R.~Agostinho, G.~C.~Branco, P.~M.~F.~Pereira, M.~N.~Rebelo and J.~I.~Silva-Marcos,
%``Can one have significant deviations from leptonic 3 $\times $ 3 unitarity in the framework of type I seesaw mechanism?,''
Eur. Phys. J. C \textbf{78}, no.11, 895 (2018)
%doi:10.1140/epjc/s10052-018-6347-2
[arXiv:1711.06229 [hep-ph]].

\bibitem{Dao:2021vqp}
T.~N.~Dao, M.~M\"uhlleitner and A.~V.~Phan,
%``Loop-corrected Higgs masses in the NMSSM with inverse seesaw mechanism,''
Eur. Phys. J. C \textbf{82} (2022) no.8, 667
%doi:10.1140/epjc/s10052-022-10590-9
[arXiv:2108.10088 [hep-ph]].

\bibitem{Biggio:2019eeo}
C.~Biggio, E.~Fernandez-Martinez, M.~Filaci, J.~Hernandez-Garcia and J.~Lopez-Pavon,
%``Global Bounds on the Type-III Seesaw,''
JHEP \textbf{05}, 022 (2020)
%doi:10.1007/JHEP05(2020)022
[arXiv:1911.11790 [hep-ph]].


%------experiment data cLFV

\bibitem{MEG:2016leq} 
A.~M.~Baldini \textit{et al.} [MEG],
%``Search for the lepton flavour violating decay $\mu ^+ \rightarrow \mathrm {e}^+ \gamma $ with the full dataset of the MEG experiment,''
Eur. Phys. J. C \textbf{76}, no.8, 434 (2016)
%doi:10.1140/epjc/s10052-016-4271-x
[arXiv:1605.05081 [hep-ex]].

\bibitem{BaBar:2009hkt}
B.~Aubert \textit{et al.} [BaBar],
%``Searches for Lepton Flavor Violation in the Decays tau+- ---\ensuremath{>} e+- gamma and tau+- ---\ensuremath{>} mu+- gamma,''
Phys. Rev. Lett. \textbf{104}, 021802 (2010)
%doi:10.1103/PhysRevLett.104.021802
[arXiv:0908.2381 [hep-ex]].

%--------
\bibitem{Jegerlehner:2009ry}
F.~Jegerlehner and A.~Nyffeler,
%``The Muon g-2,''
Phys. Rept. \textbf{477}, 1-110 (2009)
%doi:10.1016/j.physrep.2009.04.003
[arXiv:0902.3360 [hep-ph]].


\bibitem{Coutinho:2013lta}
Y.~A.~Coutinho, V.~Salustino Guimar\~aes and A.~A.~Nepomuceno,
%``Bounds on Z' from 3-3-1 model at the LHC energies,''
Phys. Rev. D \textbf{87}, no.11, 115014 (2013)
%doi:10.1103/PhysRevD.87.115014
[arXiv:1304.7907 [hep-ph]].

\bibitem{Salazar:2015gxa}
C.~Salazar, R.~H.~Benavides, W.~A.~Ponce and E.~Rojas,
%``LHC Constraints on 3-3-1 Models,''
JHEP \textbf{07}, 096 (2015)
%doi:10.1007/JHEP07(2015)096
[arXiv:1503.03519 [hep-ph]].


\bibitem{Nepomuceno:2019eaz}
A.~Nepomuceno and B.~Meirose,
%``Limits on 331 vector bosons from LHC proton collision data,''
Phys. Rev. D \textbf{101}, 035017 (2020)
%doi:10.1103/PhysRevD.101.035017
[arXiv:1911.12783 [hep-ph]].

%%%%%%%%%%%%%%%%%%%%%%%%%%





%%%%new relevant 
\bibitem{LHeCStudyGroup:2012zhm}
J.~L.~Abelleira Fernandez \textit{et al.} [LHeC Study Group],
%``A Large Hadron Electron Collider at CERN: Report on the Physics and Design Concepts for Machine and Detector,''
J. Phys. G \textbf{39}, 075001 (2012)
%doi:10.1088/0954-3899/39/7/075001
[arXiv:1206.2913 [physics.acc-ph]].

%----------Arindam  
\bibitem{Das:2012ze}
A.~Das and N.~Okada,
%``Inverse seesaw neutrino signatures at the LHC and ILC,''
Phys. Rev. D \textbf{88}, 113001 (2013)
%doi:10.1103/PhysRevD.88.113001
[arXiv:1207.3734 [hep-ph]].

\bibitem{Das:2014jxa}
A.~Das, P.~S.~Bhupal Dev and N.~Okada,
%``Direct bounds on electroweak scale pseudo-Dirac neutrinos from $\sqrt s=8$ TeV LHC data,''
Phys. Lett. B \textbf{735}, 364-370 (2014)
%doi:10.1016/j.physletb.2014.06.058
[arXiv:1405.0177 [hep-ph]].

\bibitem{Das:2015toa}
A.~Das and N.~Okada,
%``Improved bounds on the heavy neutrino productions at the LHC,''
Phys. Rev. D \textbf{93}, no.3, 033003 (2016)
%doi:10.1103/PhysRevD.93.033003
[arXiv:1510.04790 [hep-ph]].



\bibitem{Das:2016hof}
A.~Das, P.~Konar and S.~Majhi,
%``Production of Heavy neutrino in next-to-leading order QCD at the LHC and beyond,''
JHEP \textbf{06}, 019 (2016)
%doi:10.1007/JHEP06(2016)019
[arXiv:1604.00608 [hep-ph]].



\bibitem{Das:2018usr}
A.~Das, S.~Jana, S.~Mandal and S.~Nandi,
%``Probing right handed neutrinos at the LHeC and lepton colliders using fat jet signatures,''
Phys. Rev. D \textbf{99}, no.5, 055030 (2019)
%doi:10.1103/PhysRevD.99.055030
[arXiv:1811.04291 [hep-ph]].

%%---------------- updeated reference--------

\bibitem{Tully:2000kk}
M.~B.~Tully and G.~C.~Joshi,
%``Generating neutrino mass in the 331 model,''
Phys. Rev. D \textbf{64}, 011301 (2001)
%doi:10.1103/PhysRevD.64.011301
[arXiv:hep-ph/0011172 [hep-ph]].

\bibitem{Chang:2006aa}
D.~Chang and H.~N.~Long,
%``Interesting radiative patterns of neutrino mass in an SU(3)(C) x SU(3)(L) x U(1)(X) model with right-handed neutrinos,''
Phys. Rev. D \textbf{73}, 053006 (2006)
%doi:10.1103/PhysRevD.73.053006
[arXiv:hep-ph/0603098 [hep-ph]].

\bibitem{CarcamoHernandez:2017cwi}
A.~E.~C\'arcamo Hern\'andez, S.~Kovalenko, H.~N.~Long and I.~Schmidt,
%``A variant of 3-3-1 model for the generation of the SM fermion mass and mixing pattern,''
JHEP \textbf{07}, 144 (2018)
%doi:10.1007/JHEP07(2018)144
[arXiv:1705.09169 [hep-ph]].





\bibitem{Costantini:2020xrn}
A.~Costantini, M.~Ghezzi and G.~M.~Pruna,
%``Theoretical constraints on the Higgs potential of the general $331$ model,''
Phys. Lett. B \textbf{808}, 135638 (2020)
%doi:10.1016/j.physletb.2020.135638
[arXiv:2001.08550 [hep-ph]].


\bibitem{Tran:2020tsj}
H.~M.~Tran and Y.~Kurihara,
%``Tau $g-2$ at $e^-e^+$ colliders with momentum dependent form factor,''
Eur. Phys. J. C \textbf{81}, no.2, 108 (2021)
%doi:10.1140/epjc/s10052-021-08846-x
[arXiv:2006.00660 [hep-ph]].

\bibitem{Crivellin:2021spu}
A.~Crivellin, M.~Hoferichter and J.~M.~Roney,
``Towards testing the magnetic moment of the tau at one part per million,''
[arXiv:2111.10378 [hep-ph]].


\bibitem{Baker:2020vkh}
M.~J.~Baker, P.~Cox and R.~R.~Volkas,
%``Has the Origin of the Third-Family Fermion Masses been Determined?,''
JHEP \textbf{04}, 151 (2021)
%doi:10.1007/JHEP04(2021)151
[arXiv:2012.10458 [hep-ph]].

\bibitem{Baker:2021yli}
M.~J.~Baker, P.~Cox and R.~R.~Volkas,
%``Radiative muon mass models and $(g-2)_\mu$,''
JHEP \textbf{05}, 174 (2021)
%doi:10.1007/JHEP05(2021)174
[arXiv:2103.13401 [hep-ph]].

\bibitem{OPAL:1998dsa}
K.~Ackerstaff \textit{et al.} [OPAL],
%``An Upper limit on the anomalous magnetic moment of the tau lepton,''
Phys. Lett. B \textbf{431}, 188-198 (1998)
%doi:10.1016/S0370-2693(98)00520-6
[arXiv:hep-ex/9803020 [hep-ex]].

\bibitem{L3:1998lhr}
M.~Acciarri \textit{et al.} [L3],
%``Measurement of the weak dipole moments of the tau lepton,''
Phys. Lett. B \textbf{426}, 207-216 (1998)
%doi:10.1016/S0370-2693(98)00290-1

\bibitem{DELPHI:2003nah}
J.~Abdallah \textit{et al.} [DELPHI],
%``Study of tau-pair production in photon-photon collisions at LEP and limits on the anomalous electromagnetic moments of the tau lepton,''
Eur. Phys. J. C \textbf{35}, 159-170 (2004)
%doi:10.1140/epjc/s2004-01852-y
[arXiv:hep-ex/0406010 [hep-ex]].

\bibitem{Gonzalez-Sprinberg:2000lzf}
G.~A.~Gonzalez-Sprinberg, A.~Santamaria and J.~Vidal,
%``Model independent bounds on the tau lepton electromagnetic and weak magnetic moments,''
Nucl. Phys. B \textbf{582}, 3-18 (2000)
%doi:10.1016/S0550-3213(00)00275-3
[arXiv:hep-ph/0002203 [hep-ph]].

\bibitem{Eidelman:2016aih}
S.~Eidelman, D.~Epifanov, M.~Fael, L.~Mercolli and M.~Passera,
%``$\tau$ dipole moments via radiative leptonic $\tau$ decays,''
JHEP \textbf{03}, 140 (2016)
%doi:10.1007/JHEP03(2016)140
[arXiv:1601.07987 [hep-ph]].






%------------------------------------
\bibitem{DeConto:2015eia}
G.~De Conto, A.~C.~B.~Machado and V.~Pleitez,
%``Minimal 3-3-1 model with a spectator sextet,''
Phys. Rev. D \textbf{92}, no.7, 075031 (2015)
%doi:10.1103/PhysRevD.92.075031
[arXiv:1505.01343 [hep-ph]].

\bibitem{Faro:2019vcd}
F.~S.~Faro and I.~P.~Ivanov,
%``Boundedness from below in the $U(1)\times U(1)$ three-Higgs-doublet model,''
Phys. Rev. D \textbf{100} (2019) no.3, 035038
%doi:10.1103/PhysRevD.100.035038
[arXiv:1907.01963 [hep-ph]].

\bibitem{Kannike:2012pe}
K.~Kannike,
%``Vacuum Stability Conditions From Copositivity Criteria,''
Eur. Phys. J. C \textbf{72}, 2093 (2012)
%doi:10.1140/epjc/s10052-012-2093-z
[arXiv:1205.3781 [hep-ph]].

\bibitem{Maniatis:2006fs}
M.~Maniatis, A.~von Manteuffel, O.~Nachtmann and F.~Nagel,
%``Stability and symmetry breaking in the general two-Higgs-doublet model,''
Eur. Phys. J. C \textbf{48}, 805-823 (2006)
%doi:10.1140/epjc/s10052-006-0016-6
[arXiv:hep-ph/0605184 [hep-ph]].

\end{thebibliography}
%
%
\section{ \label{app_cHiggs}Masses and mixing of the singly charged Higgs bosons}
From the three relations corresponding to the minimal conditions of the Higgs potential \eqref{eq_hpo1}, three parameters $\mu_{1,2,3}$ are written in terms of the remaining Higgs potential couplings and non-zero vevs of the neutral Higgs components, namely
\begin{align}
	\label{eq_minEq}	
	\mu _1^2&= -\frac{f c_{\beta } u}{s_{\beta }}-\frac{1}{2} c_{\beta }^2 \lambda _{12} v^2-\frac{\lambda _{13} u^2}{2}-\lambda _1 s_{\beta }^2 v^2,
	\crn \mu _2^2&= -\frac{f s_{\beta } u}{c_{\beta }}-c_{\beta }^2 \lambda _2 v^2-\frac{\lambda _{23} u^2}{2}-\frac{1}{2} \lambda _{12} s_{\beta }^2 v^2,
	\crn \mu _3^2& -\frac{f c_{\beta } s_{\beta } v^2}{u}-\frac{1}{2} c_{\beta }^2 \lambda _{23} v^2-\lambda _3 u^2-\frac{1}{2} \lambda _{13} s_{\beta }^2 v^2. 
\end{align}
Inserting these relations into the Higgs potential \eqref{eq_hpo1}, we obtain the squared mass matrix of the singly charged Higgs bosons in the basis $(\rho^\pm,\; \eta^\pm,\; h^\pm)^T$ as follows 
\begin{align}
	\mathcal{M}^2_c=	\left(
	\begin{array}{ccc}
		\frac{c_{\beta } \left(c_{\beta } \tilde{\lambda }_{12} s_{\beta } v^2-2 f u\right)}{2 s_{\beta }} & \frac{1}{2} c_{\beta } \tilde{\lambda }_{12} s_{\beta } v^2-f u & \frac{c_{\beta } f_h v}{\sqrt{2}} \\
		\frac{1}{2} c_{\beta } \tilde{\lambda }_{12} s_{\beta } v^2-f u & \frac{s_{\beta } \left(c_{\beta } \tilde{\lambda }_{12} s_{\beta } v^2-2 f u\right)}{2 c_{\beta }} & \frac{f_h s_{\beta } v}{\sqrt{2}} \\
		\frac{c_{\beta } f_h v}{\sqrt{2}} & \frac{f_h s_{\beta } v}{\sqrt{2}} & \frac{1}{2} \left(\lambda _3^h u^2+\left(\lambda _2^h c_{\beta }^2+s_{\beta }^2 \lambda _1^h\right) v^2+2 \mu _4^2\right) \\
	\end{array}
	\right).
\end{align}
Diagonalizing this matrix will result in a zero eigenvalue and two massive ones denoted as $m^2_{H^\pm_{12}}$, which correspond to a goldstone boson $\phi^\pm_W$ and two physical singly charged Higgs bosons $H^\pm_{1,2}$.  The mixing matrix $C$ used to diagonalize $C\mathcal{M}^2_cC^T= \mathrm{diag}\left( 0,\; m^2_{H^\pm_1},\;m^2_{H^\pm_2}\right)$ can be written as a product of the  two unitary transformations   $C=C_2C_1$, where
\begin{align}
	C_1\equiv \left(
	\begin{array}{ccc}
		-s_{\beta } & c_{\beta } & 0 \\
		c_{\beta } & s_{\beta } & 0 \\
		0 & 0 & 1 \\
	\end{array}
	\right),\; C_2\equiv \left(
	\begin{array}{ccc}
		1 & 0 & 0 \\
		0 & c_{\alpha } & -s_{\alpha } \\
		0 & s_{\alpha } & c_{\alpha } \\
	\end{array}
	\right).
\end{align}
The  $C_1$ was introduced  previously \cite{Diaz:2004fs, Buras:2012dp} corresponding to the decoupling limit between $h^\pm$ and two Higgs triplets $\rho^\pm$ and $\eta^\pm$, namely the matrix 
\begin{align}
	\label{eq_Mc1}
	C_1	\mathcal{M}^2_cC_1^T=\mathcal{M}^2_{c,1}=\left(
	\begin{array}{ccc}
		0 & 0 & 0 \\
		0 & \frac{\tilde{\lambda }_{12} v^2}{2}-\frac{f u}{c_{\beta } s_{\beta }} & \frac{f_h v}{\sqrt{2}} \\
		0 & \frac{f_h v}{\sqrt{2}} & \frac{1}{2} \left(\lambda _3^h u^2+\left(\lambda _2^h c_{\beta }^2+s_{\beta }^2 \lambda _1^h\right) v^2+2 \mu _4^2\right) \\
	\end{array}
	\right),
\end{align}
which is diagonal when triple coupling $f_h=0$.  In contrast, $C_2$ presents the mixing between the singlet $h^\pm$ and two singly charged Higgs components of the two Higgs triplets,  $C_2	\mathcal{M}^2_{c,1}C_2^T= \mathrm{diag}\left( 0,\; m^2_{H^\pm_1},\;m^2_{H^\pm_2}\right)$.
It can be proved that
\begin{align} \label{eq_alpha}
	\tan(2\alpha)=  \frac{2 \sqrt{2} c_{\beta } f_h s_{\beta } v}{2 f u+c_{\beta }^3 s_{\beta } \lambda _2^h v^2+c_{\beta } s_{\beta } \left[v^2 \left( s_{\beta }^2 \lambda _1^h-\tilde{\lambda
		}_{12}\right)+\lambda _3^h u^2+2 \mu _4^2\right]},
\end{align}
and  $m_{H^\pm_{1,2}}$ are functions of parameters included in the matrix $\mathcal{M}^2_{c,1}$ including $f_h,\;f$, and $ \mu_4$. On the other hand, we can choose  $m_{H^\pm_{1,2}}$ and $\alpha$ as free parameters while $f_h,\;f$, and $ \mu_4$ are dependent parameters,  their analytic  formulas are given in Eqs. \eqref{eq_Higgscouplings}.

\end{document}